\pdfoutput=1
\documentclass[PAPER, cernpreprint, atlasdraft=false, UKenglish, texmf, orcidlogo]{atlasdoc}
\usepackage[subcaption]{atlaspackage}
\usepackage{atlasbiblatex}

\usepackage{atlasphysics}

\usepackage{longtable}

\usepackage{changepage}
\usepackage{color, colortbl}
\usepackage{pdflscape}
\usepackage[final]{listings}
\usepackage{multirow}
\usepackage{enumitem}

\usepackage{floatrow}
\crefname{figure}{Figure}{Figures}
\Crefname{figure}{Figure}{Figures}
\crefname{equation}{Eq.}{Eqs.}
\Crefname{equation}{Eq.}{Eqs.}

\graphicspath{{logos/}{figures/}}

\usepackage{ANA-TOPQ-2020-14-PAPER-defs}

\AtlasTitle{Measurements of differential cross-sections of $WbWb$ production in the dilepton channel in $pp$ collisions at $\sqrt{s}$ = 13 TeV using the ATLAS detector}

\AtlasAbstract{%
At the Large Hadron Collider, the $WbWb$ final state is expected to be dominated by $t\bar{t}$ production with a contribution from single-top processes. Differential cross-sections for $WbWb$ production in the dilepton decay channel are measured at the particle level as a function of various kinematic variables. The analysis is based on data from proton--proton collisions at a centre-of-mass energy of $\sqrt{s} = 13$ TeV, recorded by the ATLAS detector at the Large Hadron Collider over the period from 2015 to 2018, corresponding to an integrated luminosity of 140 fb$^{-1}$. Measurements are performed within the fiducial phase-space defined by the presence of two $b$-jets and one electron and one muon of opposite charges. The differential cross-sections are corrected for detector effects and unfolded to the particle level. Results are compared with predictions from Monte Carlo event generators at next-to-leading order in perturbative quantum chromodynamics; overall the measurements are in reasonable agreement with several generator setups, although no single prediction is able to describe all measured distributions simultaneously. These measurements provide valuable constraints on the modelling of $WbWb$ production and the interference between doubly resonant and singly resonant $WbWb$ production.
}

\AtlasRefCode{TOPQ-2020-14}

\PreprintIdNumber{CERN-EP-2025-094}

\AtlasJournal{JHEP}
\AtlasJournalRef{\JHEP 02 (2026) 153}
\AtlasDOI{10.1007/JHEP02(2026)153}

\hypersetup{pdftitle={ATLAS document},pdfauthor={The ATLAS Collaboration}}

\begin{document}

\maketitle

\tableofcontents

\clearpage

\clearpage

\clearpage
%

%
\section{Introduction}%
\begin{figure}[bh]
\centering
\begin{subfigure}[b]{0.45\linewidth}
\centering
\includegraphics[width=\linewidth]{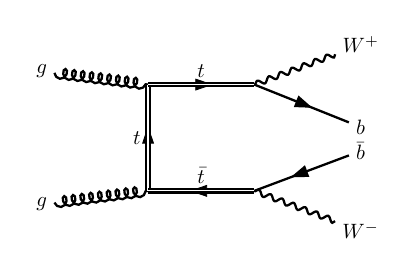}
\caption{}\label{fig:ttbar_diagram}
\end{subfigure}
\begin{subfigure}[b]{0.45\linewidth}
\centering
\includegraphics[width=\linewidth]{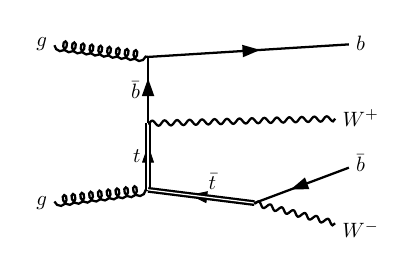}
\caption{}\label{fig:tWb_diagram}
\end{subfigure}
\caption{Representative Feynman diagrams corresponding to gluon-initiated \protect\subref{fig:ttbar_diagram} \ttbar\ and \protect\subref{fig:tWb_diagram} $tWb$ single-top-quark processes. The top quarks are indicated by double fermion lines.}
\label{fig:basic_feynman_diagrams}
\end{figure}

\label{sec:introduction}
Top-quark pair (\ttbar) production is one of the most extensively studied processes at the Large Hadron Collider (LHC)~\cite{Evans:2008zzb} and constitutes a critical background for many searches for physics beyond the Standard Model (BSM). The differential cross-sections for \ttbar\ production have been both measured~\cite{TOPQ-2018-15,TOPQ-2018-26,TOPQ-2020-09,CMS-TOP-20-001,CMS-TOP-18-004} and calculated~\cite{Beneke:2011mq,Cacciari:2011hy,Baernreuther:2012ws,Czakon:2012zr,Czakon:2012pz,Czakon:2013goa,Czakon:2011xx,Catani:2019hip} with high precision across a broad kinematic range. However, these typically rely on the narrow-width approximation for top-quark decays, effectively separating \ttbar\ production from the production of a single top quark in association with a $W$ boson and a $b$-quark ($tWb$). Due to their identical \wbwb\ final states, processes with one or two time-like top-quark propagators (referred to as singly and doubly resonant, respectively) interfere. This interference becomes evident when comparing leading-order (LO) \ttbar\ production with next-to-leading-order (NLO) $tW$ production, as illustrated in \cref{fig:basic_feynman_diagrams}.

This quantum interference feature has been extensively studied from a theoretical perspective~\cite{White:2009yt,Frixione:2008yi,Hollik:2012rc,Demartin:2016axk,ATL-PHYS-PUB-2016-020} and provides a significant source of uncertainty for measurements of many Standard Model (SM) processes and BSM searches. Fixed-order calculations of the full next-to-leading-order $pp \rightarrow l^{+} \nu l^{-} \bar{\nu} b \bar{b}$ process, where $l$ denotes a lepton of any flavour, include the correct treatment of the interference between \ttbar{} and $tW$ and have set the stage for corresponding predictions matched to a parton shower~\cite{Ravindran:2002dc,Denner:2012yc,Bevilacqua:2010qb,Cascioli:2013wga,Frederix:2013gra,Jezo:2016ujg}. While ATLAS previously measured this interference-sensitive region using \SI{36.1}{\ifb} of data collected at $\sqrt{s}=\SI{13}{\TeV}$~\cite{TOPQ-2017-05}, the present work significantly enhances the sensitivity by leveraging the full Run 2 data sample of \SI{140}{\ifb} collected by the ATLAS detector between 2015 and 2018. This larger data sample, coupled with optimised event selection, improved calibrations, and refined modelling of uncertainties, achieves a substantial reduction in the total measurement uncertainty.
The analysis focuses on events with one electron and one muon, which provide a clean signature for studying these processes.

Several observables are measured for the first time for the combined \ttbar\ and $tW$ signal process. These measurements are crucial for validating the modelling of fundamental kinematic distributions in Monte Carlo (MC) generators that account for interference effects in \wbwb\ final states.

\clearpage

%

%
\section{The ATLAS detector}
\label{sec:detector}

The ATLAS detector~\cite{PERF-2007-01} at the LHC covers nearly the entire solid angle around the collision point.\footnote{\AtlasCoordFootnote}
It consists of an inner tracking detector surrounded by a thin superconducting solenoid, electromagnetic and hadronic calorimeters,
and a muon spectrometer incorporating three large superconducting air-core toroidal magnets.

The inner-detector system (ID) is immersed in a \qty{2}{\tesla} axial magnetic field
and provides charged-particle tracking in the range \(|\eta| < 2.5\).
The high-granularity silicon pixel detector covers the vertex region and typically provides four measurements per track,
the first hit generally being in the insertable B-layer installed before Run~2~\cite{ATLAS-TDR-19,PIX-2018-001}.
It is followed by the SemiConductor Tracker, which usually provides eight measurements per track.
These silicon detectors are complemented by the transition radiation tracker (TRT),
which enables radially extended track reconstruction up to \(|\eta| = 2.0\).
The TRT also provides electron identification information
based on the fraction of hits (typically 30 in total) above a higher energy-deposit threshold corresponding to transition radiation.

The calorimeter system covers the pseudorapidity range \(|\eta| < 4.9\).
Within the region \(|\eta|< 3.2\), electromagnetic calorimetry is provided by barrel and
endcap high-granularity lead/liquid-argon (LAr) calorimeters,
with an additional thin LAr presampler covering \(|\eta| < 1.8\)
to correct for energy loss in material upstream of the calorimeters.
Hadronic calorimetry is provided by the steel/scintillator-tile calorimeter,
segmented into three barrel structures within \(|\eta| < 1.7\), and two copper/LAr hadronic endcap calorimeters.
The solid angle coverage is completed with forward copper/LAr and tungsten/LAr calorimeter modules
optimised for electromagnetic and hadronic energy measurements, respectively.

The muon spectrometer comprises separate trigger and
high-precision tracking chambers measuring the deflection of muons in a magnetic field generated by the superconducting air-core toroidal magnets.
The field integral of the toroids ranges between \num{2.0} and \qty{6.0}{\tesla\metre}
across most of the detector.
Three layers of precision chambers, each consisting of layers of monitored drift tubes, cover the region \(|\eta| < 2.7\),
complemented by cathode-strip chambers in the forward region, where the background is highest.
The muon trigger system covers the range \(|\eta| < 2.4\) with resistive-plate chambers in the barrel, and thin-gap chambers in the endcap regions.

The luminosity is measured mainly by the LUCID--2~\cite{LUCID2} detector that records Cherenkov light produced in the quartz windows of photomultipliers located close to the beampipe.

Events were selected by the first-level trigger system implemented in custom hardware,
followed by selections made by algorithms implemented in software in the high-level trigger~\cite{TRIG-2016-01}.
The first-level trigger accepted events from the \qty{40}{\MHz} bunch crossings at a rate close to \qty{100}{\kHz},
which the high-level trigger further reduced in order to record complete events to disk at about \qty{1.25}{\kHz}.

A software suite~\cite{SOFT-2022-02} is used in data simulation, in the reconstruction
and analysis of real and simulated data, in detector operations, and in the trigger and data acquisition
systems of the experiment.


%

%
\section{Data and Monte Carlo samples}%
\label{sec:datamc}
This analysis uses the proton--proton (\(pp\)) collision data at a centre-of-mass energy of \SI{13}{\TeV} collected by the ATLAS detector during Run 2 operations (2015--2018). This data was recorded during stable beam LHC operations with the ATLAS detector fully functioning. After all quality criteria are applied~\cite{DAPR-2018-01}, the recorded data corresponds to an integrated luminosity of \SI{140}{\ifb}~\cite{DAPR-2021-01}.

\FloatBarrier
\subsection{MC simulated samples}
\label{sec:datamc:mc}

Various MC generators were employed to produce simulated events, which are used to estimate contributions from signal and background processes, determine detector resolution and acceptance correction factors, and evaluate systematic uncertainties. Additionally, MC generators facilitate comparisons between unfolded data and theoretical predictions. All samples were normalised according to the best available theoretical cross-section predictions, as detailed in the following sections.

The responses of the ATLAS detector and trigger were simulated~\cite{SOFT-2010-01} based on the detailed model implemented in the \GEANT\ framework~\cite{Agostinelli:2002hh}. The ATLAS fast-simulation package \textsc{AtlFast-II}, which parametrised hadronic showers in the electromagnetic and hadronic calorimeters~\cite{ATL-SOFT-PUB-2014-001}, was employed to estimate some of the signal modelling uncertainties.

The effect of multiple interactions in the same and neighbouring bunch crossings (\pileup) was modelled by overlaying the simulated hard-scattering event with inelastic proton--proton  events generated with \PYTHIA[8.186]~\cite{Sjostrand:2007gs} using the \NNPDF[2.3lo] set of parton distribution functions (PDF)~\cite{Ball:2012cx} and the A3 set of tuned parameters~\cite{ATL-PHYS-PUB-2016-017} (tune). The MC events were weighted to reproduce the distribution of the average number of interactions per bunch crossing (\(\left<\mu \right>\)) observed in the data. The \(\left<\mu \right>\) value in data was rescaled by a factor of \(1.03\pm 0.04\) to improve agreement between data and simulation in both charged-particle track distributions and the visible inelastic \(pp\) cross-section~\cite{STDM-2015-05}.

All simulated events were processed using the same reconstruction algorithms and analysis chain as the data, and corrections were applied so that the object reconstruction and identification efficiencies, energy scales and energy resolutions in simulation matched those determined from data.

If not stated otherwise, some common settings were defined for the different MC simulations and theoretical calculations. The top-quark mass $\mtop$ was set to $172.5\,\GeV$. The \PYTHIA[8] parton shower (PS) used the A14 set of tuned parameters~\cite{ATL-PHYS-PUB-2014-021} and the \NNPDF[2.3lo] PDF set. \EVTGEN~\cite{Lange:2001uf} was used for the decays of the bottom and charm hadrons for the samples showered with \PYTHIA[8] and \HERWIG[7]~\cite{Bellm:2015jjp,Bahr:2008pv}.

\subsubsection{Nominal signal sample}
\label{sec:datamc:mc_nominal_signal}
MC samples produced with specific settings and parameters are used as the nominal signal sample for the analysis. The signal samples are defined as a combination of the \ttbar{} and $tW$ samples from the same generators and parton showers. Contributions from non-resonant $WWbb$ production are negligible~\cite{Demartin:2016axk} and, if not explicitly stated, are not included in the simulated samples. This is cross-checked with the \POWPY[8] $bb4l$ sample described below.

At NLO, it is crucial to avoid double-counting $tWb$ events with $m_{bW} \sim \mtop$, whose diagrams are already included, at NLO, in the definition of the \ttbar\ sample. The default scheme employed in this analysis is diagram removal (DR)~\cite{Frixione:2008yi}, where all doubly resonant amplitudes are excluded from the $tW$ sample. Alternative methods~\cite{Frixione:2008yi,Hollik:2012rc,ATL-PHYS-PUB-2021-042} include diagram subtraction (DS), which subtracts gauge-invariant terms corresponding to doubly resonant contributions rather than removing them. This approach preserves a more accurate representation of interference effects between the $\ttbar$ and $tW$ processes.  In the rest of the paper, unless explicitly stated, the $tW$ samples were generated using the DR scheme.

In this analysis, the nominal \ttbar\ and $tW$ were both generated with \POWHEGBOX[v2]~\cite{Frixione:2007nw,Nason:2004rx,Frixione:2007vw,Alioli:2010xd} and interfaced to \PYTHIA[8]~\cite{Sjostrand:2014zea}, with the settings and parameters described below, and are referred to as `PWG+PY8'.

\paragraph{\ttbar\ production}
The production of \ttbar\ events was simulated using the
\POWHEGBOX[v2] generator, which provides matrix elements at next-to-leading
order in the strong coupling constant \alphas with the
\NNPDF[3.0nlo]~\cite{Ball:2014uwa} PDF set.
The \hdamp\ parameter, which controls the matching in \POWHEG and
effectively regulates the high transverse momentum (\pt) radiation against which the
\ttbar\ system recoils, was set to 1.5~\mtop~\cite{ATL-PHYS-PUB-2016-020}.
The functional form of the renormalisation scale \muR\ and factorisation scale \muF\  was
set to the default scale $\sqrt{\mtop^2 + \pTX[2]}$, where $\pTX[2]$ is evaluated  with the kinematics of the underlying $2\rightarrow2$ \POWHEG\ process.
The events were interfaced to \PYTHIA[8.230] for the parton shower and hadronisation. The \pthard\ parameter, which regulates how the \PYTHIA[8] radiation phase space is determined to avoid overlaps with the regions of the phase space already covered by \POWHEG~\cite{Bierlich:2022pfr}, was set to $\pthard=0$. The modelling of second and subsequent gluon emission in the $t \to Wb$ process~was performed by recoiling the gluon against $b$-quarks, which is the default strategy in \POWHEG\cite{Brooks:2019xso}.

The nominal \ttbar\ sample was normalised to the cross-section prediction at next-to-next-to-leading order (NNLO)
in perturbative quantum chromodynamics (QCD) including the resummation of next-to-next-to-leading logarithmic (NNLL) soft-gluon terms calculated using
\textsc{Top++2.0}~\cite{Beneke:2011mq,Cacciari:2011hy,Baernreuther:2012ws,Czakon:2012zr,Czakon:2012pz,Czakon:2013goa,Czakon:2011xx} with the \PDFforLHC[21] PDF set~\cite{Ball:2022oua} in the five-flavour scheme.
For $pp$ collisions at a centre-of-mass energy of \rts~=~\SI{13}{\TeV}, this cross-section is
$\sigma(\ttbar)_{\textrm{NNLO+NNLL}} = 834^{\,+21}_{-30}(\mathrm{scale})\pm21(\mathrm{PDF+}\alphas)\pm23(\mtop)$~pb.
The uncertainties in the cross-section due to the PDF and \alphas\ are calculated using the \PDFforLHC\ prescription
with the \MSHT~\cite{Bailey:2020ooq}, \CT[18]~\cite{Hou:2019efy} and \NNPDF[3.1]~\cite{NNPDF:2017mvq} PDF sets. The uncertainties in the cross-section due to the choice of the renormalisation and factorisation scales are determined by varying the scales independently up and down by a factor of two, while ensuring that they do not differ from each other by more than a factor of two. This procedure is used consistently throughout the analysis whenever scale uncertainties are evaluated. The scale uncertainty is defined as the envelope of the resulting cross-section values. The uncertainties due to the top-quark mass \mtop\ are evaluated by varying \mtop\ by $\pm \SI{1.0}{\GeV}$.

\paragraph{$tW$ production}
The nominal single-top-quark $tW$ associated production was simulated using the
\POWHEGBOX[v2]~\cite{Re:2010bp}
generator that provides matrix elements at NLO in QCD in the five-flavour
scheme, with the \NNPDF[3.0nlo] PDF set. The functional form of the
renormalisation and factorisation scales was set to $H_\mathrm{T}/2$, where $H_\mathrm{T}$ is defined as the scalar sum of the \pT\ of all outgoing partons.
The DR scheme was employed to treat the interference with \ttbar\ production. The events were interfaced to \PYTHIA[8.230]. As in the nominal \ttbar\ sample, the \pthard\ parameter was set to $\pthard=0$.

The predicted cross-section of $tW$ production in $pp$ collisions at a centre-of-mass energy of \SI{13}{\TeV} is $\sigma(tW)=79.3\pm2.8$~pb and was computed at NLO in QCD with the addition of third-order corrections to soft-gluon emissions by resumming NNLL terms~\cite{Kidonakis:2021vob}. The \PDFforLHC[21] PDF set was used. The quoted uncertainty includes the uncertainty due to the choice of \muR\ and  \muF\ and the uncertainty in the PDFs. The uncertainty from the scale choice is determined by varying the scales simultaneously up and down by a factor of two. The PDF uncertainties are based on the Hessian method and include the uncertainty in $\alphas$.

The perturbative orders at which the total cross-sections are calculated differ between the \ttbar\ and $tW$ processes. The \ttbar\ cross-section is calculated at NNLO+NNLL, while the $tW$ cross-section is calculated at NLO+NNLL\@. The impact of this difference on the measurement is included as a source of systematic uncertainty. The impact on the predictions of the normalised differential cross-sections is found to be negligible by comparing the predictions obtained with the nominal settings (with the \ttbar\ and $tW$ processes normalised to NNLO+NNLL and NLO+NNLL, respectively) to the predictions obtained without the higher-order corrections (with the \ttbar\ and $tW$ processes normalised to NLO).

\subsubsection{Alternative signal samples}
\label{sec:datamc:mc_alt_signal}

Several additional \ttbar\ and $tW$ samples are considered for evaluating modelling systematic uncertainties and for comparison with data.
\paragraph{Samples for systematic uncertainties}
For the estimation of signal modelling uncertainties different samples are used, obtained with either different settings for the nominal \POWPY[8] generator or with different generators. The predictions obtained with these samples, with the exception of the samples obtained by varying the top-quark mass, are also used in the comparisons with the measured differential cross-sections. These samples are summarised below.
\begin{description}[style=unboxed,leftmargin=0cm]
\item[$tW$ PWG+PY8 (DS)] This sample is similar to the nominal PWG+PY8 sample but with DS scheme used for the modelling of the $tW$ component. In the rest of the paper, this sample is referred to as `PWG+PY8 (DS)'.
\item[\ttbar\ and $tW$ PWG+PY8 ($h_{\mathrm{damp}} = 3 \mtop$)] This sample was generated with the same settings as the nominal PWG+PY8 sample but with the $h_{\mathrm{damp}}$ parameter set to $3\mtop$. This sample is referred to as `PWG+PY8 ($h_{\mathrm{damp}}$)'.
\item[\ttbar\ and $tW$ PWG+H7] This sample consists of \ttbar\ events simulated using the same \POWHEGBOX[v2] set-up with the same settings as the nominal sample but interfaced to \HERWIG[7.1.3], which uses an angular-ordered parton-shower model, with the \MMHT[lo] PDF set~\cite{Harland-Lang:2014zoa}. This sample used the \HERWIG[7.1] default set of tuned parameters. For the $tW$ process, events simulated with the \POWHEGBOX[v2] generator using the same settings as the nominal sample were interfaced to \HERWIG\ with the same configuration set-up as the \ttbar\ sample. The DR scheme was employed to treat the interference with \ttbar production. This sample is referred to as `PWG+H7'.
\item[\ttbar\ and $tW$ PWG+PY8 ($\pthard=1$)] This sample is similar to the nominal PWG+PY8 sample but used $\pthard = 1$ instead of $\pthard = 0$~\cite{Hoche:2021mkv,ATL-PHYS-PUB-2023-029}. It is referred to as `PWG+PY8 ($\pthard$)'.

\item[\ttbar\ and $tW$ PWG+PY8 (mass variations)] These samples are based on the nominal PWG+PY8 set-up but incorporate variations in the top-quark mass by $\pm3.5$~\GeV around the nominal mass of $172.5$~\GeV. These variations are used to assess the impact on the measured differential cross-sections of the assumed top-quark mass in the simulated \ttbar\ and $tW$ processes. These samples are utilised solely for evaluating systematic uncertainties, as described in \cref{sec:syst:modelling}, and are not used for direct comparisons with data.
\end{description}
\paragraph{Samples only for comparison with data}

Additional samples are considered solely for comparison with the measured differential cross-sections. These samples are summarised below.
\begin{description}[style=unboxed,leftmargin=0cm]
\item[\ttbar\ PWG+PY8 (recoil)]         This sample is similar to the nominal \ttbar\ PWG+PY8 sample but uses a different recoil strategy. In this variation, the gluon recoils against the top-quark itself during the second and subsequent gluon emissions in the $t \to Wb$ process~\cite{TOPQ-2017-17}. The $tW$ samples were kept at their nominal settings. This sample is referred to as `PWG+PY8 (recoil)'.
\item[PWG+PY8 $bb4l$] This sample was produced using a generalisation of \POWHEG\ that generates $l^{+} \nu l^{-} \bar{\nu} b \bar{b}$ final states. This approach accounts for quantum interference effects between the \ttbar\ and $tW$ production modes, and off-shell and non-resonant effects~\cite{Jezo:2016ujg}. Events were simulated using the  \POWHEGBOXRES\ framework~\cite{Jezo:2018yaf}, with matrix elements at NLO in QCD with the \NNPDF[3.0nlo] PDF set in the four-flavour scheme. A general NLO+PS matching technique that allows a consistent treatment of resonances, referred to as `resonance-aware matching'~\cite{Jezo:2015aia,Jezo:2023rht}, was employed. The renormalisation and factorisation scales were set using the functional form:
$$\left[\left(\mtop^2 + p_{\text{T},t}^2\right) \left(\mtop^2 + p_{\text{T},\bar{t}}^2\right)\right]^{\frac{1}{4}},$$
where the masses and transverse momenta of the (anti)top-quarks are defined in the underlying Born phase space, based on the final state (off-shell) decay products. For diagrams involving an intermediate $Z$ boson, the scales were set to $\frac{\sqrt{p_Z^2}}{2}$, where $\mathbf{p}_Z = \mathbf{p}_{l^+} + \mathbf{p}_{l^-} + \mathbf{p}_{\nu} + \mathbf{p}_{\bar{\nu}}$. The \hdamp\ parameter was set to $1.5 \, \mtop$. For the \POWHEG--\PYTHIA matching, a special UserHook (\texttt{PowhegBB4ltms}) was employed~\cite{ATL-PHYS-PUB-2021-042}. This generator is still under review within the ATLAS experiment, and a complete set of systematic uncertainties is not yet available; therefore, it cannot be used as a nominal sample.
\item[\ttbar\ NNLO reweighting] To assess the impact of NNLO corrections on the agreement between data and simulation, the nominal PWG+PY8 \ttbar\ sample was reweighted at parton level to match the NNLO QCD + NLO electroweak (EW) parton-level predictions presented in Ref.~\cite{Czakon:2017wor}. The reweighting was applied as a function of three variables: $\pt(t)$, $m(\ttbar)$, and $\pt(\ttbar)$, using the kinematic distributions of the top quarks in the MC samples after initial- and final-state radiation. The predictions for $\pt(t)$ and $m(\ttbar)$ were calculated at NNLO in QCD with NLO EW corrections using the \NNPDF[3.0qed] PDF set and dynamic renormalisation and factorisation scales of $\mT(t)/2$ for $\pt(t)$ and $\HT/4$ for $m(\ttbar)$, as proposed in Ref.~\cite{Czakon:2017wor}. The prediction for $\pt(\ttbar)$ was calculated at NNLO in QCD~\cite{Catani:2019hip,Grazzini:2017mhc} with the \NNPDF[3.0] PDF set and \muR\ and \muF\ set to $\HT/4$. All predictions assume $\mtop=173.3$~\GeV. The reweighting was performed iteratively, ensuring that the reweighted MC sample agrees well with the higher-order predictions for each variable. The reweighted predictions are not equivalent to full NNLO+PS calculations but are used to estimate the effect of NNLO contributions on the measured observables. Because higher-order predictions for the $tW$ process are unavailable, the NNLO reweighting cannot be applied to $tW$, and the analysis therefore uses the nominal $tW$ PWG+PY8 sample produced with the DR scheme. For this reason, this sample, along with the \minnlo\ sample described below, improves the modelling accuracy of the \ttbar\ component of the $WbWb$ process but it does not   enhance the $tW$ component. This sample is referred to as `PWG+PY8 (NNLO rew.)'.
\item[\ttbar\ NNLO+PS] The first matched computation of top-quark pair production at NNLO in QCD, incorporating all-order radiative corrections through parton-shower simulations, is detailed in Ref.~\cite{Mazzitelli:2020jio} and referred to as the \minnlo\ procedure. For generating this sample, the \NNPDF[3.0 nnlo] PDF set was utilised, with \PYTHIA[8.230] used for the fragmentation and underlying-event simulation. This sample was normalised to the same cross-section calculated with the Top++ 2.0 program at NNLO+NNLL in QCD, as previously mentioned. For the $tW$ process, the nominal sample is used. This sample is referred to as `PWG+PY8 \minnlo'.
\end{description}

All the alternative samples were normalised to the same total cross-section as the nominal sample, with two exceptions: the PWG+PY8 mass variation samples, which were normalised to their respective higher-order cross-section calculations, and the PWG+PY8 $bb4l$ sample, which was normalised to the cross-section of $84.2$~pb obtained directly from the MC simulation without corrections for higher-order calculations.

\subsubsection{Background samples}
\label{sec:datamc:mc_background}
Several processes share the same final state (two $b$-quarks, two charged leptons) as the \wbwb\ process. The events produced by these backgrounds need to be estimated and subsequently subtracted from the data to determine the \wbwb\ cross-sections. They are all estimated by using MC simulation. The processes considered are $Z$+jets production, diboson final states ($WW/ZZ/WZ$), $\ttbar{}$ produced in association with weak bosons ($\ttbar+W/Z/H$, collectively referred to as \ttX) and, for the background due to fake and non-prompt leptons, single-lepton \ttbar, $tW$ and $W$+jets processes.

The production of $Z$+jets and $W$+jets events was simulated with the
\SHERPA[2.2.1]~\cite{Bothmann:2019yzt} generator using NLO matrix elements for up to two partons, and LO matrix elements for up to four partons calculated with the \COMIX~\cite{Gleisberg:2008fv} and \OPENLOOPS~\cite{Buccioni:2019sur,Cascioli:2011va,Denner:2016kdg} libraries. They were matched with the \SHERPA\ parton shower~\cite{Schumann:2007mg} based on Catani--Seymour dipole factorisation using the \MEPSatNLO prescription~\cite{Hoeche:2011fd,Hoeche:2012yf,Catani:2001cc,Hoeche:2009rj} using the set of tuned parameters developed by the \SHERPA authors.
The \NNPDF[3.0nnlo] set of PDFs was used and the samples
were normalised to an NNLO
prediction~\cite{Anastasiou:2003ds}. The $Z$+jets events lead to the same final state as the signal process when the $Z$ boson decays into two $\tau$-leptons, which in turn decay into an electron and a muon; on the other hand, $W$+jets events can mimic the signal signature only via the presence of fake or non-prompt leptons. The contribution from both processes is negligible in the signal regions considered.

The diboson processes were simulated with the \SHERPA[2.2.2] and 2.2.1 generators including off-shell effects and Higgs boson contributions, where appropriate. The samples were generated by using matrix elements at NLO accuracy in QCD for up to one additional parton and at LO accuracy for up to three additional parton emissions. Samples for the loop-induced processes \(gg \to VV\) were generated by using LO-accurate matrix elements for up to one additional parton emission for both fully leptonic and
semileptonic final states.
In this set-up, multiple matrix elements (NLO for up to one additional parton and LO for up to three additional partons) were matched and merged with the \SHERPA\ parton shower. The virtual QCD corrections for matrix elements at NLO accuracy were provided by the \OPENLOOPS library. The
\NNPDF[3.0nnlo] set of PDFs was used, along with a
dedicated set of tuned parton-shower parameters developed by the
\SHERPA\ authors.

The production of $\ttbar+W/Z$ events was simulated using the
\MGNLO[2.3.3]~\cite{Alwall:2014hca} generator at NLO with the
\NNPDF[3.0nlo] PDF set and interfaced to \PYTHIA[8.210].

The production of \ttH events was simulated using the
\POWHEGBOX[v2]~\cite{Hartanto:2015uka}
generator at NLO with the \NNPDF[3.0nlo] PDF set interfaced to \PYTHIA[8.230].

The background arising from events containing fake or non-prompt leptons, referred to as `Fakes', is estimated by using MC simulations. Specifically, these are events where at least one reconstructed lepton does not match a generated lepton from the hard-scattering process. The samples used for this estimation include single-lepton \ttbar, $tW$, and $W$+jets processes, generated by using the same set-ups described earlier.


%

\section{Object reconstruction and event selection}
\label{sec:reco}

The following sections describe the detector- and particle-level objects used to characterise the final-state event topology and to define the fiducial phase-space regions for the measurements.

\subsection{Detector-level object reconstruction}\label{sec:obj_reco}
Events are required to have a primary vertex with at least two associated tracks.
The primary vertex is selected as the one with the largest \(\Sigma \pT^2\),
where the sum is over all tracks with transverse momentum \(\pT > \qty{0.5}{\GeV}\) that are associated with the vertex. Events are considered only if they were accepted by at least one of the single-muon or single-electron triggers~\cite{TRIG-2016-01,TRIG-2018-05,TRIG-2018-01}, which are highly efficient for leptons with $\pt^{l}>\SI{27}{\GeV}$.

Electrons are identified using both the tracking information from the inner detector and energy deposits in the electromagnetic calorimeter~\cite{EGAM-2018-01}. The electrons are required to satisfy the reconstruction and identification quality requirements corresponding to the tight working point (WP)~\cite{EGAM-2021-01} and to have $\left| \eta \right| < 1.37$ or $1.52 < \left| \eta \right| < 2.47$ and $\pt > \SI{28}{\GeV}$. To ensure that each reconstructed electron comes from a primary interaction vertex, tracks matched to electrons are required to satisfy \(|z_0 \sin \theta| < 0.5\,\text{mm}\) and \(|d_0|/\sigma(d_0) < 5\), where \(z_0\) and \(d_0\) are the longitudinal and transverse impact parameters of the electron track, respectively. Finally, each reconstructed event must satisfy the tight working point (\textit{PLVTight}) of the isolation requirements based on boosted decision trees trained to separate prompt and non-prompt electrons and muons~\cite{HIGG-2017-02}.

Muons are identified using the tracking information from the muon detector and the inner detector. The muons are required to satisfy the reconstruction and identification \textit{Medium} quality requirement and to have $\left| \eta \right| < 2.5$ and $\pt > \SI{28}{\GeV}$~\cite{MUON-2018-03,MUON-2022-01}. Muons are required to come from a primary vertex, by requiring the ID tracks to pass \(|z_0 \sin \theta| < 0.5\,\text{mm}\) and \(|d_0|/\sigma(d_0) < 3\), and to fulfil the \textit{PLVTight} isolation requirement.

Jets are reconstructed using the anti-$k_t$ algorithm~\cite{Cacciari:2008gp} with radius parameter $R = 0.4$ implemented in FastJet~\cite{Fastjet} using particle flow objects (PFlow)~\cite{PERF-2015-09}. To reduce the number of jets originating from \pileup, an additional selection criterion based on a jet-vertex tagging (JVT) technique is applied. The JVT is a likelihood discriminant that combines information from several track-based variables~\cite{PERF-2014-03}, and is only applied to jets with $\pt < \SI{60}{\GeV}$ and $\left|\eta\right| < 2.4$. The jet energy and direction are calibrated using an energy- and $\eta$-dependent simulation-based calibration scheme with in situ corrections based on data~\cite{JETM-2018-05}. Jets are required to have $\pt >\SI{25}{\GeV}$ and $\left|\eta\right| < 2.5$.

The DL1r flavour tagging algorithm~\cite{FTAG-2019-07} is employed to identify jets containing $b$-hadrons. This tagger uses a deep neural network to exploit the distinctive features of $b$-hadrons, such as the impact parameters of tracks and the displaced vertices reconstructed in the inner detector. It also incorporates discriminating variables from a recurrent neural network, which capture spatial and kinematic correlations among tracks originating from the same $b$-hadron. For each jet, the algorithm computes a multivariate $b$-tagging discriminant, classifying the jet as $b$-tagged if this discriminant exceeds a predefined threshold. Operating points corresponding to $b$-tag efficiencies of 70\% and 85\% in \ttbar\ MC events are used~\cite{FTAG-2018-01}.

To avoid the reconstruction of the same energy deposits in the detector as multiple objects, overlap removal is applied first between electrons and muons that share a track, with the electron being removed in such cases. Next, if an electron is within $\Delta R = 0.2$ of one or more jets, the closest jet to the electron is removed. If there are still jets within $\Delta R = 0.4$ of the electron, the electron is removed. Jets with fewer than three tracks within $\Delta R = 0.4$ of muons are removed, whereas if the jet has at least three tracks, the muon is removed and the jet is retained.

The missing transverse energy \etmiss\ in the event is the magnitude of the missing transverse momentum vector $\mathbf{\pt}^{\mathrm{miss}}$, defined as the negative vector sum of the \pt\ of all selected and calibrated physics objects in the event, with an extra track-based term added to account for soft energy in the event that is not associated with any of the selected objects~\cite{JETM-2020-03}.

\subsection{Detector-level event selection}
\label{sec:evtsel:reco}
The event selection comprises a set of requirements based on the general event quality and on the reconstructed objects, defined in \cref{sec:obj_reco}, that characterise the final-state event topology.

The analysis focuses on the $e\mu$ final state of the dileptonic decay channel, excluding the $ee$ and $\mu\mu$ channels. The large number of events collected in the $e\mu$ channel allow for precise measurements while significantly reducing the impact of the $Z$+jets background, which is more prominent in the $ee$ and $\mu\mu$ channels due to the presence of same-flavour leptons and contributes to the $e\mu$ channel through the $Z$ boson decay into a pair of $\tau$-leptons.
Each event is required to contain exactly one electron and one muon, of opposite electric charges, both with $\pt>28$\,\GeV. In addition, at least one lepton must be matched to a trigger object of the same flavour.

Finally, at least two jets with $\pt > \SI{25}{\GeV}$, $b$-tagged using the 70\% WP, are required. This requirement is referred to as $N_{b\text{-tags}} \geq 2$, while the region defined by these requirements is referred to as the $2b$-inclusive signal region.

\begin{figure}[tb]
\centering
\begin{subfigure}[b]{0.45\textwidth}
\centering
\includegraphics[width=\linewidth]{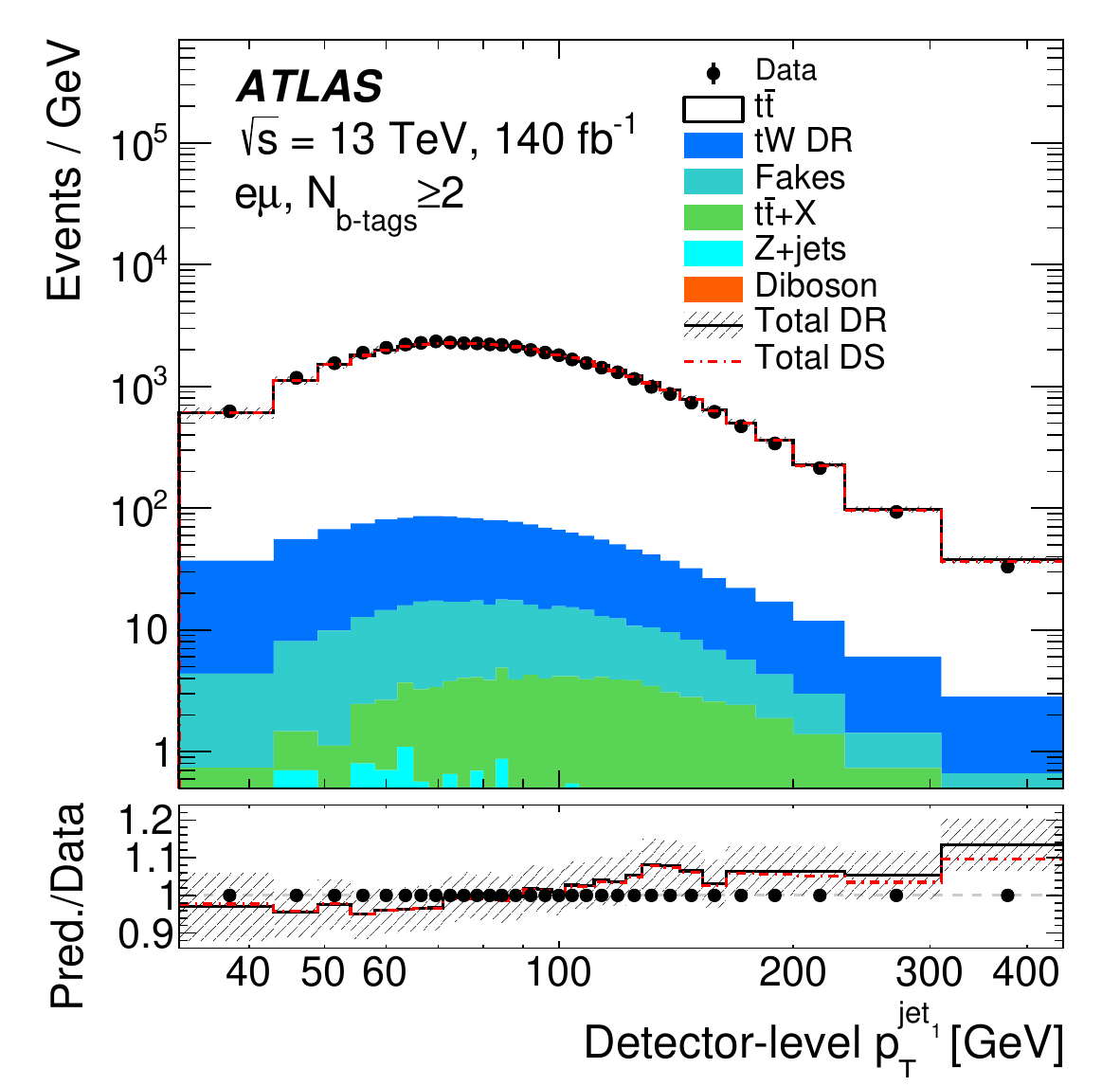}
\caption{}\label{fig:control_inclusive_pT_jet1}
\end{subfigure}
\begin{subfigure}[b]{0.45\textwidth}
\centering
\includegraphics[width=\linewidth]{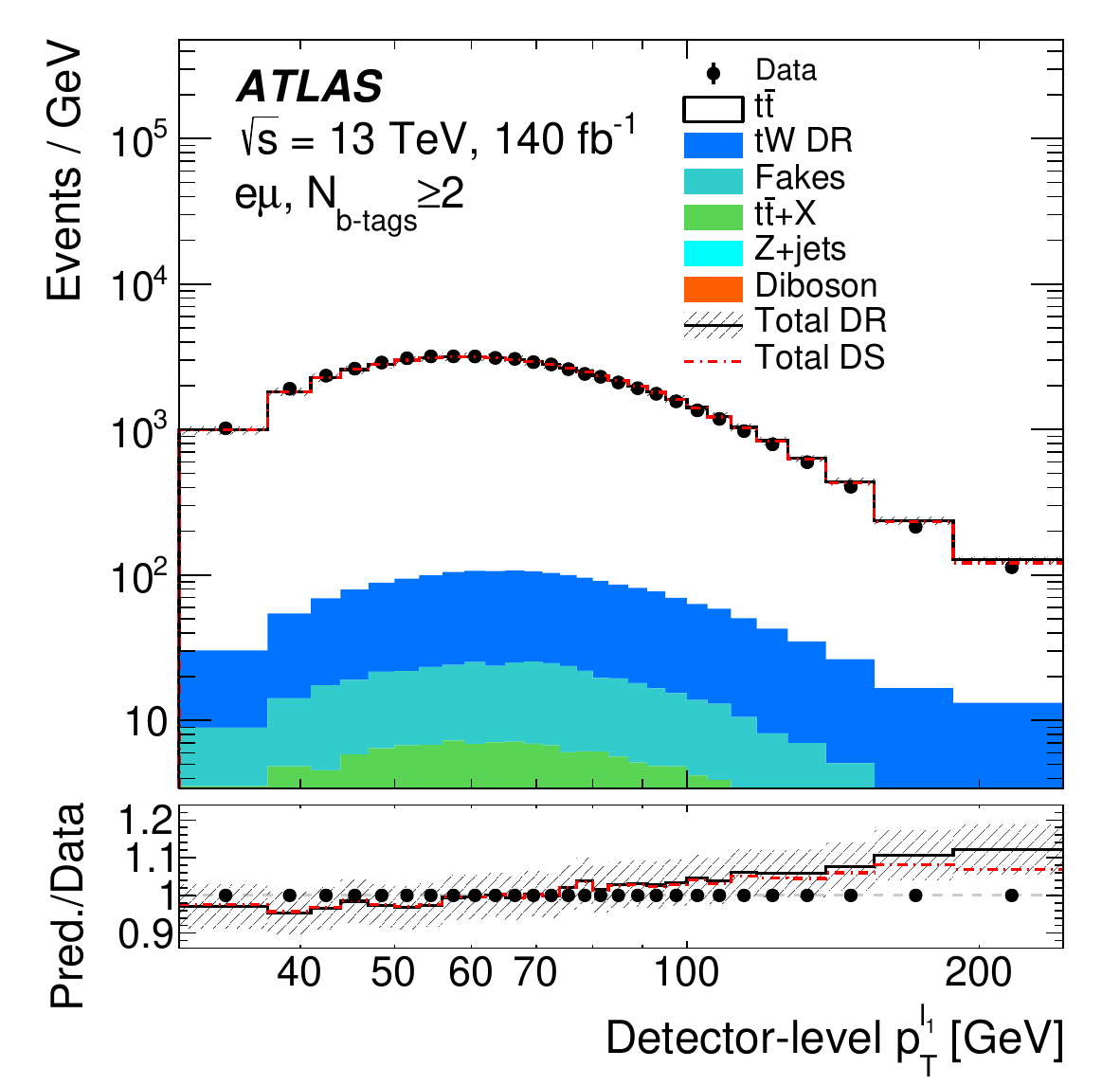}
\caption{}\label{fig:control_inclusive_pT_lep1}
\end{subfigure} \\
\begin{subfigure}[b]{0.45\textwidth}
\centering
\includegraphics[width=\linewidth]{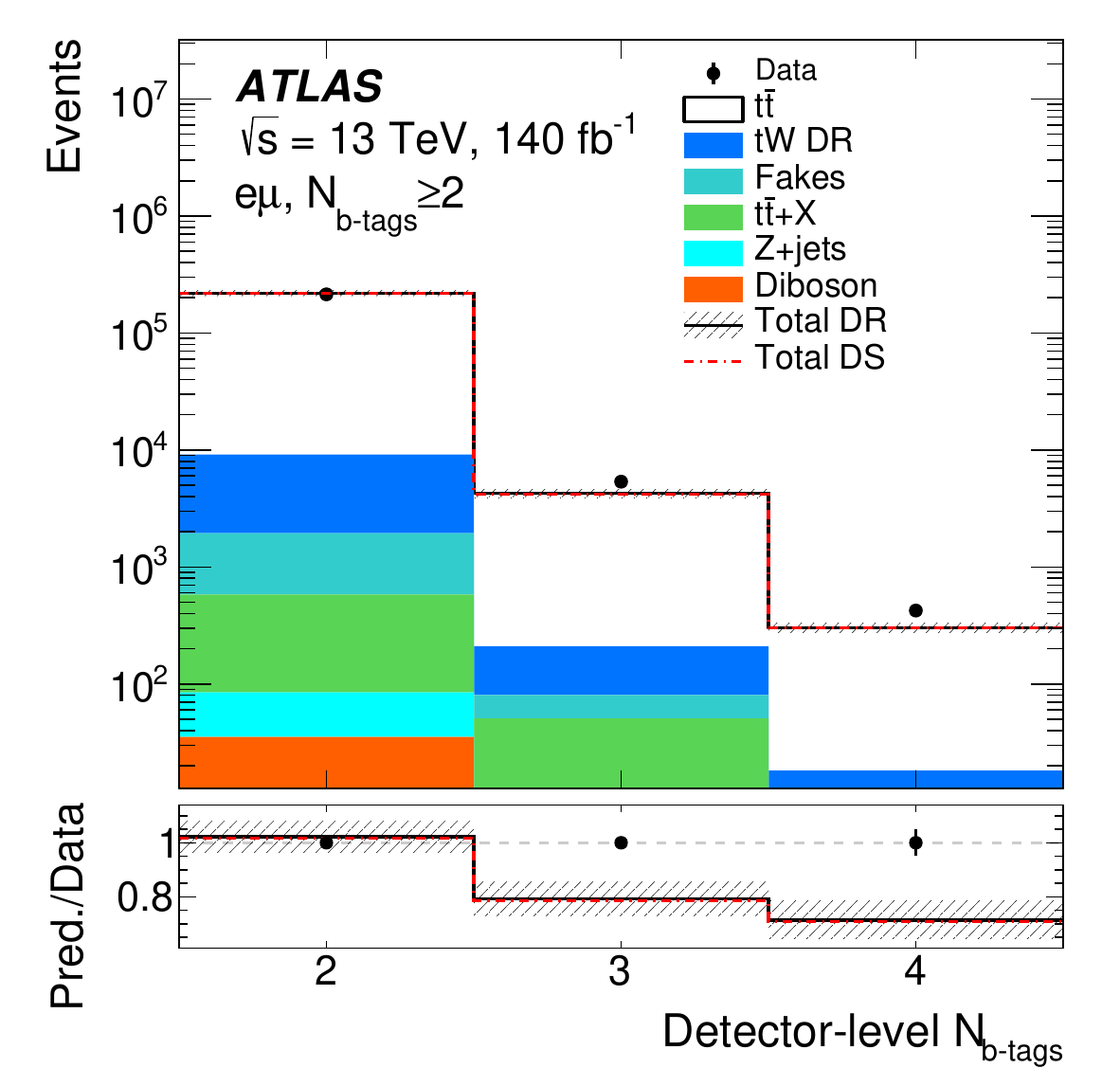}
\caption{}\label{fig:control_inclusive_nbjets}
\end{subfigure}
\begin{subfigure}[b]{0.45\textwidth}
\centering
\includegraphics[width=\linewidth]{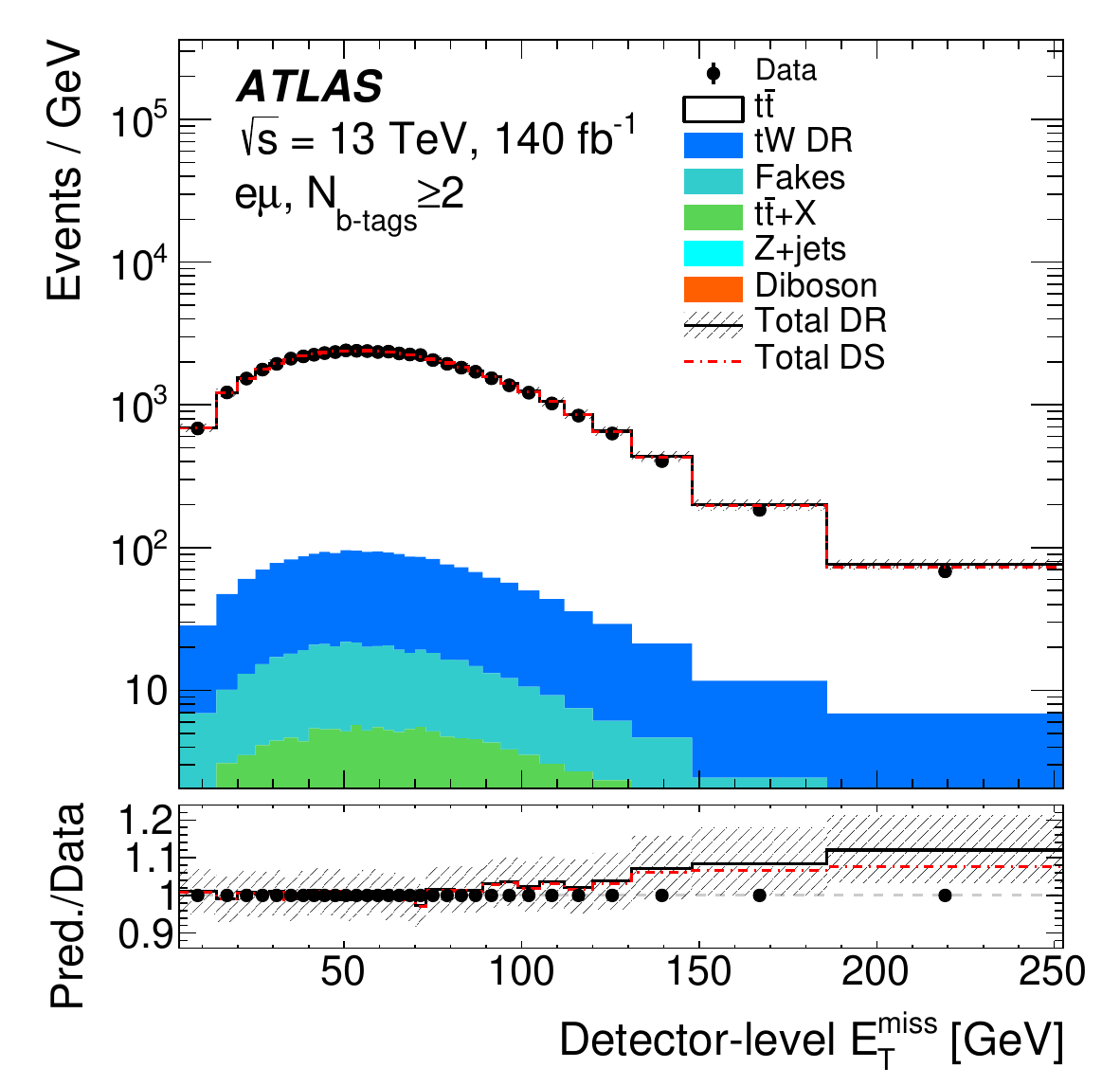}
\caption{}\label{fig:control_inclusive_met}
\end{subfigure} \\
\caption{Comparison of data and MC predictions for  the $\pt$ of \protect\subref{fig:control_inclusive_pT_jet1} the leading jet and  \protect\subref{fig:control_inclusive_pT_lep1} the leading lepton, \protect\subref{fig:control_inclusive_nbjets} the number of $b$-tagged jets and \protect\subref{fig:control_inclusive_met} the \etmiss\ in the  $2b$-inclusive signal region. The total MC predictions are obtained by employing either the DR or DS schemes for the generation of the $tW$ sample. The hatched area indicates the combined statistical and systematic uncertainties in the total prediction, excluding systematic uncertainties related to the signal modelling and normalisation. Events beyond the range of the horizontal axis are included in the last bin and their yield normalised to the bin width, for presentation purposes. The lower panels show the ratios of the predictions to the data.}
\label{fig:control_inclusive}
\end{figure}

A selection of detector-level event distributions after the full event selection is shown in \cref{fig:control_inclusive}. The data and simulated distributions (obtained using both the DR and DS schemes for the simulation of the $tW$ process) generally show good agreement. One exception is the lepton transverse momentum distribution, which appears softer in data compared to the nominal signal and background simulation. Similar discrepancies were observed in previous measurements at $\sqrt{s} = 13$~\TeV~\cite{TOPQ-2018-26,CMS-TOP-17-014,TOPQ-2019-03}. Another exception is the $b$-tagged jet multiplicity distribution, which shows an excess in data compared to the simulation at higher multiplicity values, as already observed in previous measurements \cite{TOPQ-2019-03,CMS-TOP-22-009}. The impact of this mismodelling has been tested for all measured observables in this phase space and found to be negligible. For most distributions, no significant differences are observed between the DR and DS schemes. To enhance sensitivity to the interference between the \ttbar\ and $tW$ processes, the variable \mbl\ is defined in a region with a veto on additional $b$-tagged jets, as described in the next section.

\FloatBarrier
\subsection{Sensitivity to \ttbar\ and $tW$ interference}

The observable that is most sensitive to the interference between double and single top production is \mbl\ and is defined as:

\begin{displaymath}
\mbl\equiv \min \left\{\max \left(m^{b_1 l_1}, m^{b_2 l_2}\right), \max \left(m^{b_1 l_2}, m^{b_2 l_1}\right)\right\}\mbox{,}
\end{displaymath}
where $b_i$ and $l_i$ represent the two $b$-jets and two leptons, respectively. This definition ensures that, for doubly resonant events, the value of \mbl\ is constrained to be smaller than the mass of the top quark. For this reason, this variable is sensitive to the interference for high mass values (above about 180~\GeV) due to the suppression of doubly resonant contributions. The sensitivity of \mbl\ to interference effects was first explored in Ref.~\cite{TOPQ-2017-05}. However, the presence of additional $b$-jets in doubly resonant events can spoil the constraint, resulting in a high value of \mbl\ due to the combinatorics between $b$-jets and leptons. Moreover, given the sensitivity of \mbl\ to the number of $b$-tagged jets, limiting the signal region to events with exactly two $b$-tagged jets reduces the impact of the mismodelling of the $b$-tagged jet multiplicity at higher values, as shown in \cref{fig:control_inclusive_nbjets}. Therefore, a veto on additional $b$-jets is used to suppress the contributions of the combinatorics and enhance sensitivity to interference effects. Following a dedicated optimisation of the phase space, the best results are achieved by discarding events with additional $b$-tagged jets using the 85\% WP\@. This additional requirement is referred to as $N_{b\text{-tags}} = 2$, while the region defined by these requirements is referred to as the $2b$-exclusive signal region. This region is only used to measure the differential cross-section as a function of \mbl, while the differential cross-sections as a function of all other variables are measured in the $2b$-inclusive region.

\begin{figure}[tb]
\centering
\begin{tabular}{c}
\includegraphics[width=.65\textwidth]{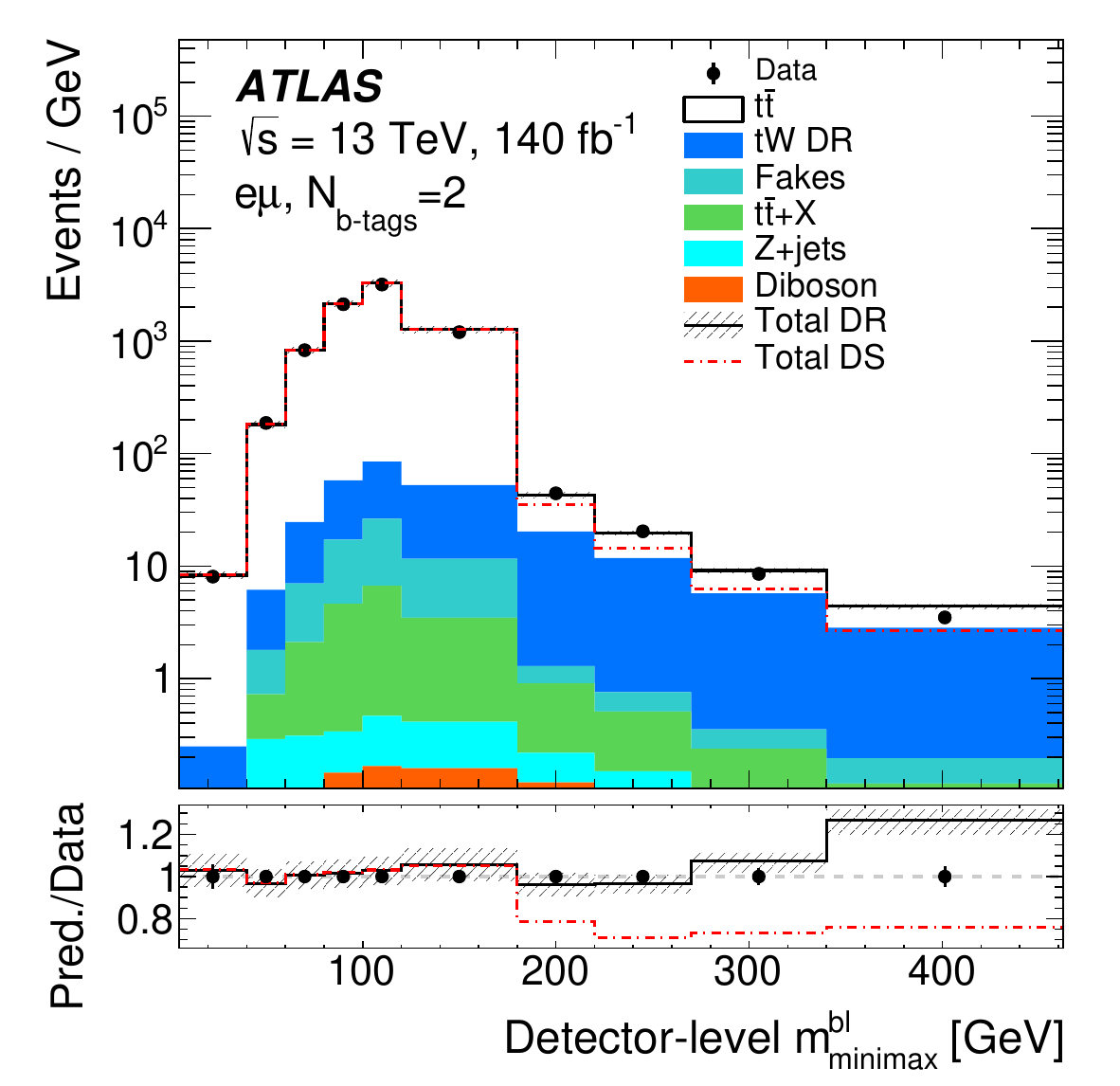} \\
\end{tabular}
\caption{Comparison of data and MC predictions for the \mbl\ variable in the $2b$-exclusive signal region. The total MC predictions are obtained by employing either the DR or DS schemes for the generation of the $tW$ sample. The hatched area indicates the combined statistical and systematic uncertainties in the total prediction, excluding systematic uncertainties related to the signal modelling and normalisation. Events beyond the range of the horizontal axis are included in the last bin and their yield normalised to the bin width, for presentation purposes. The lower panel shows the ratio of the predictions to the data.}
\label{fig:control_exclusive_minimax}
\end{figure}

\begin{figure}[tb]
\centering
\begin{subfigure}[b]{0.45\textwidth}
\centering
\includegraphics[width=\linewidth]{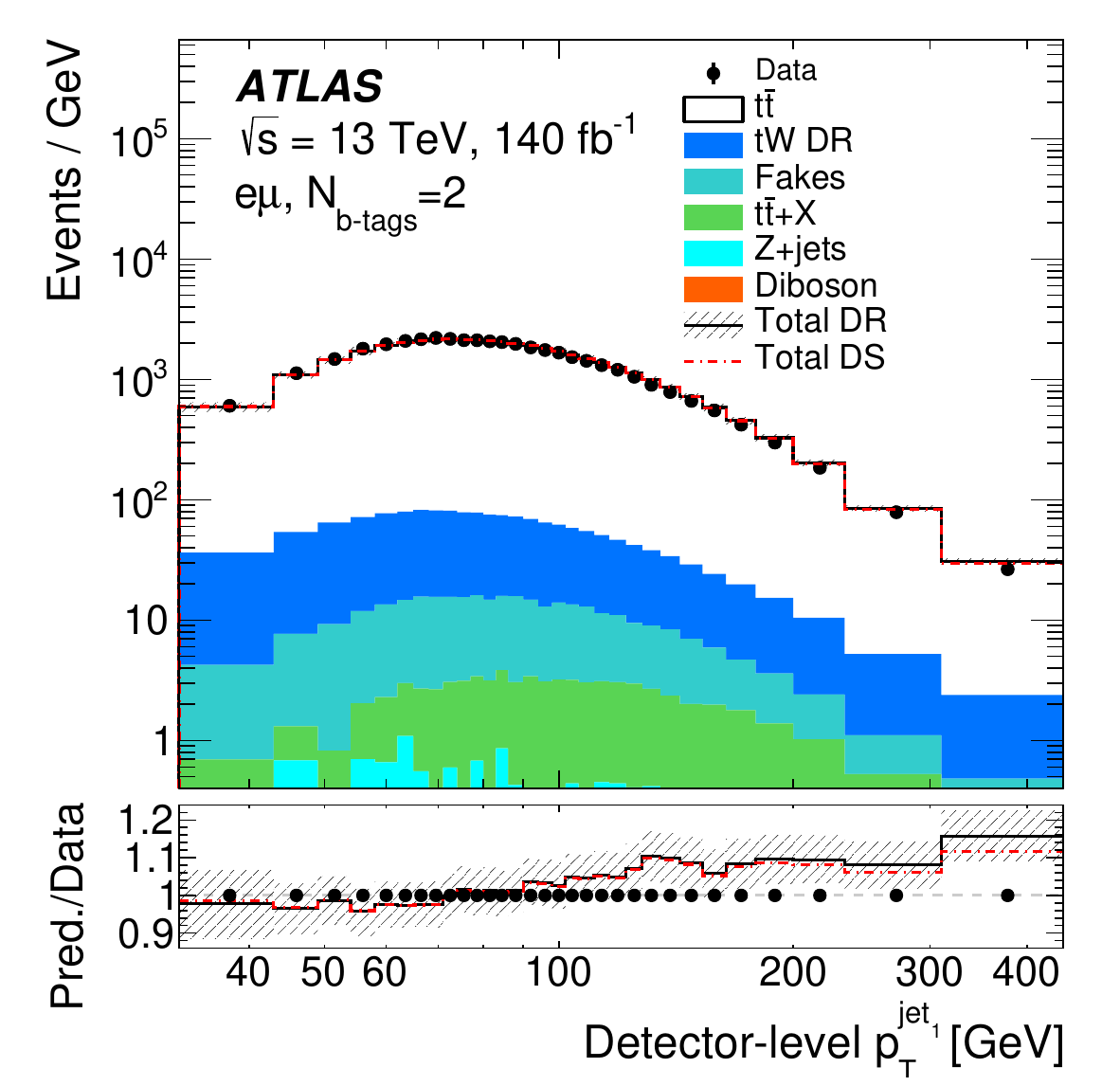}
\caption{}\label{fig:control_exclusive_pT_jet1}
\end{subfigure}
\begin{subfigure}[b]{0.45\textwidth}
\centering
\includegraphics[width=\linewidth]{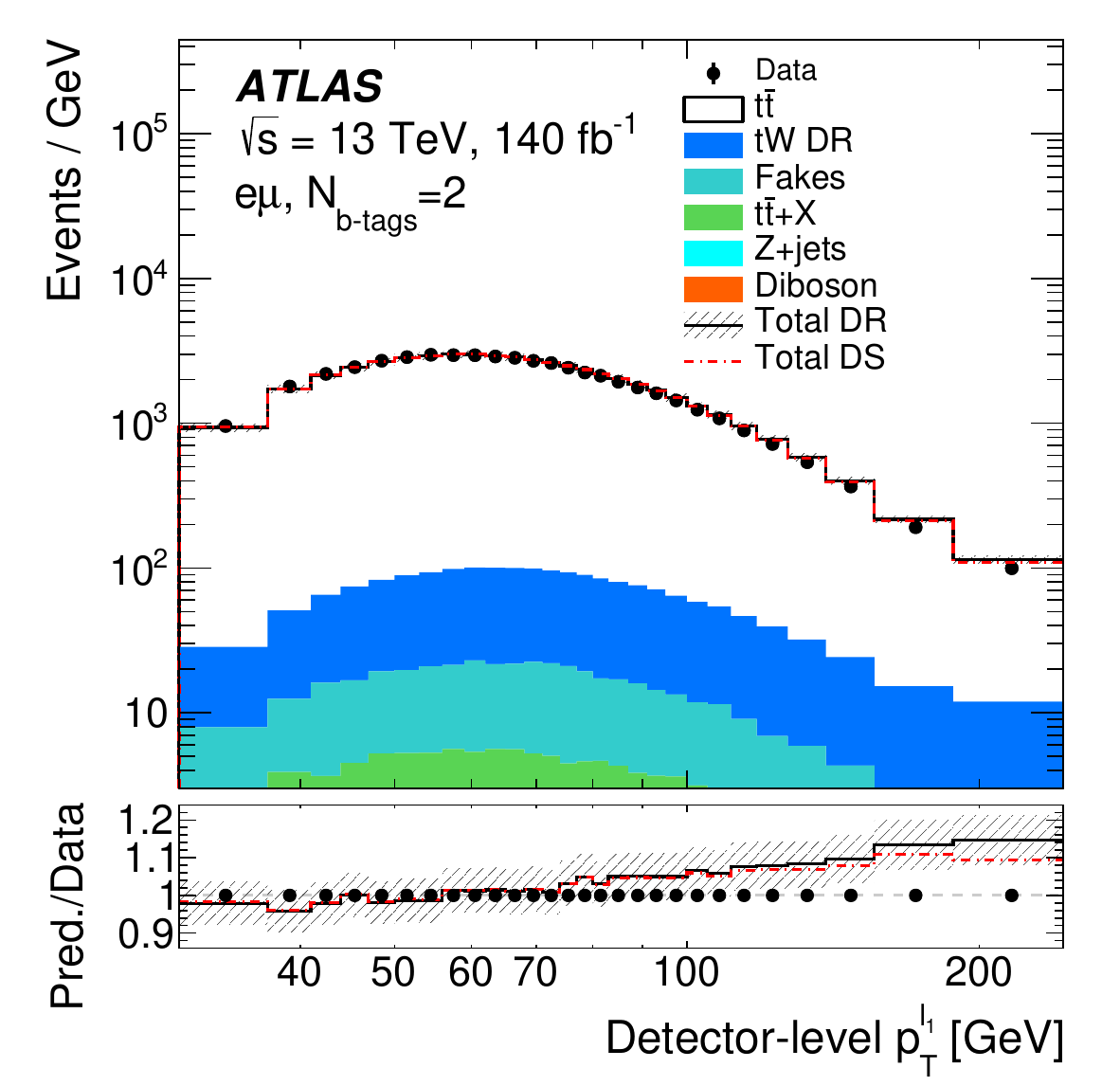}
\caption{}\label{fig:control_exclusive_pT_lep1}
\end{subfigure} \\
\begin{subfigure}[b]{0.45\textwidth}
\centering
\includegraphics[width=\linewidth]{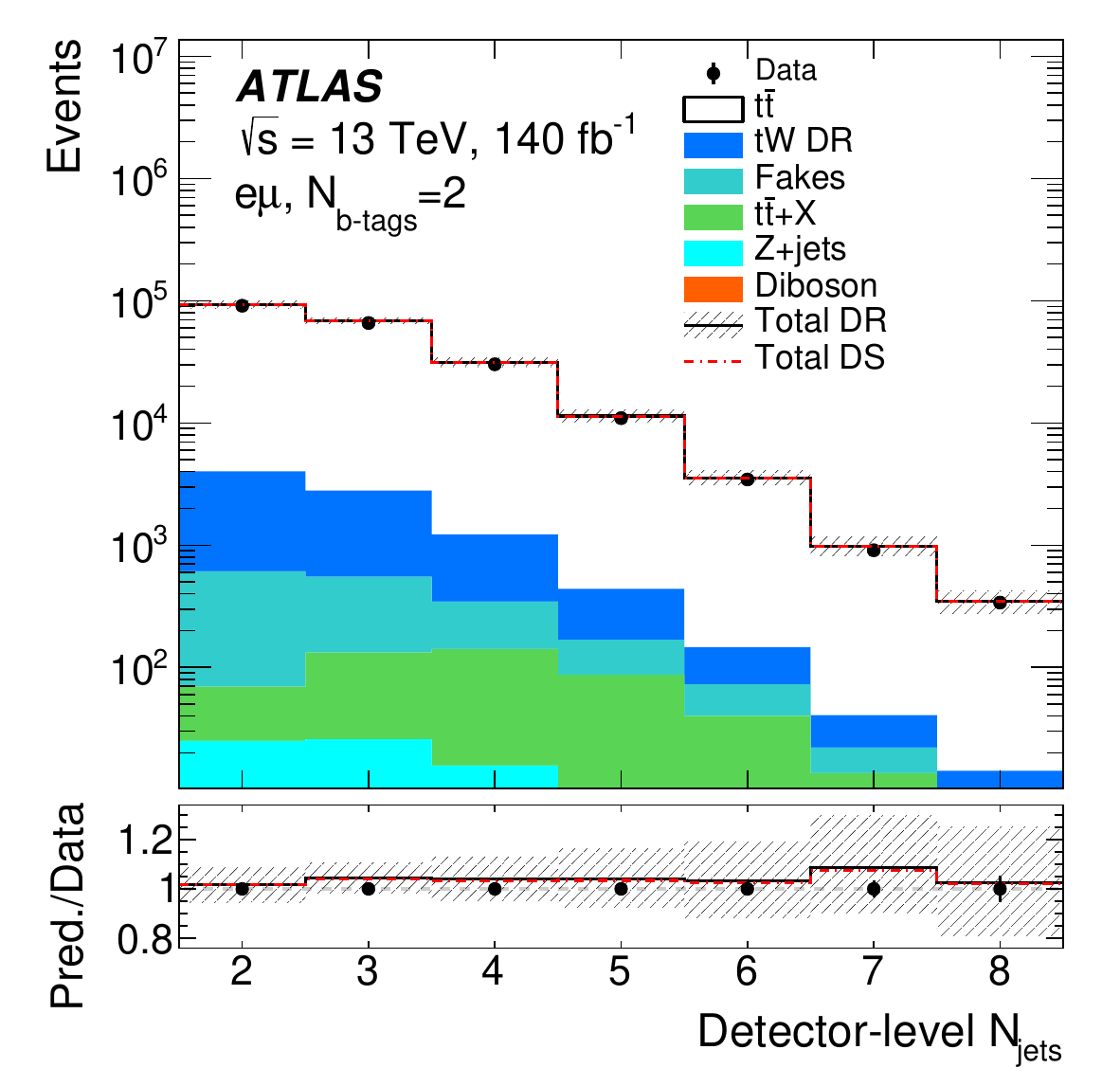}
\caption{}\label{fig:control_exclusive_njets}
\end{subfigure}
\begin{subfigure}[b]{0.45\textwidth}
\centering
\includegraphics[width=\linewidth]{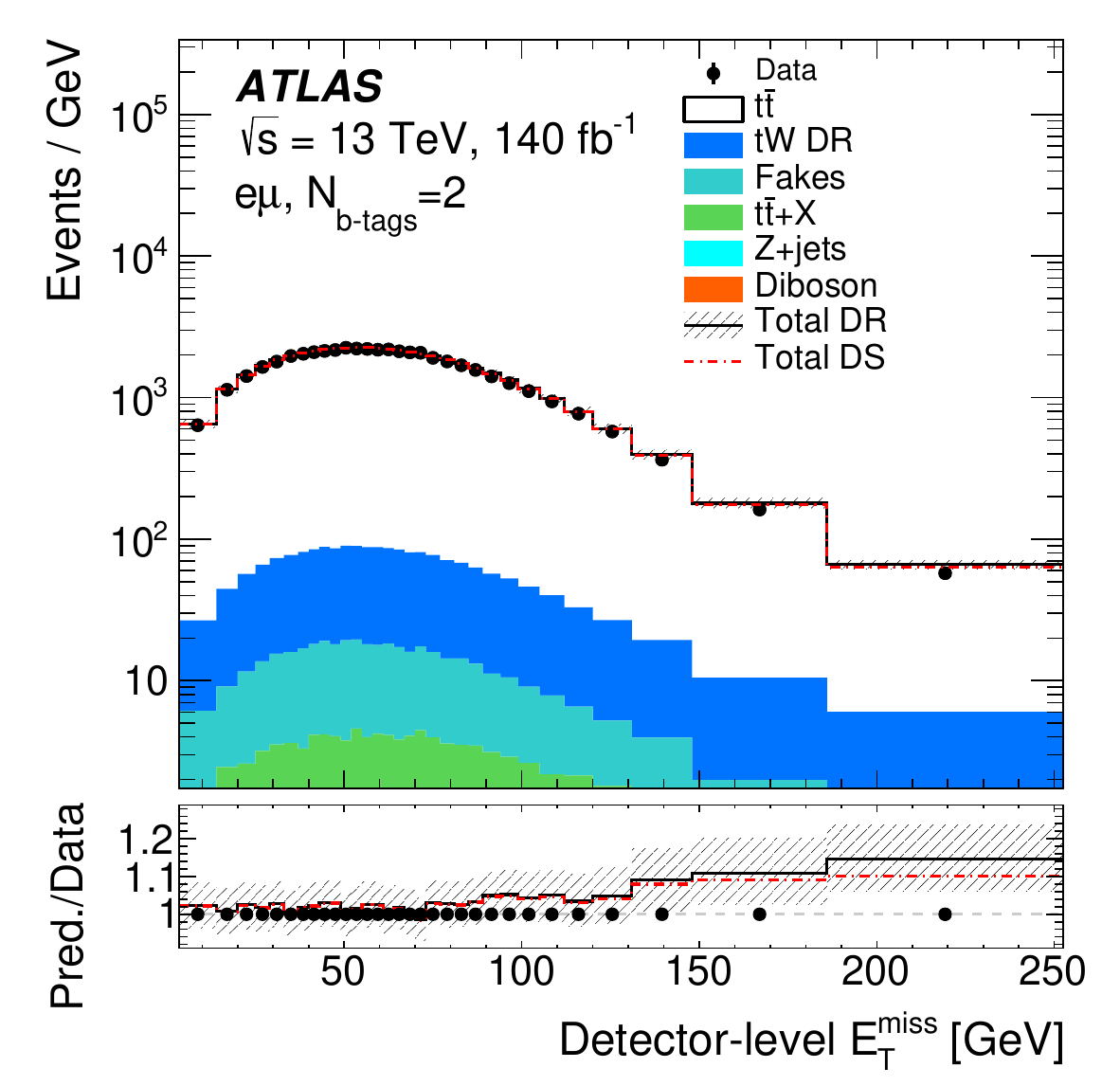}
\caption{}\label{fig:control_exclusive_met}
\end{subfigure} \\
\caption{Comparison of data and MC predictions for the $\pt$ of \protect\subref{fig:control_exclusive_pT_jet1} the leading jet and  \protect\subref{fig:control_exclusive_pT_lep1} the leading lepton, \protect\subref{fig:control_exclusive_njets} the number of jets and \protect\subref{fig:control_exclusive_met} the \etmiss\ in the $2b$-exclusive signal region. The total MC predictions are obtained by employing either the DR or DS schemes for the generation of the $tW$ sample. The hatched area indicates the combined statistical and systematic uncertainties in the total prediction, excluding systematic uncertainties related to the signal modelling and normalisation. Events beyond the range of the horizontal axis are included in the last bin and their yield normalised to the bin width, for presentation purposes. The lower panels show the ratios of the predictions to the data.}
\label{fig:control_exclusive}
\end{figure}

The detector-level distribution for the \mbl\ variable is shown in \cref{fig:control_exclusive_minimax} for events satisfying the $2b$-exclusive selection. The comparison between predictions obtained with the DR and DS schemes shows the region of the distribution that is sensitive to the different schemes used to model the interference, while a selection of detector-level event distributions in the $2b$-exclusive signal region are shown in \cref{fig:control_exclusive} for some object kinematic variables.  Unlike \mbl, these distributions do not exhibit a noticeable difference between the DR and DS predictions. The level of agreement between data and MC for these variables is comparable to that seen in the $2b$-inclusive signal region.

The observed and expected numbers of events after the two event selections are shown in \cref{tab:event_yields:reco}. With these selections the total detector-level signal yields are composed of 90\% fiducial events (defined with the requirements described \cref{sec:objects:particle,sec:selection:particle}) and correspond to 23\% of the total particle-level fiducial events.

\begin{DIFnomarkup}
\begin{table}[!ht]
\centering\begin{tabular}{c|r@{}@{}r@{}@{}rr|r@{}@{}r@{}@{}rr}\hline
\toprule
\hline
\multicolumn{1}{c|}{Sample} & \multicolumn{4}{@{}c@{}|}{$e\mu$, $N_{b\text{-tags}}=2$} & \multicolumn{4}{@{}c@{}}{$e\mu$, $N_{b\text{-tags}}\geq2$}    \\
\hline
$t\bar{t}$   & $200000$    & $\,\pm\,$     & $12000$   & ($\pm 6\%$)  & $214000$ & $\,\pm\,$ & $13000$   & ($\pm 6\%$)  \\
$tW$ (DR)  & $6900$    & $\,\pm\,$     & $400$   & ($\pm 6\%$)  & $7300$ & $\,\pm\,$ & $400$   & ($\pm 6\%$)  \\
$t\bar{t}X$  & $500$     & $\,\pm\,$     & $70$    & ($\pm 13\%$) & $800$  & $\,\pm\,$ & $110$   & ($\pm 13\%$) \\
Diboson    & $32$      & $\,\pm\,$     & $13$    & ($\pm 40\%$) & $36$   & $\,\pm\,$ & $15$    & ($\pm 40\%$) \\
$Z$+jets   & $47$      & $\,\pm\,$     & $18$    & ($\pm 40\%$) & $51$   &     & $^{\,+20}_{-19}$     & ($\pm 40\%$) \\
Fakes    & $1300$    & $\,\pm\,$     & $700$   & ($\pm 50\%$) & $1400$ & $\,\pm\,$ & $700$   & ($\pm 50\%$) \\
\hline
Total expected & $209000$    & $\,\pm\,$     & $12000$   & ($\pm 6\%$)  & $224000$ & $\,\pm\,$ & $13000$   & ($\pm 6\%$)  \\
Purity   & $0.9910$    & $\,\pm\,$     & $0.0031$  & ($\pm 0.32\%$) & $0.9898$ & $\,\pm\,$ & $0.0032$  & ($\pm 0.32\%$) \\
\hline
Data     & 202738    &    &  & & 219882 &     &  & \\
MC/Data    & $1.03$    & $\,\pm\,$     & \multicolumn{1}{@{}l}{$0.06$} & ($\pm 6\%$)  & $1.02$ & $\,\pm\,$ & \multicolumn{1}{@{}l}{$0.06$} & ($\pm 6\%$)  \\
\hline

\bottomrule
\end{tabular}
\caption{Observed and expected number of events after both detector-level event selections. The uncertainties include the MC statistical uncertainties and all systematic uncertainties except for the signal modelling and normalisation. The purity is defined as the ratio of the  expected number of signal ($\ttbar+tW$) events to the  expected number of total events.}
\label{tab:event_yields:reco}
\end{table}
\end{DIFnomarkup}


\FloatBarrier
\subsection{Particle-level object definition}\label{sec:objects:particle}
In addition to the detector-level objects, `particle-level' objects are also defined. These objects are identified using generator-level information from stable particles in the generator record, defined as those that satisfy $c \tau > 10$ nm, with $\tau$ being the particle lifetime, without any simulation of their interaction  with the detector components. The particle-level objects are used to define the fiducial phase-space regions and variables of interest for the measurements.

Particle-level electrons and muons are required to not originate from a hadron, either directly or via an intermediate $\tau$-lepton decay. With this requirement, a $\tau$-lepton is accepted as a parent of the electron or a muon only if the $\tau$-lepton itself comes directly from a $W$. Leptons that come from a quark or hadron, including hadronically decaying $\tau$-lepton, are rejected. With this definition, all particle-level leptons originate from the decays of EW bosons without requiring a direct match to a $W$ boson. Their momentum is calculated to include the radiation loss due to final-state radiation (FSR) photons (`dressing'), classified as photons within $\Delta R = 0.1$ that do not originate from hadron decays. The transverse momentum and $\eta$ requirements are the same as the detector-level signal objects.

Particle-level jets are reconstructed by clustering generated particles (excluding those identified as leptons or neutrinos from EW boson decays), with the anti-$k_t$ algorithm with a radius parameter of $R = 0.4$. %
The ghost-association procedure~\cite{Cacciari:2008gn} is used to match $b$-hadrons with $\pT > 5\,\GeV$ to jets where generator-level jet clustering is run as usual, except that the energy of all $b$-hadrons is set to a negligible value. All the particle jets containing a constituent $b$-hadron are considered $b$-tagged after this procedure.

The particle-level missing transverse momentum is calculated from the four-vector sum of all neutrinos, except those coming from hadron decays, either directly or via $\tau$-lepton decays.

Particle-level objects are subject to different overlap removal criteria relative to detector-level objects, after dressing and jet reconstruction. Muons and electrons within $\Delta R = 0.4$ of a jet are excluded. No electron--muon overlap removal is applied at the particle level.

\subsection{Particle-level selection}%
\label{sec:selection:particle}

Using particle-level objects, two fiducial regions are defined by applying selections that replicate the detector-level ones as closely as possible:

\begin{itemize}
\item one electron and one muon with $\pt$ $> \SI{28}{\GeV}$ and $\left| \eta \right| < 2.5$;
\item at least two jets with $\pt$ $> \SI{25}{\GeV}$ and $\left| \eta \right| < 2.5$;
\item at least two jets tagged as $b$-jets with the $b$-hadron ghost-association matching;
\item \emph{only for \mbl}: exactly two $b$-jets, tagged with the $b$-hadron ghost-association matching.
\end{itemize}

These requirements define the particle-level $2b$-inclusive and $2b$-exclusive fiducial regions, labelled as \fidInclusive\ and \fidExclusive\  respectively. The measured observables are unfolded to these phase spaces. In \cref{tab:event_yields:particle}, the expected numbers of events after the two particle-level event selections are shown.

\begin{DIFnomarkup}
\begin{table}[!ht]
\centering\begin{tabular}{c|r@{}@{}r@{}@{}rr|r@{}@{}r@{}@{}rr}\hline
\toprule
\hline
\multicolumn{1}{c|}{Sample} & \multicolumn{4}{@{}c@{}|}{$e\mu$, $N_{b\text{-jets}}=2$} & \multicolumn{4}{@{}c@{}}{$e\mu$, $N_{b\text{-jets}}\geq2$}    \\
\hline
$t\bar{t}$   & $806640$    & $\,\pm\,$     & $180$   &      & $823250$ & $\,\pm\,$ & $180$   &      \\
$tW$ (DR)  & $27386$    & $\,\pm\,$     & $20$   &      & $27918$ & $\,\pm\,$ & $21$   &      \\
\hline
Total signal & $834030$    & $\,\pm\,$     & $180$   &      & $851160$ & $\,\pm\,$ & $180$   &      \\

\hline

\bottomrule
\end{tabular}
\caption{Expected event yields after applying both particle-level selections. The uncertainties shown are statistical only, originating from the Monte Carlo simulation.}
\label{tab:event_yields:particle}
\end{table}
\end{DIFnomarkup}


\FloatBarrier


%

\section{Cross-section extraction}%
\label{sec:analysis_strategy}
The differential cross-sections are derived from detector-level events, which are corrected for detector and reconstruction effects—a process commonly referred to as `unfolding'. This correction accounts for efficiency, acceptance, and resolution effects, with the latter handled using an iterative Bayesian unfolding procedure~\cite{DAgostini:1994fjx}, implemented via the RooUnfold package~\cite{Adye:2011gm}.

For each observable, the unfolding procedure starts from the number of events observed in data at detector level in bin $j$ of the distribution ($N_\mathrm{data}^j$), from which the background event yield ($N_\mathrm{bkg}^j$), estimated as described in~\cref{sec:datamc:mc_background}, is subtracted. Then the unfolding corrections, described below, are applied. All corrections are evaluated using the nominal \POWPY[8] MC simulation of the $t\bar{t} + tW$ signal sample.

As the first step, an acceptance correction ($f^j_\mathrm{acc}$) is applied. The acceptance correction in bin $j$ is defined as the fraction of signal events reconstructed in this bin that also satisfy the particle-level selection:
\begin{equation}
f^j_\mathrm{acc}  \equiv \frac{ N_{ \mathrm{reco} \land \mathrm{part} }^j }{N_\mathrm{reco}^j }{,}\label{eq:acc}
\end{equation}
where $N_{ \mathrm{reco} \land \mathrm{part} }^j$ is the number of events that satisfy both the detector-level selection and the particle-level selections  in a given detector-level bin $j$, and $N_\mathrm{reco}^j$ is the number of events that satisfy the detector-level selection in the same bin. This correction is a bin-by-bin factor that accounts for events that are generated outside the fiducial phase-space region but satisfy the detector-level selection.

The resulting detector-level distribution is then corrected for the effects of the detector resolution to produce the particle-level distribution. The procedure uses as input a migration matrix $\mathcal{M}$ derived from the simulated signal sample that maps the particle-level bin $i$ in which an event is generated to the bin $j$ in which it is reconstructed. The probability for particle-level events to be reconstructed in the same bin is represented by the diagonal elements, and the off-diagonal elements describe the fraction of particle-level events that migrate into other bins. Therefore, the elements of each row sum to unity. The procedure is performed via the iterative Bayesian unfolding, using four iterations to balance the stability relative to the previous iteration and the growth of the statistical uncertainty. The effect of varying the number of iterations by one is found to be negligible.

Finally, an efficiency correction $\varepsilon$ is applied to the unfolded spectrum, correcting the result by a bin-by-bin factor to the fiducial phase space. The efficiency correction in bin $i$ is defined as the fraction of the events generated in a particle-level bin $i$ that satisfy the detector-level selection:
\begin{displaymath}
\varepsilon^i \equiv \frac{ N_{\mathrm{part} \land \mathrm{reco}}^i }{N_\mathrm{part}^i }{,}
\end{displaymath}
where $N_{\mathrm{part} \land \mathrm{reco}}^i$ is the number of events that satisfy both the particle-level selection and the detector-level selection in a given particle-level bin $i$, and $N_\mathrm{part}^i$ is the number of events that satisfy the particle-level selection in the same bin. This factor corrects for the inefficiency of the event selection and reconstruction.

The absolute differential cross-section is then obtained by dividing the unfolded distribution by the integrated luminosity and the bin width. The extraction of the absolute differential cross-section for an observable $X$ at particle-level is then given by the following expression:
\begin{equation}
\frac{\mathrm{d}\sigma^i_\mathrm{fid}}{\mathrm{d}X^i} \equiv \frac{1}{\mathcal{L} \cdot \Delta X^i} \cdot  \frac{1}{\varepsilon^i} \cdot \sum_j \mathcal{M}^{-1} \cdot  f_\mathrm{acc}^j \cdot \left(N_\mathrm{obs}^j - N_\mathrm{bkg}^j\right)\hbox{,}\label{eq:unfolding}
\end{equation}
where the index $j$ iterates over bins of observable $X$ at detector level while the index $i$ labels bins at particle-level, $\Delta X^i$ is the bin width, $\mathcal{L}$ is the integrated luminosity and $\mathcal{M}^{-1}$ is the inverted migration matrix obtained from the iterative unfolding procedure.
The integrated fiducial cross-section is obtained by integrating the unfolded cross-section over the bins, and its value is used to compute the normalised differential cross-sections:
\begin{displaymath}
\frac{1}{\sigma_\mathrm{fid}^i}\cdot\frac{\mathrm{d}\sigma^i_\mathrm{fid}}{\mathrm{d}X^i}\hbox{.}
\end{displaymath}
This definition is applied for all uncertainties affecting the measurement, ensuring that the correlation between the uncertainties in the integrated fiducial cross-section ($\sigma_\mathrm{fid}$) and the differential cross-section in each bin ($\mathrm{d}\sigma_\mathrm{fid}^i$) is accounted for. This correlation can lead to a partial cancellation of uncertainties when computing the normalised differential cross-sections, thereby reducing their overall impact compared to the absolute differential cross-sections.

\subsection{Measured distributions at the particle level}
A total of 11 observables are chosen for the differential cross-section measurement.

The differential cross-section as a function of the \mbl\ variable is measured in the $2b$-exclusive fiducial phase space, while differential cross-sections as a function of variables related to the number of jets and the kinematics of jets, leptons and \etmiss, are measured in the $2b$-inclusive fiducial phase space. Specifically, these are:

\begin{itemize}
\item \textbf{$\pt^{\text{jet}_1}$}: transverse momentum of the leading jet;
\item \textbf{$\pt^{\text{jet}_2}$}: transverse momentum of the subleading jet;
\item \textbf{$\pt^{l_1}$}: transverse momentum of the leading lepton;
\item \textbf{$\pt^{l_2}$}: transverse momentum of the subleading lepton;
\item \textbf{$\mT^{bb4l}$}: transverse mass $\mT$ of the $bb4l$ system, with $\mT=\sqrt{E^2-p_z^2}=\sqrt{\pt^2+m^2}$. The $bb4l$ system is built as the vector sum of the four-momenta of the leading and subleading $b$-jets, the two charged leptons, and the \etmiss\ as a proxy for the neutrinos;
\item \textbf{$\PT{bb4l}$}: transverse momentum of the $bb4l$ system;
\item \textbf{$\PT{bb}$}: transverse momentum of the $b$-jet pair, built using the leading and subleading $b$-jets;
\item \textbf{$\PT{bbll}$}: transverse momentum of the $bbll$ system, built using the leading and subleading $b$-jets and the two charged leptons;
\item \textbf{$m^{bbll}$}: invariant mass of the $bbll$ system;
\item \textbf{$\Njets$}: number of jets.
\end{itemize}
These observables are selected to cover several aspects of the modelling of the \ttbar\ and $tW$ processes, including their interference effects. Some of the observables are chosen based on their sensitivity to specific theoretical uncertainties or discrepancies observed in previous measurements, while others are selected because they show significant differences among the predictions from various MC models. In particular, \mbl\ is sensitive to the treatment of the \ttbar/$tW$ interference, while \PT{bb4l} and \Njets\ are sensitive to the modelling of additional radiation. The \pt\ of leptons, jets, and the $b$-jet pair probe the modelling of the underlying hard process.

For each observable, the binning is chosen to guarantee good diagonality of the migration matrix and to ensure a sufficient number of events in each bin. The binning is optimised to balance statistical precision and resolution effects, and it is validated using the same procedures described in Ref.~\cite{TOPQ-2018-15}. For all the observables, the first and last bins do not contain under- and overflow events, except for the number of jets, where the last bin includes overflow events.
\FloatBarrier

\subsection{Unfolding corrections}
\begin{figure}[tb]
\centering
\begin{subfigure}[b]{0.45\textwidth}
\centering
\includegraphics[width=\linewidth]{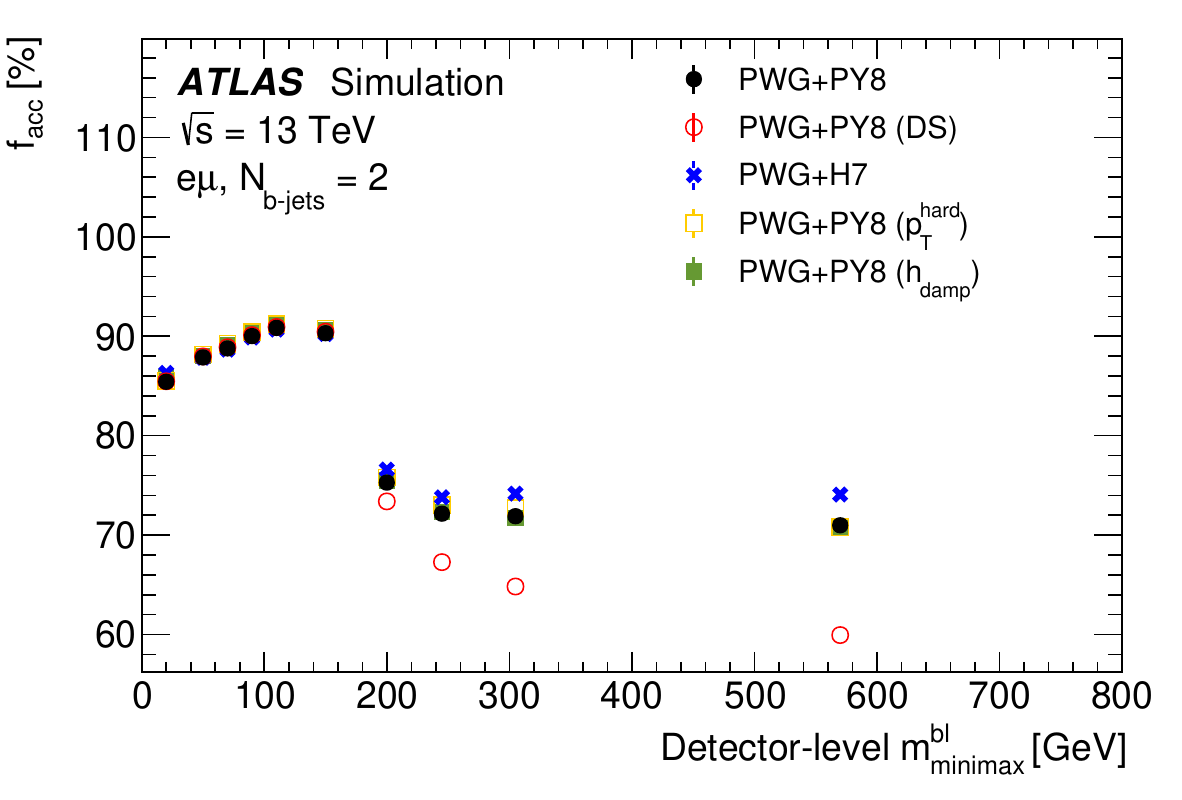}
\caption{}\label{fig:minimax:acc}
\end{subfigure}
\begin{subfigure}[b]{0.45\textwidth}
\centering
\includegraphics[width=\linewidth]{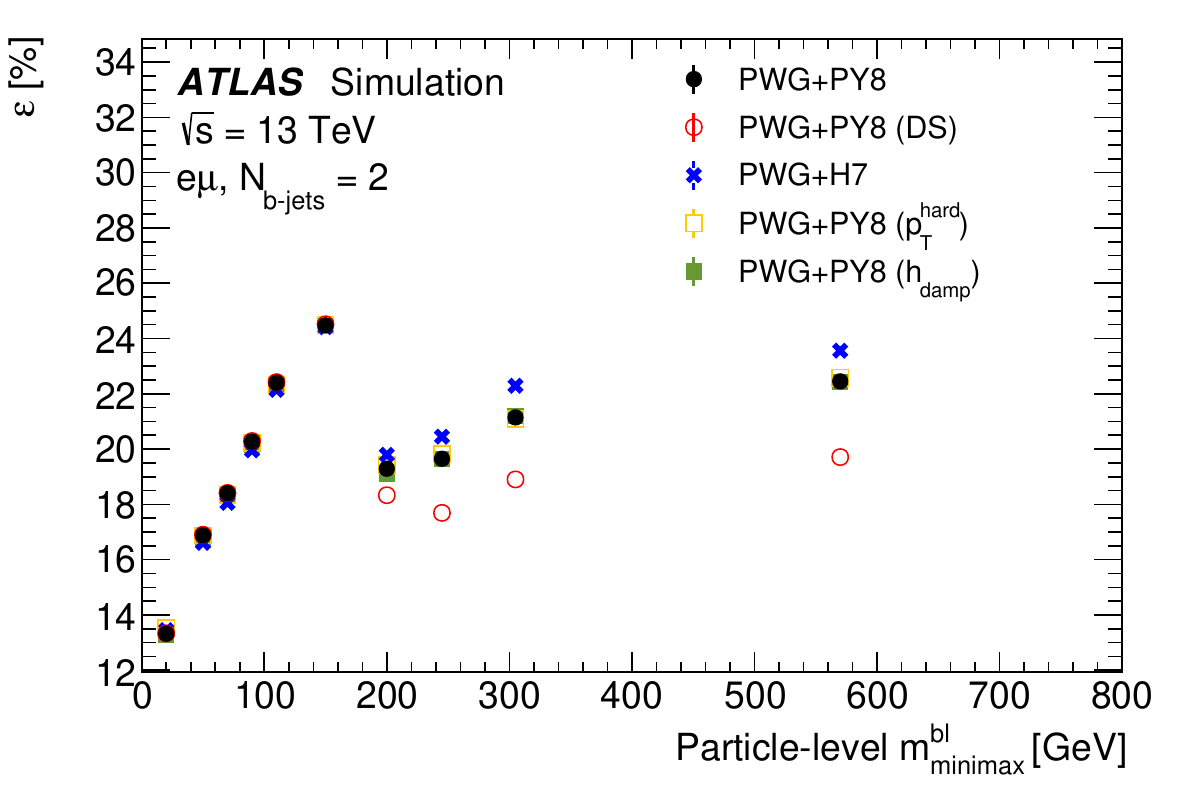}
\caption{}\label{fig:minimax:eff}
\end{subfigure}\\
\begin{subfigure}[b]{0.45\textwidth}
\centering
\includegraphics[width=\linewidth]{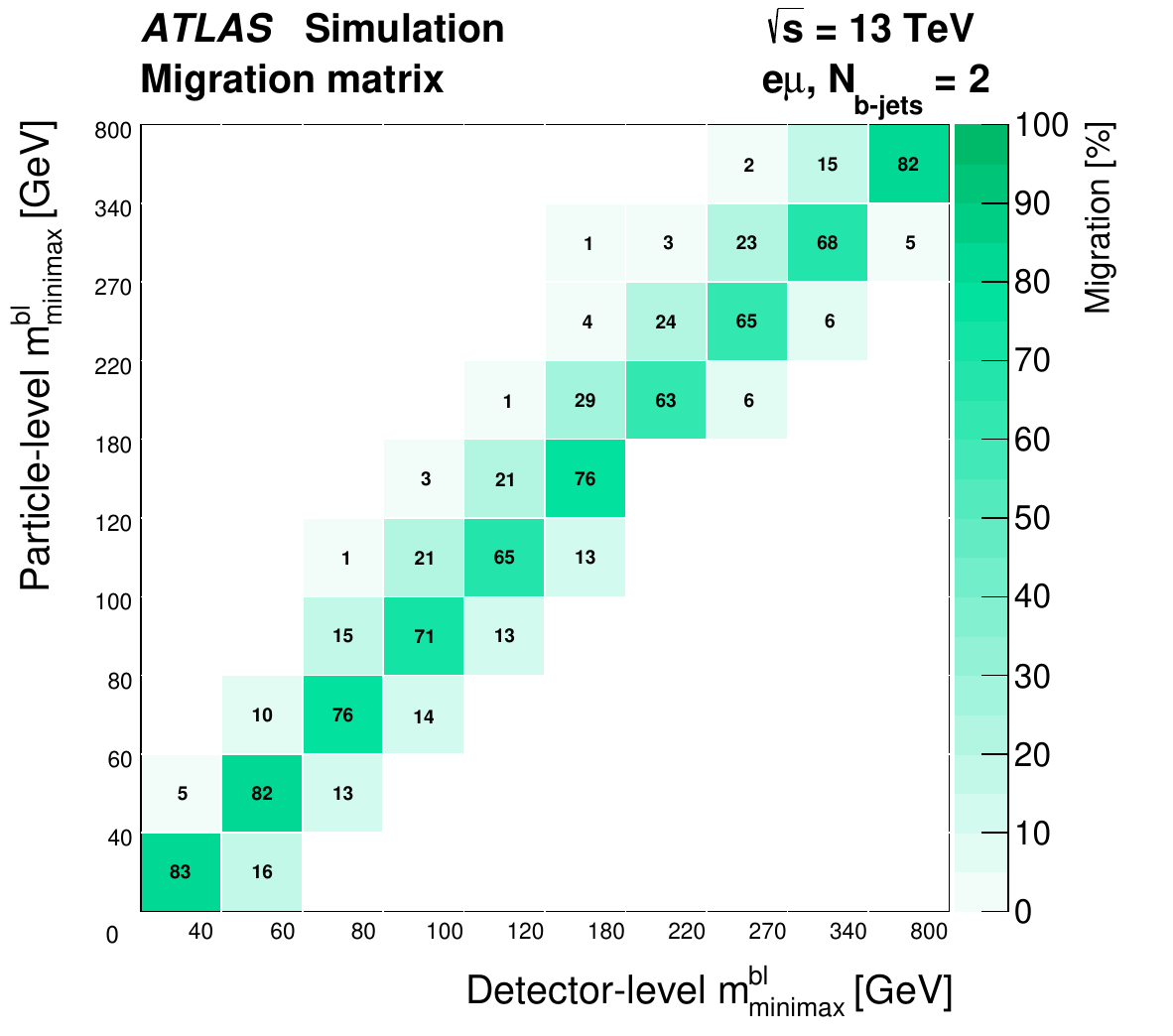}
\caption{}\label{fig:minimax:migration}
\end{subfigure}

\caption{The \protect\subref{fig:minimax:acc} acceptance ($f_{acc}$) and \protect\subref{fig:minimax:eff} efficiency ($\varepsilon$) corrections and \protect\subref{fig:minimax:migration} the migration matrix for the $\mbl$ variable in the 2$b$-exclusive region. The nominal acceptance and efficiency corrections, provided by \POWPY[8] with the DR scheme, are compared with the corrections obtained with some MC generators used to evaluate the signal modelling uncertainties. In the migration matrix, only bins where the migration is greater than 1\% are shown.}
\label{fig:minimaxmbl_acceptance_eff_migration}
\end{figure}

\begin{figure}[tb]
\centering
\begin{subfigure}[b]{0.45\textwidth}
\centering
\includegraphics[width=\linewidth]{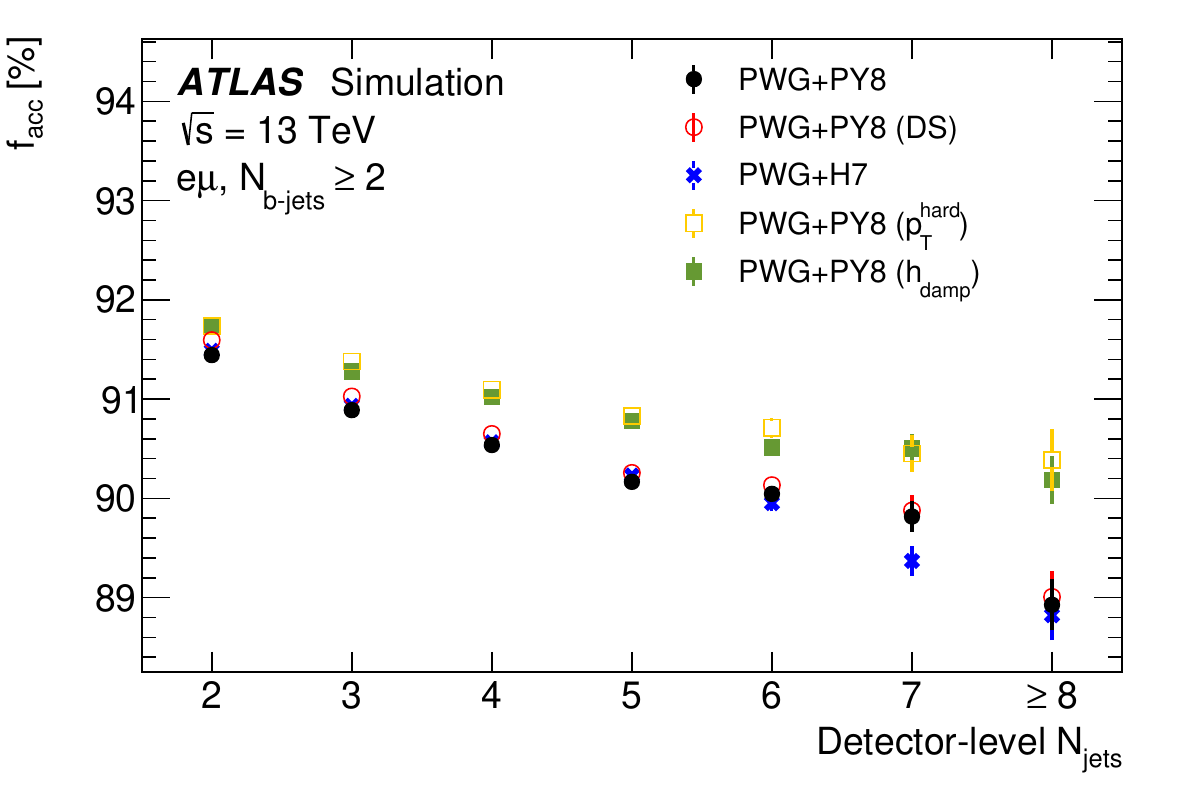}
\caption{}\label{fig:n_jets:acc}
\end{subfigure}
\begin{subfigure}[b]{0.45\textwidth}
\centering
\includegraphics[width=\linewidth]{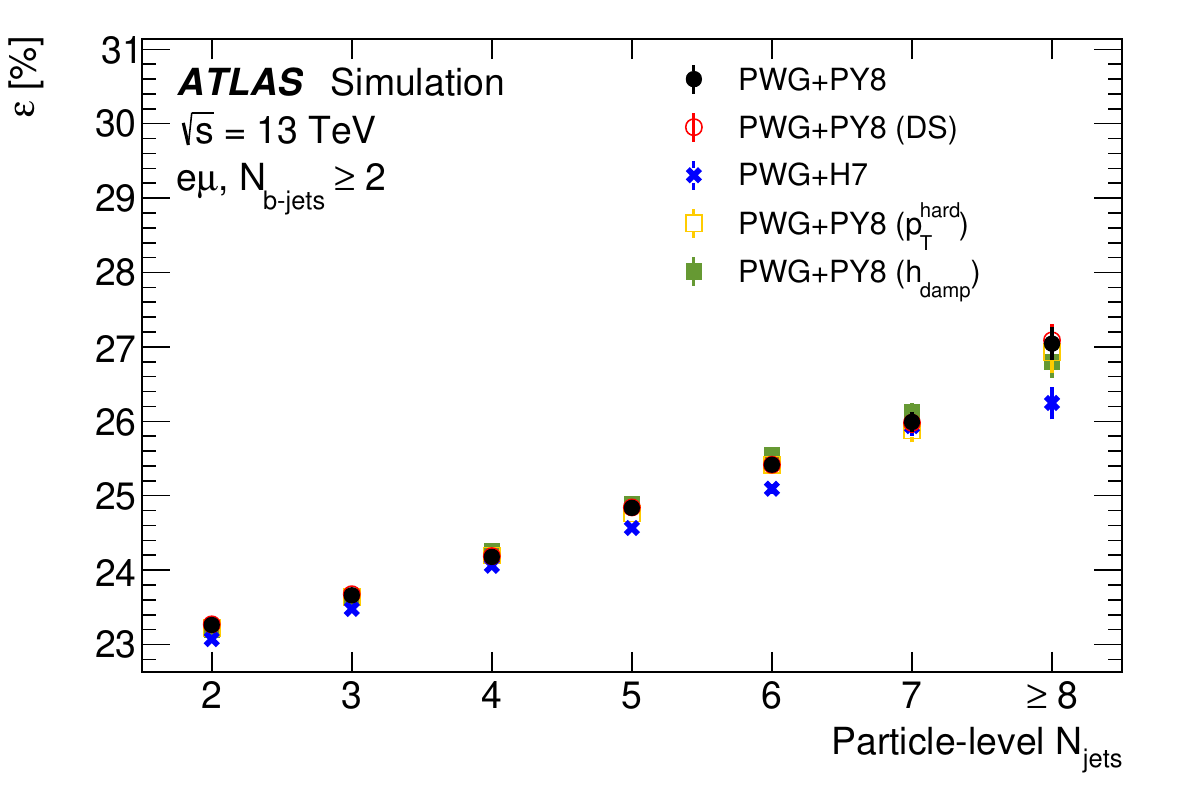}
\caption{}\label{fig:n_jets:eff}
\end{subfigure}\\
\begin{subfigure}[b]{0.45\textwidth}
\centering
\includegraphics[width=\linewidth]{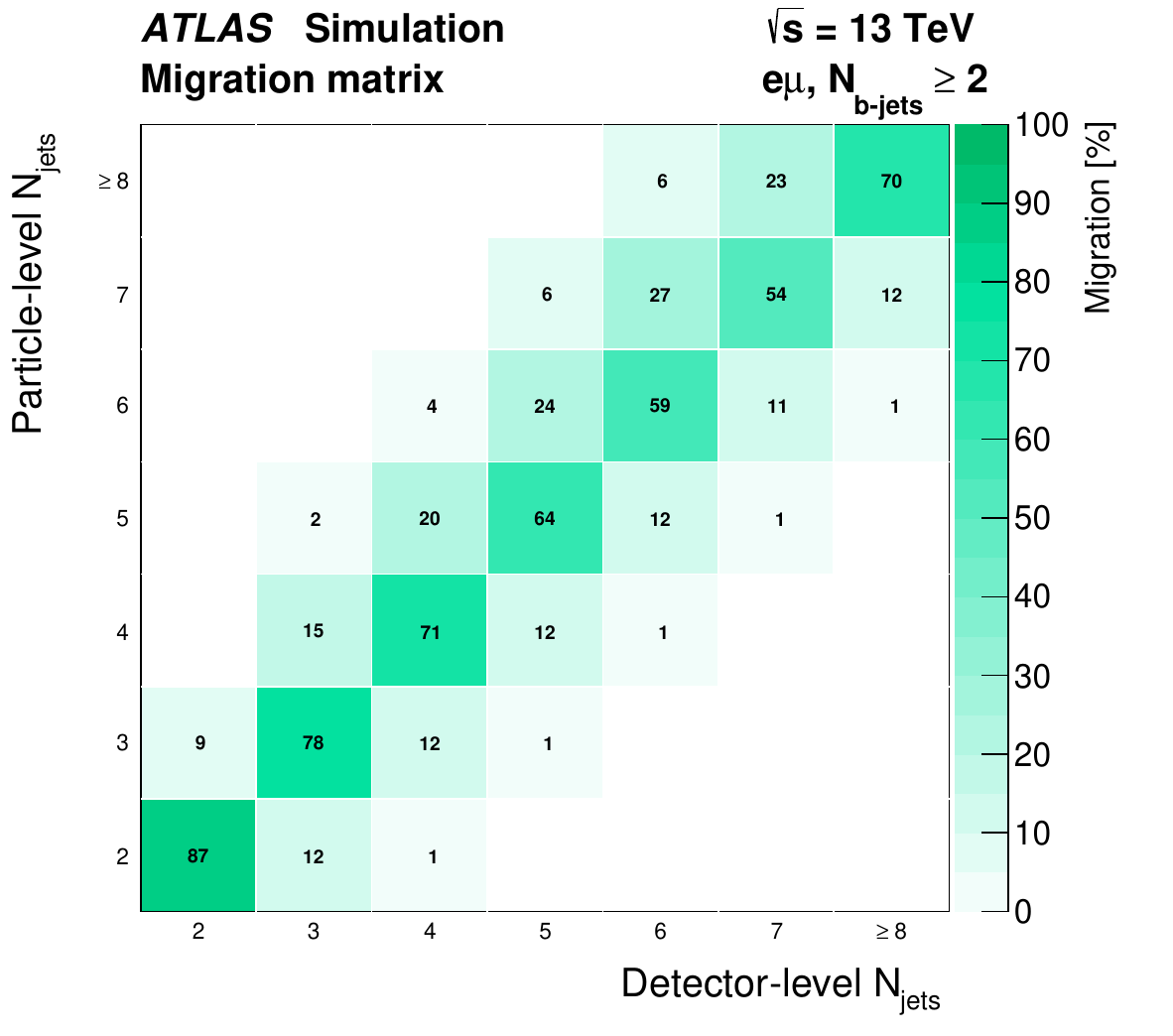}
\caption{}\label{fig:n_jets:migration}
\end{subfigure}
\caption{The \protect\subref{fig:n_jets:acc} acceptance ($f_{acc}$) and \protect\subref{fig:n_jets:eff} efficiency ($\varepsilon$) corrections and \protect\subref{fig:n_jets:migration} the migration matrix for the \Njets\ variable in the 2$b$-inclusive region. The nominal acceptance and efficiency corrections, provided by \POWPY[8] with the DR scheme, are compared with the corrections obtained with some MC generators used to evaluate the signal modelling uncertainties. In the migration matrix, only bins where the migration is greater than 1\% are shown.}
\label{fig:n_jets_acceptance_eff_migration}
\end{figure}

Examples of acceptance, efficiency, and migration matrices for the \mbl\ variable (in the 2$b$-exclusive region) are shown in \cref{fig:minimaxmbl_acceptance_eff_migration}, and for the number of jets (in the 2$b$-inclusive region) in \cref{fig:n_jets_acceptance_eff_migration}. The low values for the efficiency, in \cref{fig:minimax:eff,fig:n_jets:eff}, averaging at around 23\%, are mostly related to the requirements of having two isolated leptons and two $b$-tagged jets. The acceptance and efficiency corrections are compared with those obtained with various MC generators used to evaluate the signal modelling uncertainties. This comparison illustrates the impact of signal modelling variations on the final measurement.

In \cref{fig:minimaxmbl_acceptance_eff_migration}, a sharp drop in the acceptance and efficiency corrections at high values of \mbl\ is observed. This drop is primarily due to the \ttbar\ component of the signal, whose contribution to the high-\mbl\ region amounts to half of the total signal yield and is primarily due to events where at least one of the two $b$-tagged jets does not originate from the decay of a top quark. For such events to enter the 2$b$-exclusive region, a third $b$-jet from radiation must be present, and one of the $b$-jets from the top-quark decay must not be selected. The latter requirement significantly reduces the probability of these events satisfying the selection criteria at both particle- and detector-level, leading to the observed low acceptance and efficiency. For the rest of the variables, where the separation between \ttbar\ and $tWb$ is less pronounced, the acceptance and efficiency corrections are more stable across the phase space, as shown in \cref{fig:n_jets_acceptance_eff_migration} for the number of jets.

\FloatBarrier


%

%
\subsection{Systematic uncertainties}
\label{sec:Systematics}

This section describes the estimation of systematic uncertainties related to object reconstruction and calibration, signal modelling, and background estimation. %

\subsubsection{Uncertainty propagation}
\label{sec:syst:propagation}
To evaluate the impact of each uncertainty after the unfolding, the reconstructed signal and background distributions in simulation are varied and unfolded using corrections from the nominal \POWPY[8] signal sample. The unfolded distribution is compared with the corresponding particle-level spectrum, and the relative difference is assigned as the uncertainty in the measured distribution. For single-sided systematic uncertainties, such as most of the signal modelling ones, the relative difference is symmetrised to obtain a two-sided uncertainty.
All detector- and background-related systematic uncertainties are evaluated using the same generator, while alternative generators and generator set-ups are employed to assess modelling systematic uncertainties. In these cases, the corrections derived from the nominal generator are used to unfold the detector-level spectra of the alternative generator, and the comparison between the unfolded distribution and the alternative particle-level spectrum is used to assess the corresponding uncertainty.

\subsubsection{Object reconstruction and calibration}%
\label{sec:syst:detector}
\paragraph{Leptons and \etmiss}
Electrons and muons both have uncertainties associated with the understanding of the energy/momentum scale and resolution~\cite{EGAM-2018-01,MUON-2022-01}. These uncertainties are calculated by modifying the momentum of the reconstructed leptons and propagating these changes through the analysis chain. The lepton reconstruction efficiency in simulated events is corrected by scale factors derived from measurements of these efficiencies in data using a control region enriched in $Z\rightarrow ee$ and $Z\rightarrow\mu\mu$ events~\cite{EGAM-2021-01,MUON-2018-03}. The calibration of the efficiency of the lepton reconstruction and identification comes with associated uncertainties, and these are propagated via changes to the scale factors for the leptons. Similarly, the scale factors for the trigger efficiencies are varied to account for uncertainties in the measurements of the efficiencies of the lepton triggers~\cite{TRIG-2018-01,TRIG-2018-05}.

The uncertainty associated with $\etmiss$ is calculated by propagating the energy scale and resolution systematic uncertainties to all jets and leptons in the $\etmiss$ calculation. Additional $\etmiss$ uncertainties arising from tracks not associated with any reconstructed objects are also included~\cite{JETM-2020-03}.
\paragraph{Jets}
The impact of the uncertainty in the jet energy scale (JES) and resolution (JER) is estimated by varying the jet energies according to the uncertainties derived from simulation and in situ calibration measurements~\cite{JETM-2018-05}, parameterised by 30 independent variations for the JES and 13 for the JER\@. The JER uncertainty is evaluated by smearing both the (pseudo-)data and MC events. The uncertainty originating from the JVT requirement on jets is assessed by varying its associated scale factor~\cite{PERF-2014-03}.

\paragraph{Flavour tagging}
Differences in the $b$-tagging efficiencies between the data and simulation are corrected using scale factors~\cite{FTAG-2018-01,FTAG-2020-08,FTAG-2019-02}. The corrections are applied for the tagging efficiency of $b$-jets and the mis-tag rates of $c$- and light-jets, with corresponding uncertainties derived for each jet flavour. A principal component analysis is performed to diagonalise the covariance matrix given the input systematic uncertainties considered in the calibration of the tagging efficiencies. This results in a number of orthogonal eigen-variations for each jet flavour, which are then used as separate two-sided uncertainties. Three additional components are considered, corresponding to the uncertainties associated with the efficiency extrapolation to high jet \pt\ ($>400$\,\GeV), derived separately for $b$-jets, $c$- and light-jets.

\paragraph{Pile-up}
Variations of the scale factors used to match the \pileup to the data are used to evaluate the uncertainty in the \pileup modelling.

\subsubsection{Signal modelling uncertainties}
\label{sec:syst:modelling}

A limited understanding of the modelling of signal processes is a significant source of uncertainty. The impact of these uncertainties is evaluated using alternative predictions, as described in \cref{sec:datamc:mc}, following the procedure outlined above. Signal modelling uncertainties affecting both the \ttbar\ and $tW$ processes are classified as `hard scattering' when they affect the matching of the matrix element and parton shower, `parton shower' when they affect the parton showering and hadronisation, and `generator parameters', when they affect other settings of the \POWPY[8] generator. All these uncertainties are considered correlated between the \ttbar\ and $tW$ processes. Uncertainties affecting only the modelling of the $\ttbar$ and $tW$ processes are classified as `\ttbar\ modelling' and `$tW$ modelling', respectively.

\paragraph{Hard scattering}
The uncertainty related to the matrix-element calculation and the parton-shower matching procedure is evaluated by comparing the nominal sample with an alternative sample obtained by setting the $\pthard$ parameter in \PYTHIA[8] to 1. This parameter regulates the definition of the vetoed region of the showering, and its variation is based on the studies described in Ref.~\cite{ATL-PHYS-PUB-2023-029}.

\paragraph{Parton shower}
The uncertainty due to the choice of the hadronisation model and other non-perturbative aspects of the parton shower is evaluated using the \POWHER[7.1.3] samples for the $\ttbar$ and $tW$ processes.

\paragraph{Generator parameters}
The uncertainty due to the choice of the \mur\ and \muf\ scales in the hard scatter is evaluated using predictions obtained with \POWPY[8], where \mur\ and \muf\ in the hard scatter are varied independently by a factor 0.5 and 2. The uncertainties in the initial-state radiation are obtained by varying the scales in the showering model according to the Var3c eigentune of the A14 tune~\cite{ATL-PHYS-PUB-2017-007}.

The uncertainties in the tuning of final-state radiation in the parton shower are obtained by changing the momentum scale settings of $\mur^{\text{FSR}}$  in \PYTHIA[8] by a factor of 0.5 or 2 relative to the nominal scale.
Another parameter affecting the modelling of the additional radiation in \POWPY[8] is \hdamp. The uncertainty due to the choice of this parameter is estimated by using a dedicated \POWPY[8] sample where \hdamp\ is multiplied by a factor of two relative to the nominal value.

The uncertainty due to the choice of the top-quark mass in the MC simulations is estimated by using two additional simulation samples, for both the \ttbar\ and $tW$ processes, with varied top-quark mass values of \SI{169}{\GeV} and \SI{176}{\GeV} which, besides the mass of the top quark, have the same set-up as the nominal sample. The final uncertainty is then scaled by 1/7 to match the effect of a $\pm\SI{0.5}{\GeV}$ shift in the top-quark mass, reflecting the precision of the top-quark mass measurements performed by ATLAS and CMS~\cite{TOPQ-2017-03,CMS-TOP-14-022}.

The impact of the choice of different PDF sets is assessed by applying an event-by-event reweighting
procedure to the nominal \POWPY[8] $\ttbar$ and $tW$ samples following the \PDFforLHC[15] prescription~\cite{Butterworth:2015oua} and using the \texttt{PDF4LHC15\_30} sets.

\paragraph{$tW$ modelling}
A significant component of the uncertainty is represented by the difference between the DR and DS approaches, used to remove the overlap between the \ttbar{} and $tW$ samples. The $tW$ (DS) sample is used instead of the $tW$ (DR) to form the alternative signal, with \ttbar{} remaining the nominal for the alternative sample.

\paragraph{Signal normalisation}
Differential cross-section measurements, performed by unfolding background-subtracted distributions in data, are typically insensitive to the overall normalisation of the signal sample. This is because the signal sample contributes to the unfolding only through efficiency, acceptance, and migration corrections. These corrections, being ratios of yields, do not depend on the total cross-section. This principle holds when the signal sample consists of a single process, as in the case of \ttbar\ differential cross-sections. However, in this analysis, the signal comprises the sum of two processes: the $tW$ and \ttbar\ production. This composition makes the unfolding corrections sensitive to the relative normalisation between the two processes. Consequently, two sources of systematic uncertainty are considered: `\ttbar\ normalisation' and `$tW$ normalisation', and assigned to the `\ttbar\ modelling' and `$tW$ modelling' categories, respectively. These uncertainties are estimated by independently varying the normalisation of the \ttbar\ and $tW$ samples, while keeping the normalisation of the other sample fixed to its nominal value. The variations are defined by rescaling the samples by  $1\pm\delta_{xs}$, where $\delta_{xs}$ is the relative error on the \ttbar\ (NNLO+NNLL) and $tW$ (NLO+NNLL) cross-sections, amounting to 5.2\% and 3.5\% respectively, as described in \cref{sec:datamc:mc_nominal_signal}.

\subsubsection{Uncertainties in the background processes}
\label{sec:systematics:background}

A global uncertainty, accounting for variations in PDFs, $\alphas$, \muF\ and \muR,  is applied to the MC prediction of the $Z$+jets background components. This uncertainty is parameterised as a function of the number of jets in the event, as calculated in Ref.~\cite{ATL-PHYS-PUB-2017-006}.

A conservative normalisation uncertainty of 50\% is applied to the diboson background, including the uncertainty in the cross-section and a contribution due to the presence of additional jets, as discussed in Ref.~\cite{STDM-2018-03}.

The uncertainties in the cross-section calculation for the processes \ttZ, \ttW\ and \ttH\ composing the \ttX\ sample are respectively 13.3$\%$,  12.5$\%$, and 9.9$\%$~\cite{deFlorian:2016spz}. A conservative uncertainty of 13.3$\%$ is assigned to all the processes contributing to the \ttX\ background.

Finally, a 50\% normalisation uncertainty is assigned to the background due to fake and non-prompt leptons to reflect the uncertainty in evaluating it from MC\@. This is validated in regions with similar kinematic requirements as the signal regions but with the two leptons of the same charge, where the contribution of events with fake/non-prompt leptons is expected to dominate, contributing around 60\% of the total event yield.

\subsubsection{Other uncertainties}

The uncertainty due to the number of data events is propagated through the unfolding by creating Poisson-smeared pseudo-experiments that are passed through the unfolding procedure. For normalised differential cross-sections, the normalisation procedure is applied for each pseudo-experiment. The procedure is replicated $10$k times, then the final statistical uncertainty is evaluated from the standard deviation in each bin over the $10$k pseudo-experiments.

To account for the limited number of events in the MC samples, pseudo-experiments are used to evaluate the impact of
finite number of events. The number of events in each bin is smeared by a Gaussian distribution with mean equal to the yield of the bin, and
standard deviation equal to the uncertainty of the bin. The resulting systematic uncertainty is found to be negligible in most cases.

The uncertainty in the combined 2015--2018 integrated luminosity is 0.83\%~\cite{DAPR-2021-01}, obtained using the LUCID-2 detector~\cite{LUCID2} for the primary luminosity measurements, complemented by measurements using the inner detector and calorimeters. This uncertainty is propagated to the unfolded results by varying the luminosity by $\pm 0.83\%$ in the reconstructed simulated spectra and unfolding them using the nominal corrections.

\subsection{Uncertainty summary}
\begin{figure}[tb]
\centering
\begin{subfigure}[b]{0.45\textwidth}
\centering
\includegraphics[width=\linewidth]{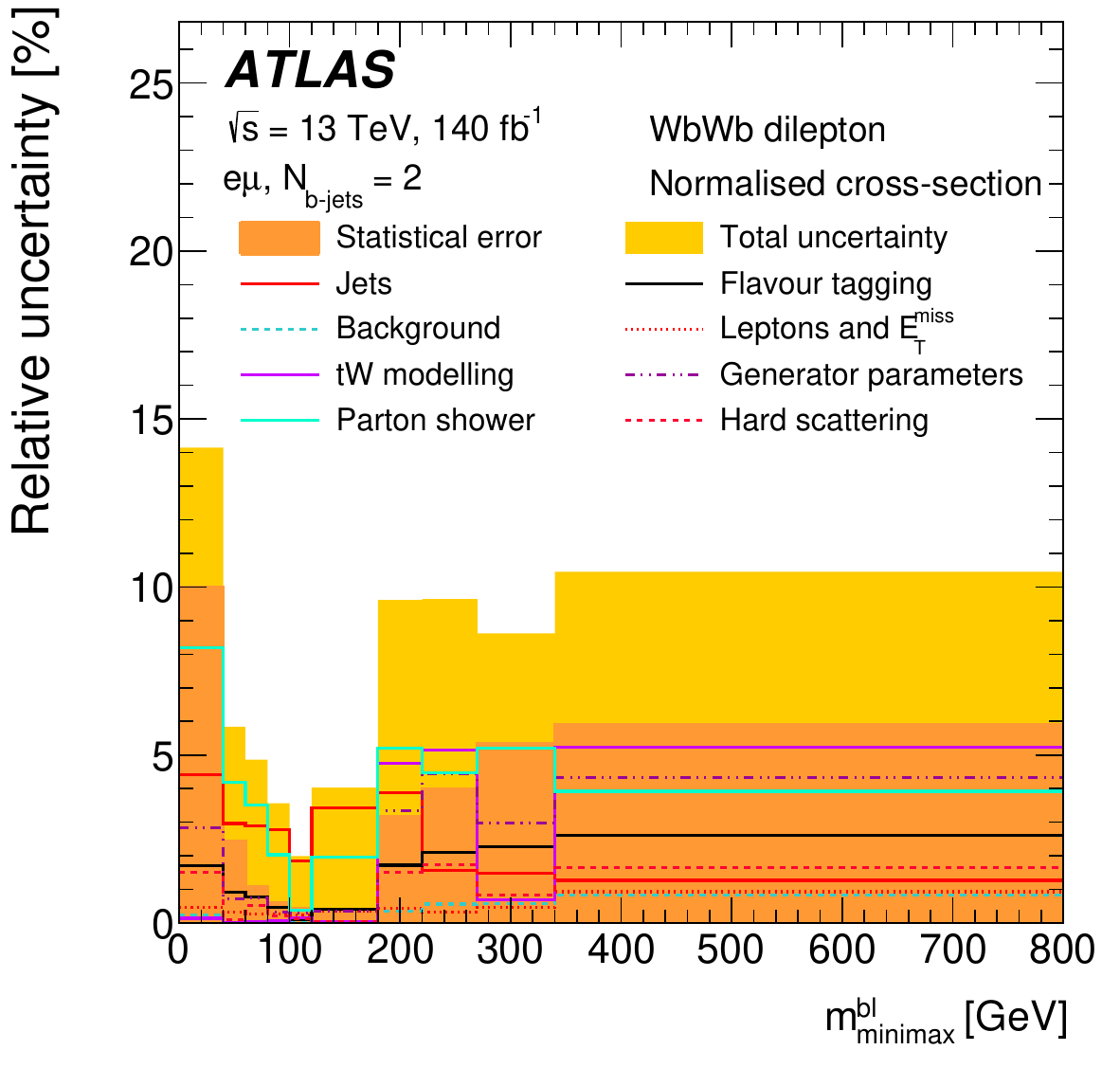}
\caption{}\label{fig:systematics:minimaxmbl}
\end{subfigure}
\begin{subfigure}[b]{0.45\textwidth}
\centering
\includegraphics[width=\linewidth]{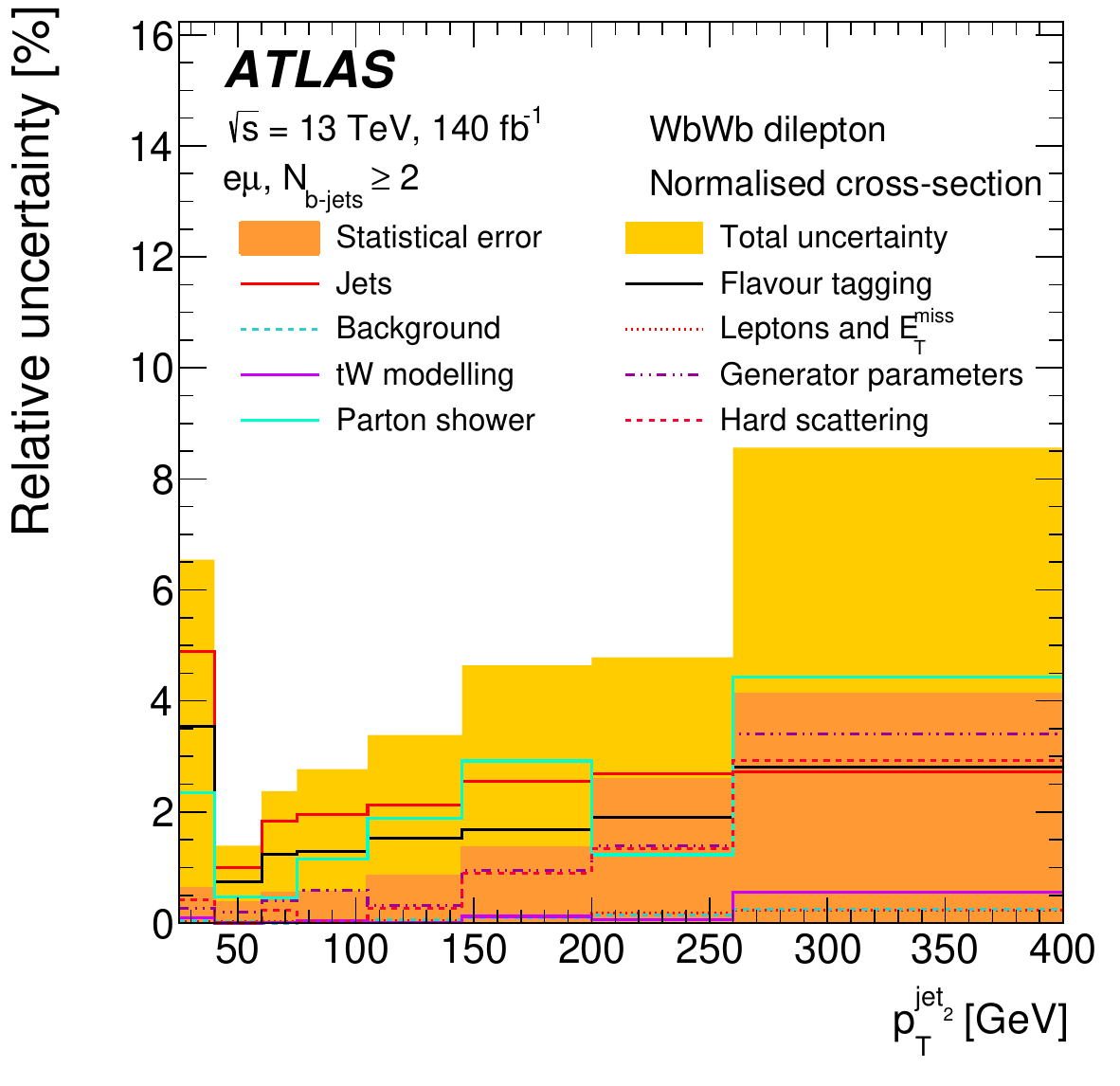}
\caption{}\label{fig:systematics:pT_jet2}
\end{subfigure}
\caption{Uncertainties in the particle-level normalised differential cross-section as a function of \protect\subref{fig:systematics:minimaxmbl}~\mbl\ and \protect\subref{fig:systematics:pT_jet2}  the subleading jet \pt. The bands represent the statistical and total uncertainty in the data.}
\label{fig:systematics}
\end{figure}

The breakdown of the impact of the different categories of uncertainties is shown in \cref{fig:systematics} for the measured normalised differential cross-section as a function of \mbl\ and the leading lepton \pt. For clarity, only the most relevant contributions are shown.

As shown in \cref{fig:systematics:minimaxmbl}, in the last four bins of the \mbl\ distribution that are most sensitive to the interference, the uncertainty is dominated by the modelling uncertainties, in particular those related to the choice of the scheme used to treat the $\ttbar/tW$ interference, the parton shower, and those related to the matrix element matching, and by the statistical uncertainty. The total uncertainty is around 10\% or lower in all bins except the first one, which is a significant reduction compared to the same region of phase space measured in Ref.~\cite{TOPQ-2017-05} where the uncertainty was above 20\%. This significant improvement is achieved due to the new modelling uncertainties and improved object calibrations and to an optimisation of the selection and binning of the distribution. In the lower mass region, the total uncertainty is dominated by the jet and parton shower uncertainties.

For the other distributions, the uncertainties are dominated by jet uncertainties, especially for jet observables, and modelling uncertainties, especially in the tails of the distributions. The statistical component can also be significant in the extreme regions of phase space, as visible in the subleading jet \pt\ breakdown plot shown in \cref{fig:systematics:pT_jet2}. For these distributions, the total uncertainties are also below 10\% in almost all the bins.

\FloatBarrier


%

\section{Results}
\label{sec:results}

The measured normalised differential cross-sections are presented in \cref{sec:results:exclusive,sec:results:inclusive}. The agreement of the different models with the measured differential cross-sections is assessed by computing the $\chi^2$, and corresponding $p$-values, of the measured differential cross-sections relative to theoretical predictions employing the full experimental covariance matrix. The $\chi^2$ values are computed for the specific sample configurations described, without including theoretical uncertainties. This approach avoids assumptions about the correlation schemes for each comparison and ensures clarity in the interpretation of the results.

To facilitate comparisons with any future MC generator predictions, all differential cross-section measurements, both normalised and absolute, together with the experimental covariance matrices and full uncertainty breakdown, will be published in HEPData~\cite{Maguire:2017ypu}.

\subsection{$\chi^2$ calculation}
\label{sec:results:chi2}

The level of agreement between the measured cross-sections and the NLO+PS predictions is evaluated using a $\chi^{2}$ test statistic. The $p$-value is extracted from the $\chi^{2}$ and number of degrees of freedom ($\text{NDF}$), with the $\text{NDF}$ equal to the number of bins ($N_b$) in the case of the absolute differential cross-sections. The $\chi^{2}$ is computed using the formula:
\begin{displaymath}
\chi^2=\mathbf{V}^\text{T} \times \mathbf{C}^{-1} \times \mathbf{V}
\end{displaymath}

where $\mathbf{V}$ represents the vector of residuals, defined as the difference between the measured differential cross-section and the prediction, and $\mathbf{C}$ is the experimental covariance matrix, which is built from the uncertainties described in the previous section. No uncertainty on the theoretical predictions is included in the covariance matrix.

For normalised differential cross-sections, the vector of differences between data and predictions $\mathbf{V}_{N_b}$ is replaced with $\mathbf{V}_{N_b-1}$, which is the same quantity but obtained by discarding the last one of the $N_b$ elements. Consequently, $\mathbf{C}_{N_b-1}$ is the $(N_b-1) \times (N_b-1)$ sub-matrix derived from the full covariance matrix of the normalised measurements by discarding the corresponding row and column. The sub-matrix obtained in this way is invertible and allows the $\chi^2$ to be computed. The $\chi^2$ value does not depend on the choice of the element discarded for the vector $\mathbf{V}_{N_b-1}$ and the corresponding sub-matrix $\mathbf{C}_{N_b-1}$. In this case, the $\text{NDF}$ becomes $N_b-1$.

\subsection{Differential cross-sections in the $2b$-exclusive region}
\label{sec:results:exclusive}
\begin{figure}[tb]
\centering
\includegraphics[width=.75\textwidth]{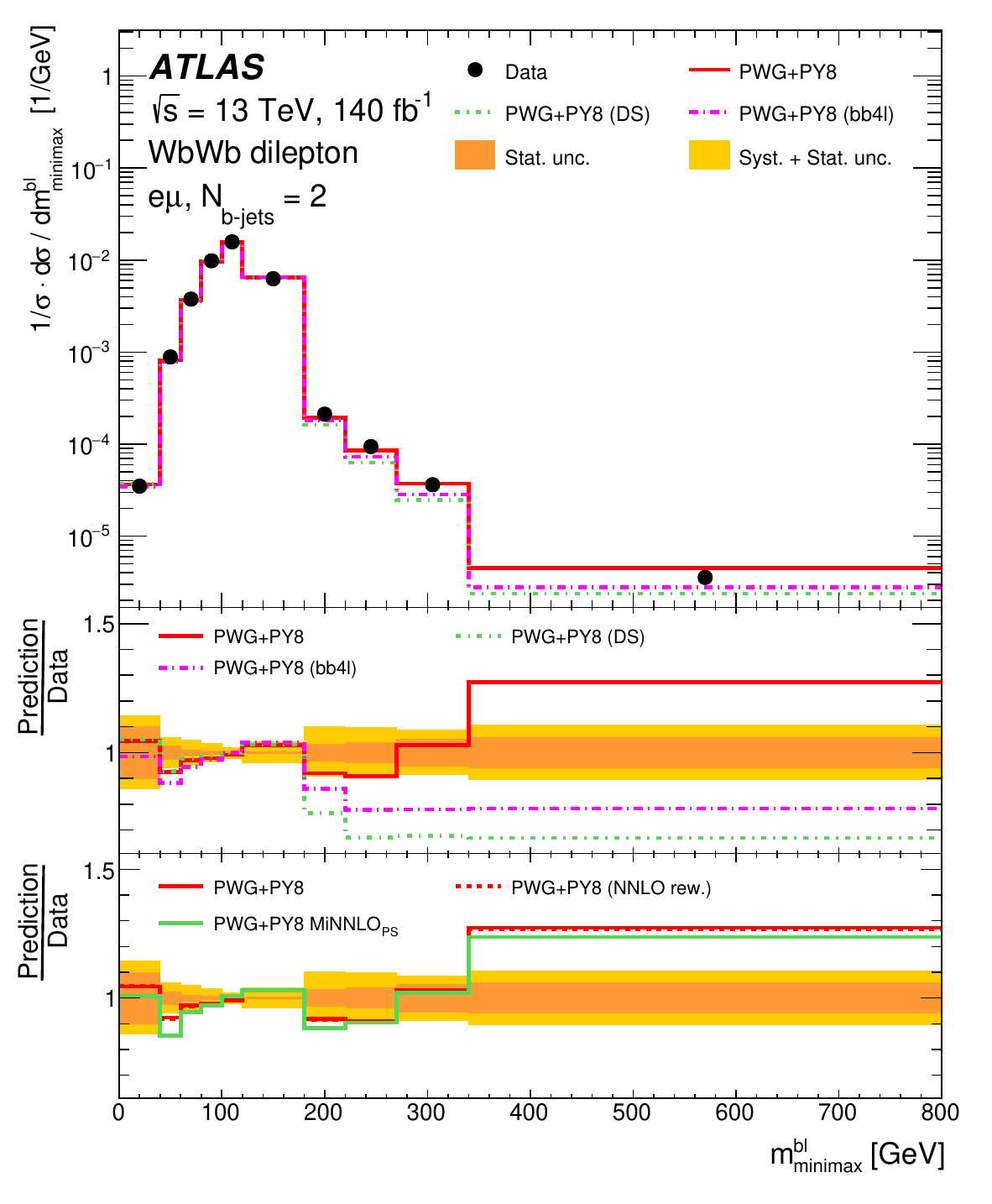}
\caption{Relative differential cross-section measurement as a function of the $\mbl$ variable. The cross-section is unfolded to particle-level in the fiducial phase space. The measurement is compared with predictions provided by \POWPY[8] with different schemes for computing the $\ttbar$/$tW$ interference and with predictions where the \ttbar\ component is simulated with an NNLO+PS MC generator or corrected to NNLO. The lower panels show the ratios of the predictions to the data with the bands showing the statistical and total uncertainties in the data.  The nominal \POWPY[8] prediction is included in all ratio panels.}
\label{fig:results:rel:minimaxmbl}
\end{figure}

The measured normalised differential cross-section as a function of $\mbl$, in the $2b$-exclusive fiducial phase space, is shown in \cref{fig:results:rel:minimaxmbl}. The measurement is compared with \POWPY[8] predictions obtained by employing different interference schemes (namely $\ttbar+tW$ DR and DS and the full $bb4l$ final state) and by including higher-order corrections for the \ttbar\ component, either by using the \minnlo\ method or the NNLO reweighting.

The $\chi^{2}$ and $p$-values for the normalised differential cross-sections are summarised in \cref{tab:ParticleLevelChi2_minimaxmbl_normalised}. Among the available predictions, the sample where the $tW$ process is simulated using the DR scheme exhibits the best agreement with the data. However, it still fails to accurately describe the shape of the tail of the \mbl\ distribution.
It is worth highlighting the PWG+PY8 ($bb4l$) sample, which, while expected to be the most accurate model for describing the interference, does not fulfil this expectation. The PWG+PY8 ($bb4l$) prediction underestimates the cross-section in the tail of the distribution where the impact of the interference is larger. This measurement will be crucial for improving future versions of this generator.

No major difference relative to the nominal \POWPY[8] sample is observed when comparing the predictions obtained with the samples where the \ttbar\ process is modelled at NNLO either via the PWG+PY8 \minnlo\ MC or NNLO reweighting.

The PWG+PY8 (DS) prediction is the least compatible with the data and shows the worst agreement with the data above 180~\GeV.

\begin{table}[t]
\centering
\small
\begin{tabular}[t]{ c| r @{/} l r  } \toprule
Models & \multicolumn{2}{c}{$\chi^{2}$/NDF} &  ~$p$-value\\
\hline
PWG+PY8 & 12.8 & 9 & 0.17 \\
PWG+PY8 (DS) & 39.1 & 9 & <0.01 \\
PWG+PY8 ($bb4l$) & 28.6 & 9 & <0.01 \\
PWG+PY8 (NNLO rew.) & 12.8 & 9 & 0.17 \\
PWG+PY8 \minnlo & 17.6 & 9 & 0.04 \\
\hline
\bottomrule
\end{tabular}
\caption{$\chi^{2}$ and $p$-values quantifying the level of agreement between the measured normalised differential cross-sections and predictions for the \mbl\ variable.}
\label{tab:ParticleLevelChi2_minimaxmbl_normalised}
\end{table}


\FloatBarrier
\subsection{Differential cross-sections in the $2b$-inclusive region}
\label{sec:results:inclusive}
\begin{figure}[tb]
\centering
\begin{subfigure}[b]{0.48\textwidth}
\centering
\includegraphics[width=\linewidth]{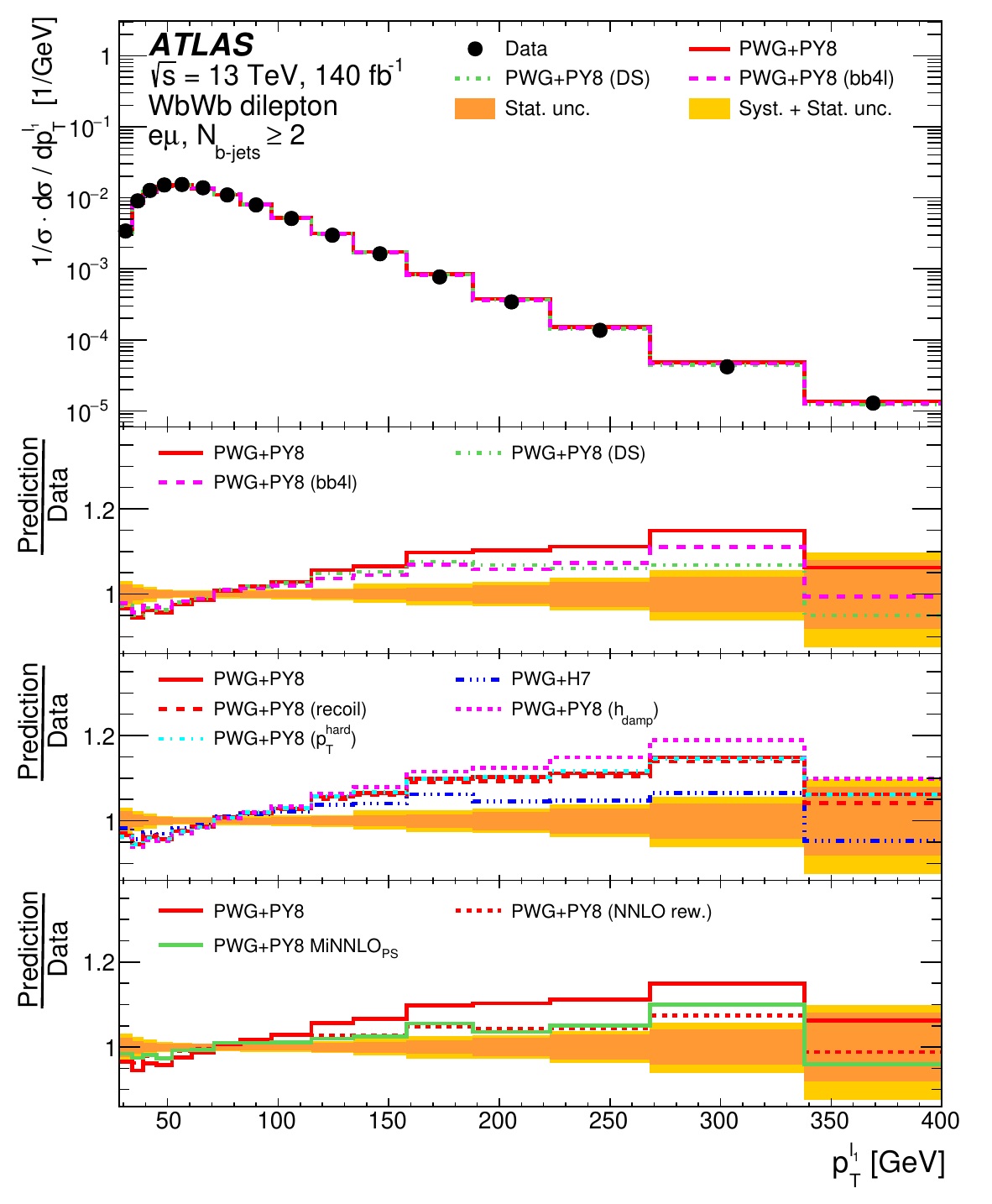}
\caption{}\label{fig:results:rel:pT_lep1}
\end{subfigure}
\begin{subfigure}[b]{0.48\textwidth}
\centering
\includegraphics[width=\linewidth]{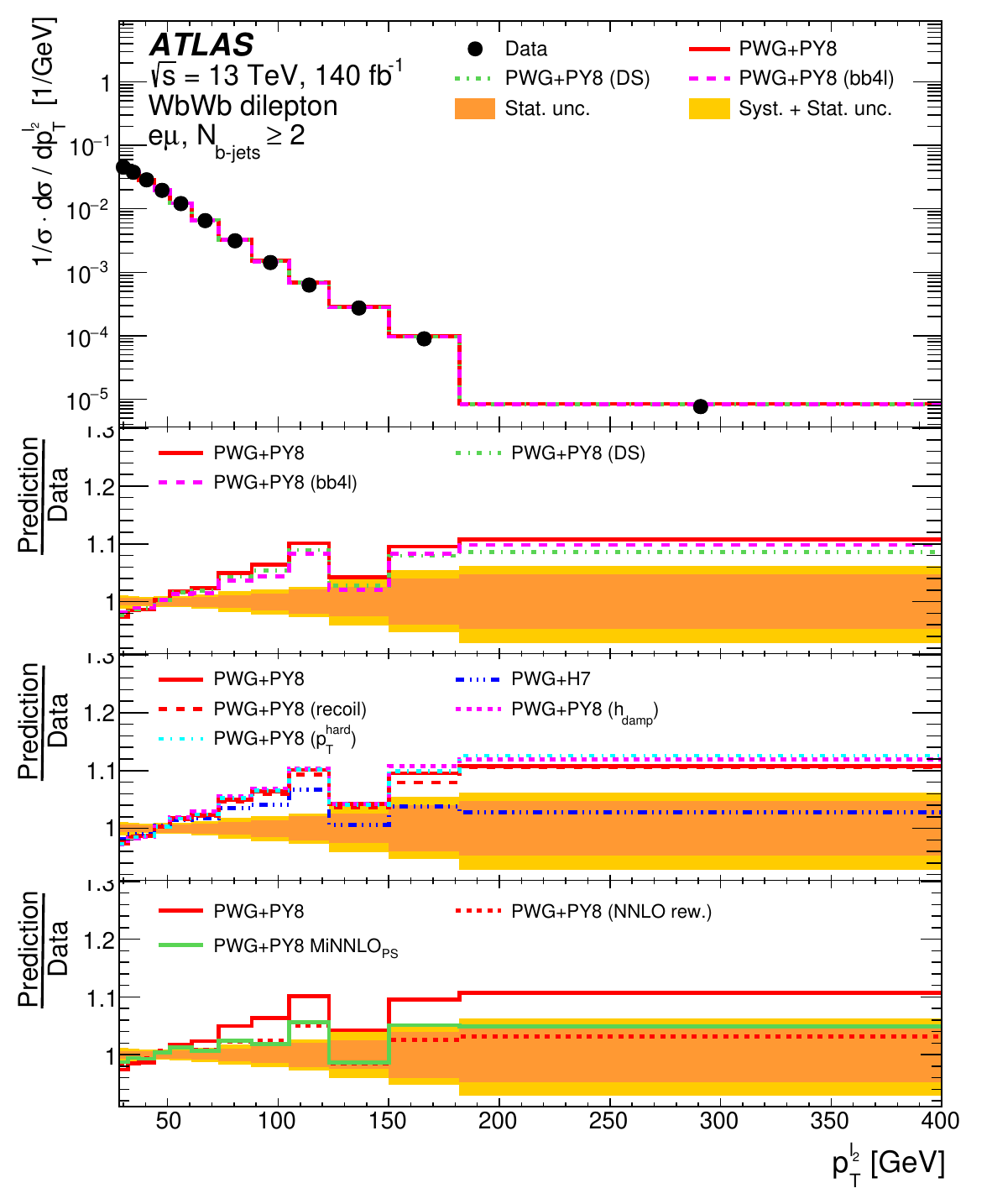}
\caption{}\label{fig:results:rel:pT_lep2}
\end{subfigure}
\caption{Relative differential cross-section measurement as a function of the \protect\subref{fig:results:rel:pT_lep1} $\PT[1]{l}$ and \protect\subref{fig:results:rel:pT_lep2} $\PT[2]{l}$ variables. The cross-section is unfolded to particle-level in the fiducial phase space. The measurement is compared with predictions provided by \POWPY[8] with different schemes for computing the $\ttbar$/$tW$ interference, with predictions obtained with different NLO+PS MC generators and predictions where the \ttbar\ component is simulated with an NNLO+PS MC generator or corrected to NNLO. The lower panels show the ratios of the predictions to the data with the bands showing the statistical and total uncertainties in the data. The nominal \POWPY[8] prediction is included in all ratio panels.}
\label{fig:results:rel:pT_lep1_pT_lep2}
\end{figure}

\begin{figure}[tb]
\centering
\begin{subfigure}[b]{0.48\textwidth}
\centering
\includegraphics[width=\linewidth]{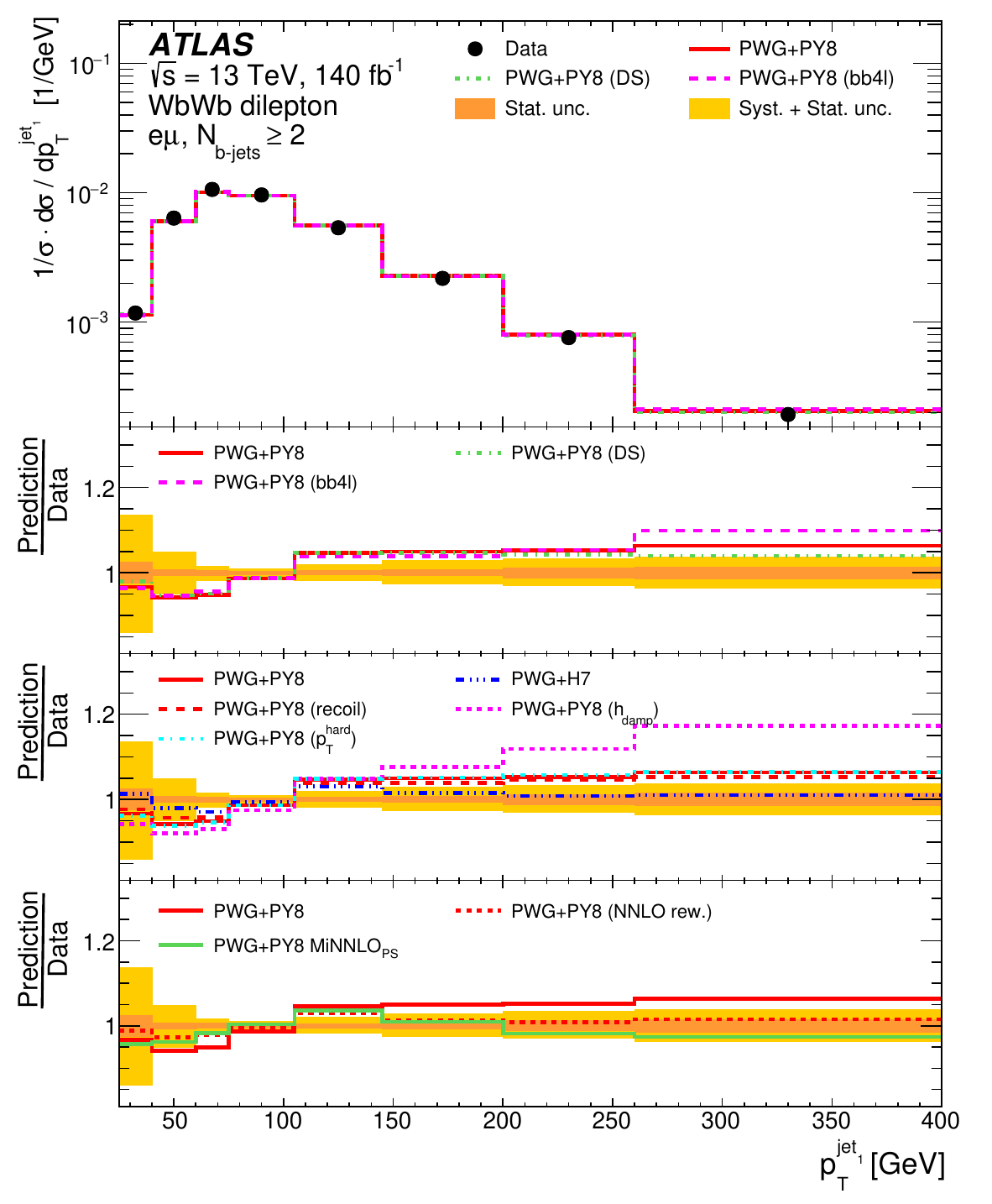}
\caption{}\label{fig:results:rel:pT_jet1}
\end{subfigure}
\begin{subfigure}[b]{0.48\textwidth}
\centering
\includegraphics[width=\linewidth]{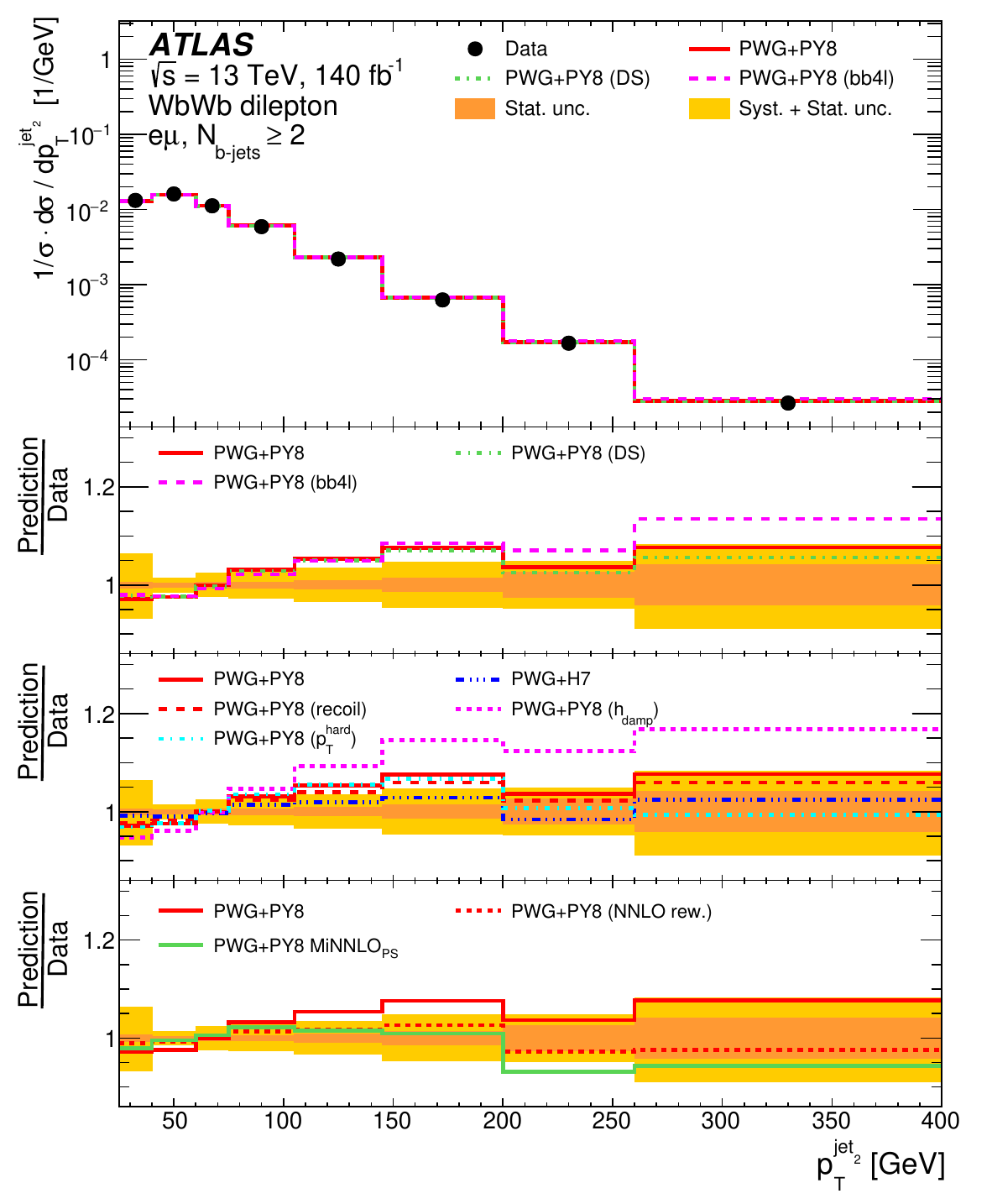}
\caption{}\label{fig:results:rel:pT_jet2}
\end{subfigure}
\caption{Relative differential cross-section measurement as a function of the \protect\subref{fig:results:rel:pT_jet1} $\PT[1]{\text{jet}}$ and \protect\subref{fig:results:rel:pT_jet2} $\PT[2]{\text{jet}}$ variables. The cross-section is unfolded to particle-level in the fiducial phase space. The measurement is compared with predictions provided by \POWPY[8] with different schemes for computing the $\ttbar$/$tW$ interference, with predictions obtained with different NLO+PS MC generators and predictions where the \ttbar\ component is simulated with an NNLO+PS MC generator or corrected to NNLO. The lower panels show the ratios of the predictions to the data with the bands showing the statistical and total uncertainties in the data. The nominal \POWPY[8] prediction is included in all ratio panels.}
\label{fig:results:rel:pT_jet1_pT_jet2}
\end{figure}

\begin{figure}[tb]
\centering
\begin{subfigure}[b]{0.48\textwidth}
\centering
\includegraphics[width=\linewidth]{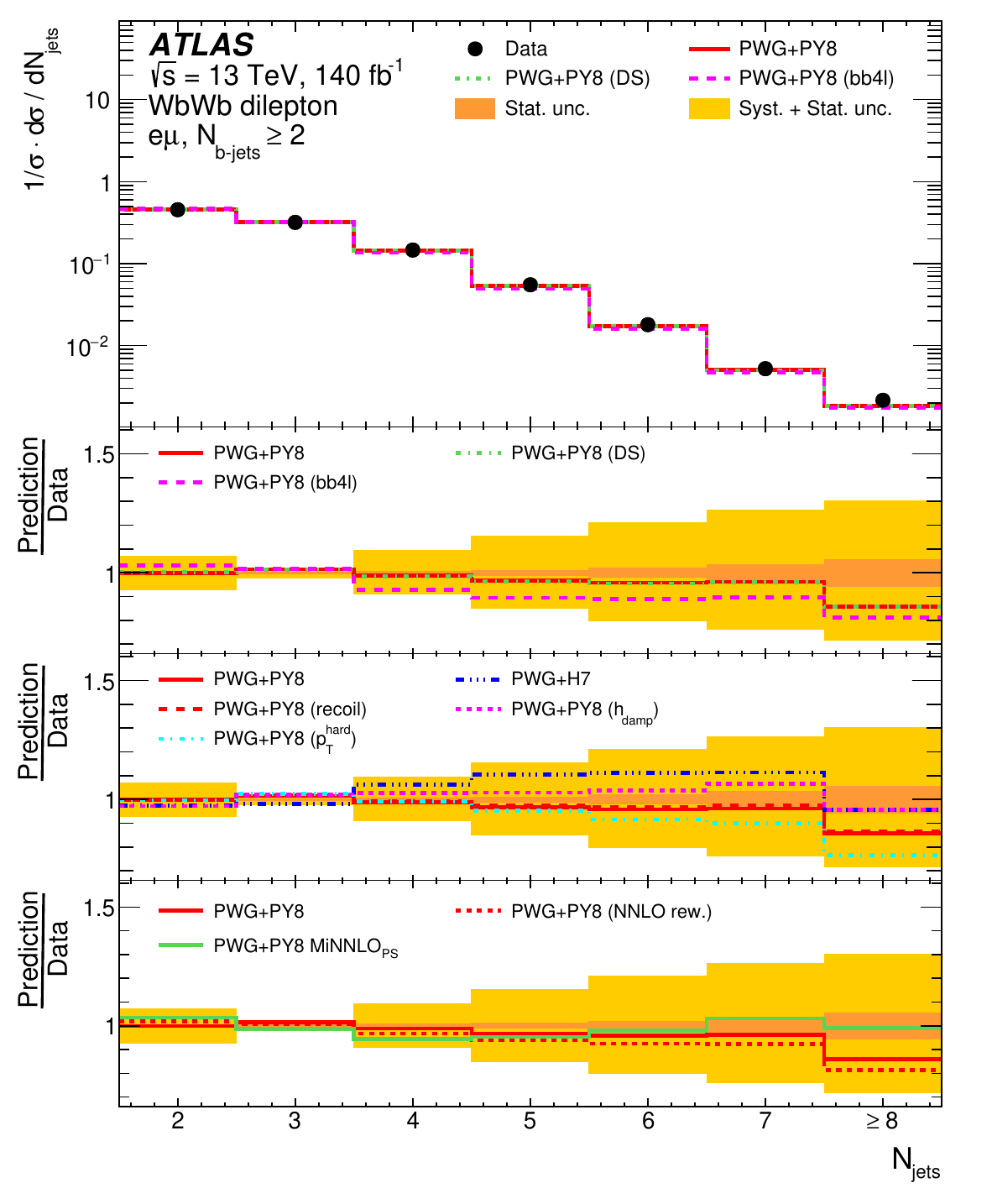}
\caption{}\label{fig:results:rel:n_jets}
\end{subfigure}
\begin{subfigure}[b]{0.48\textwidth}
\centering
\includegraphics[width=\linewidth]{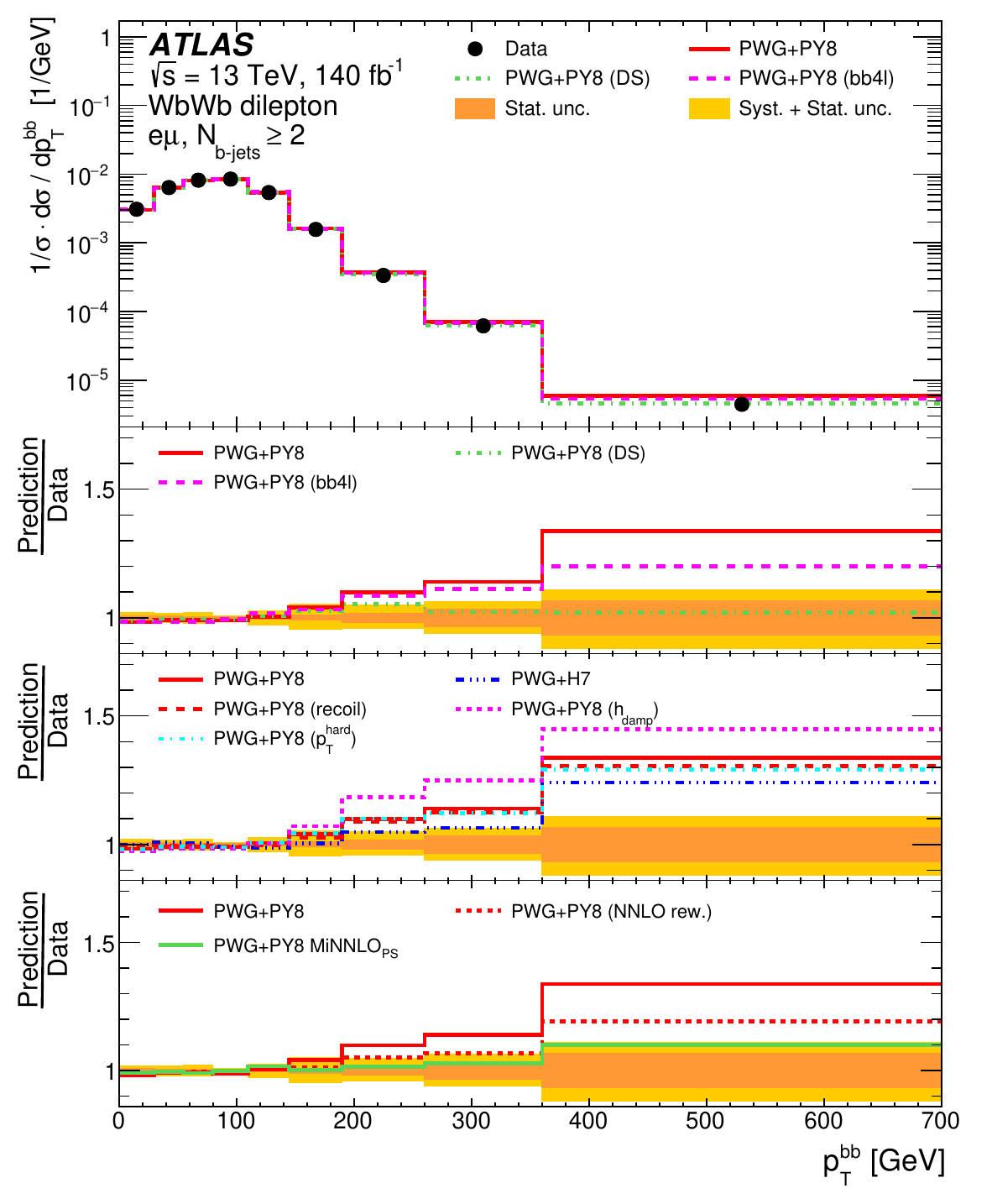}
\caption{}\label{fig:results:rel:pT_b1b2}
\end{subfigure}
\caption{Relative differential cross-section measurement as a function of the \protect\subref{fig:results:rel:n_jets} $\njets$ and \protect\subref{fig:results:rel:pT_b1b2} $\PT{bb}$ variables. The cross-section is unfolded to particle-level in the fiducial phase space. The measurement is compared with predictions provided by \POWPY[8] with different schemes for computing the $\ttbar$/$tW$ interference, with predictions obtained with different NLO+PS MC generators and predictions where the \ttbar\ component is simulated with an NNLO+PS MC generator or corrected to NNLO. The lower panels show the ratios of the predictions to the data with the bands showing the statistical and total uncertainties in the data. The nominal \POWPY[8] prediction is included in all ratio panels.}
\label{fig:results:rel:n_jets_pT_b1b2}
\end{figure}

\begin{figure}[tb]
\centering
\begin{subfigure}[b]{0.48\textwidth}
\centering
\includegraphics[width=\linewidth]{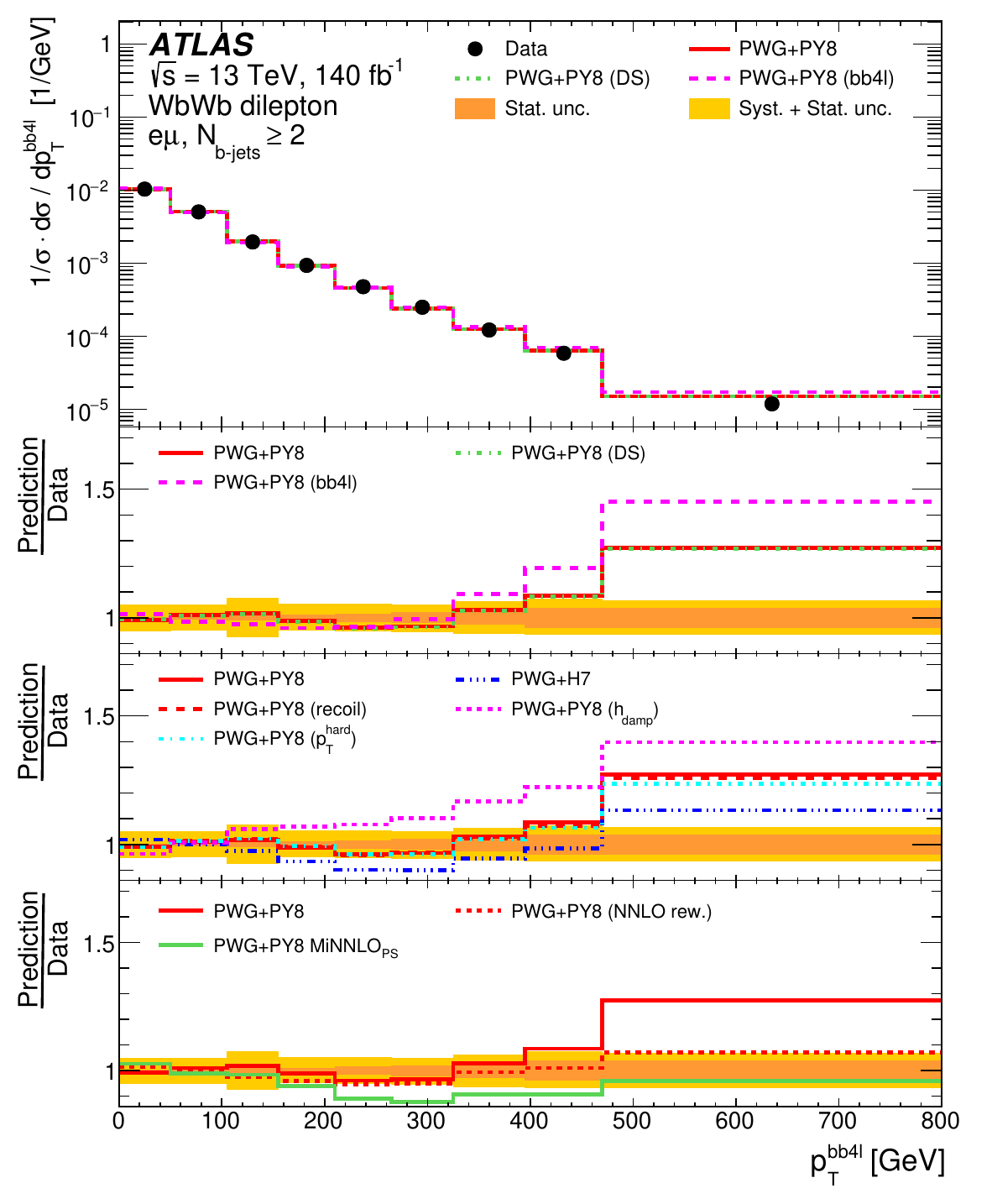}
\caption{}\label{fig:results:rel:pT_ttbar}
\end{subfigure}
\begin{subfigure}[b]{0.48\textwidth}
\centering
\includegraphics[width=\linewidth]{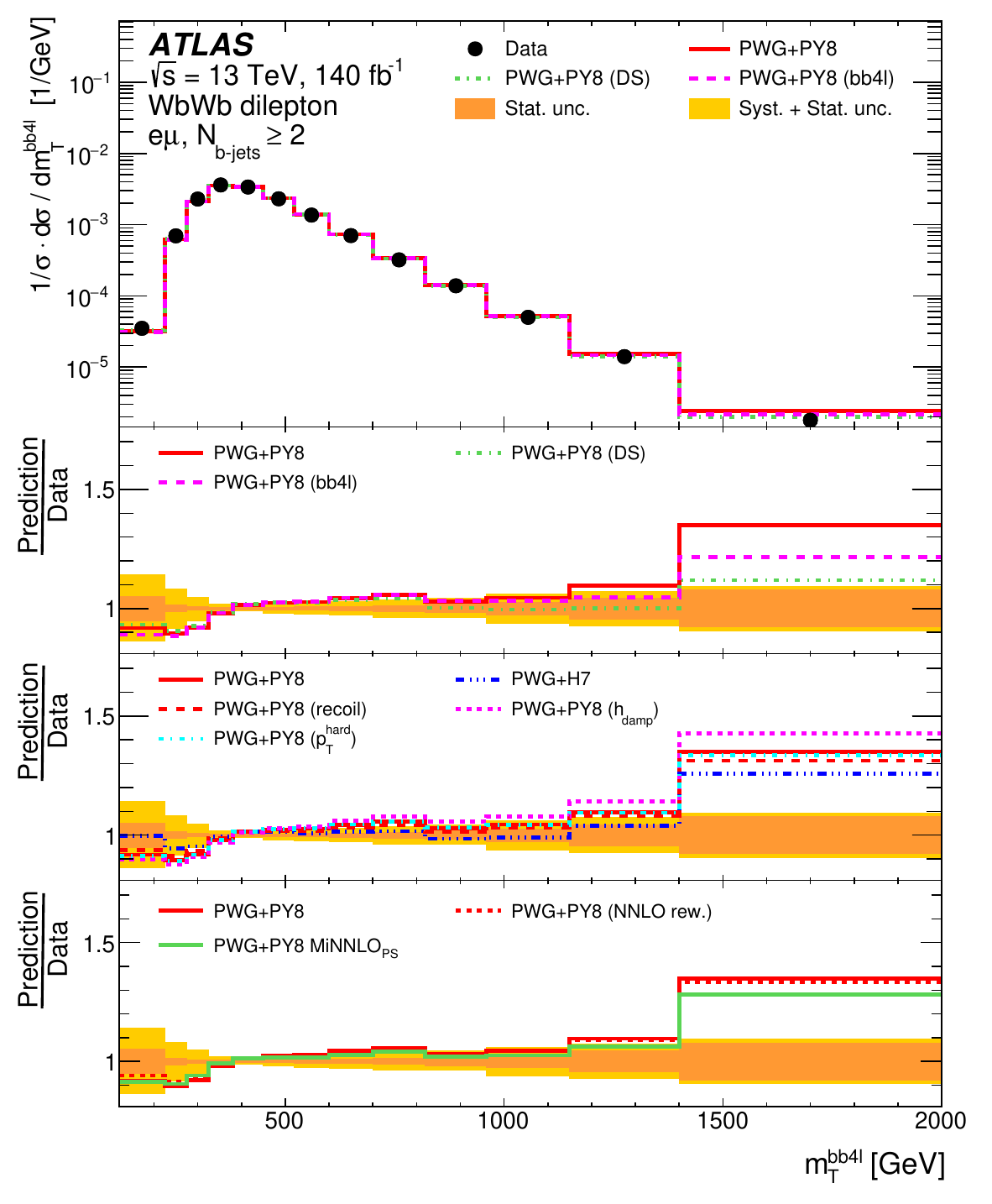}
\caption{}\label{fig:results:rel:mT_ttbar}
\end{subfigure}
\caption{Relative differential cross-section measurement as a function of the \protect\subref{fig:results:rel:pT_ttbar} $\PT{bb4l}$ and \protect\subref{fig:results:rel:mT_ttbar} $\mT^{bb4l}$ variables. The cross-section is unfolded to particle-level in the fiducial phase space. The measurement is compared with predictions provided by \POWPY[8] with different schemes for computing the $\ttbar$/$tW$ interference, with predictions obtained with different NLO+PS MC generators and predictions where the \ttbar\ component is simulated with an NNLO+PS MC generator or corrected to NNLO. The lower panels show the ratios of the predictions to the data with the bands showing the statistical and total uncertainties in the data. The nominal \POWPY[8] prediction is included in all ratio panels.}
\label{fig:results:rel:pT_ttbar_mT_ttbar}
\end{figure}

\begin{figure}[tb]
\centering
\begin{subfigure}[b]{0.48\textwidth}
\centering
\includegraphics[width=\linewidth]{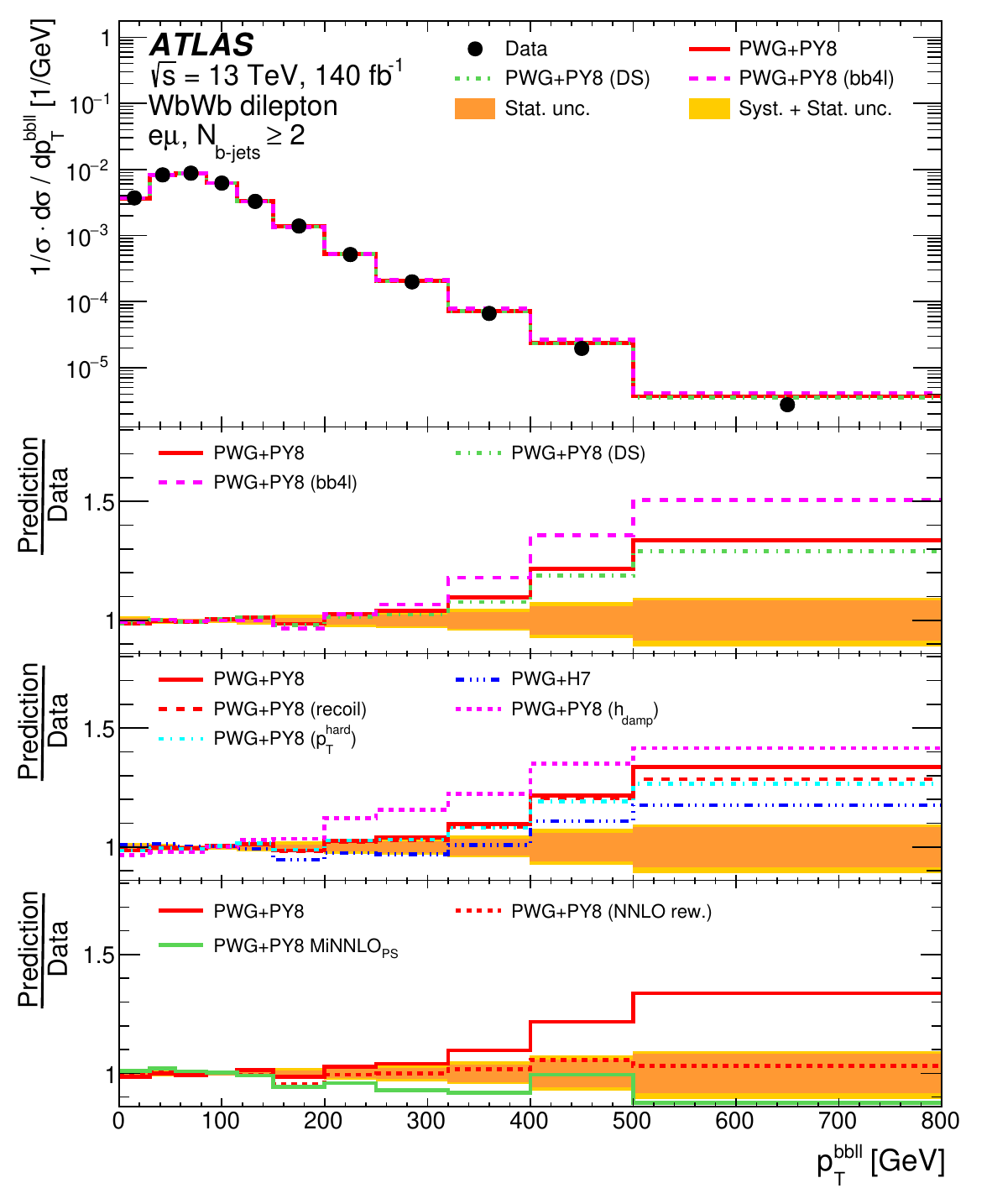}
\caption{}\label{fig:results:rel:pT_bbll}
\end{subfigure}
\begin{subfigure}[b]{0.48\textwidth}
\centering
\includegraphics[width=\linewidth]{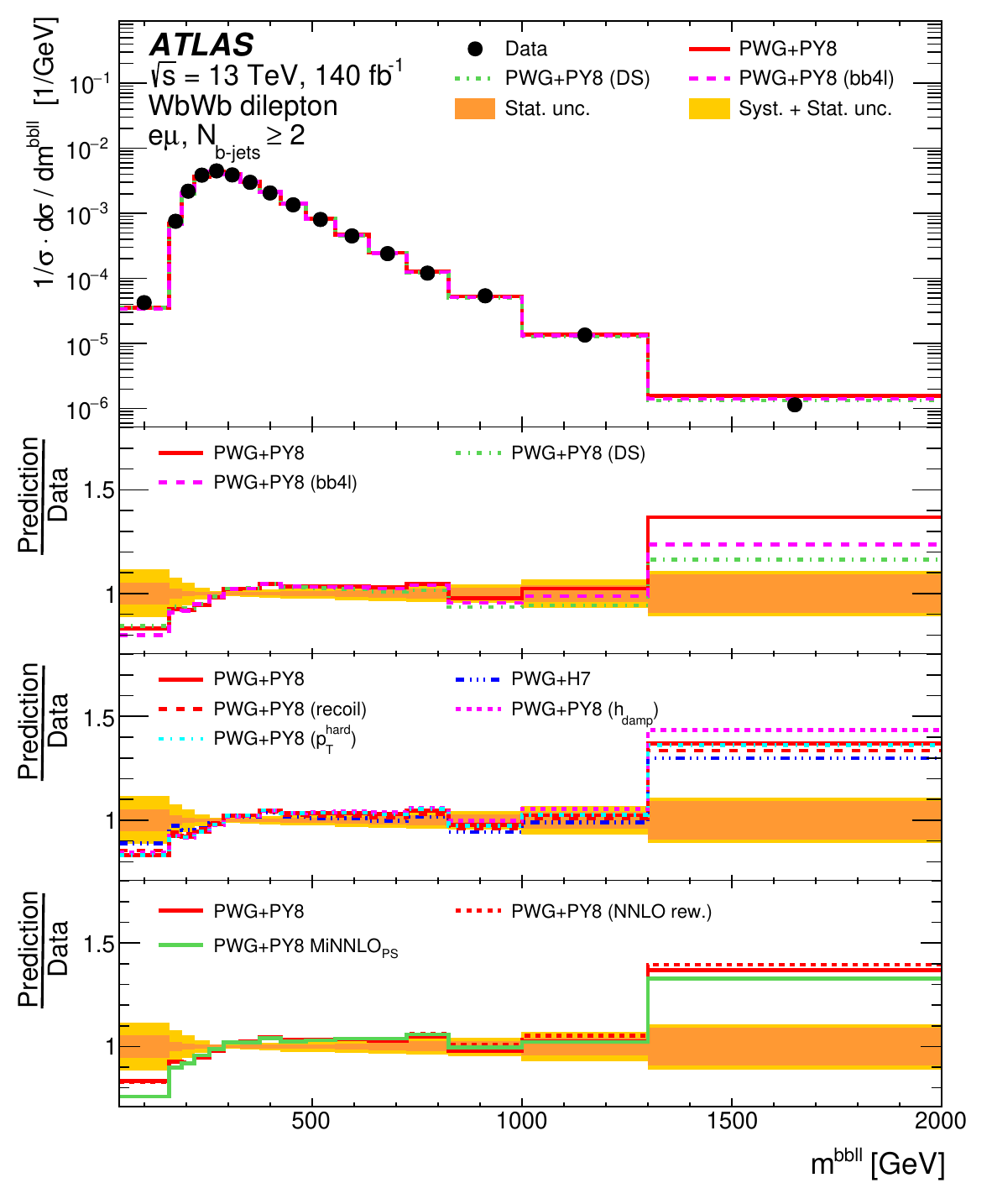}
\caption{}\label{fig:results:rel:m_bbll}
\end{subfigure}
\caption{Relative differential cross-section measurement as a function of the \protect\subref{fig:results:rel:pT_bbll} $\PT{bbll}$ and \protect\subref{fig:results:rel:m_bbll} $m^{bbll}$ variables. The cross-section is unfolded to particle-level in the fiducial phase space. The measurement is compared with predictions provided by \POWPY[8] with different schemes for computing the $\ttbar$/$tW$ interference, with predictions obtained with different NLO+PS MC generators and predictions where the \ttbar\ component is simulated with an NNLO+PS MC generator or corrected to NNLO. The lower panels show the ratios of the predictions to the data with the bands showing the statistical and total uncertainties in the data. The nominal \POWPY[8] prediction is included in all ratio panels.}
\label{fig:results:rel:pT_bbll_m_bbll}
\end{figure}

The measured normalised differential cross-sections as a function of the kinematic variables of the \wbwb\ final-state objects, measured in the $2b$-inclusive phase space, are presented in \cref{fig:results:rel:pT_lep1_pT_lep2} for the leading lepton and sub-leading lepton \pt\ and in \cref{fig:results:rel:pT_jet1_pT_jet2} for the leading jet and sub-leading jet \pt.
Differential cross-sections as a function of the jet multiplicity are shown in \cref{fig:results:rel:n_jets}.
Differential cross-sections as a function of additional variables, typically used for MC optimisation, are also measured. Specifically:
\begin{itemize}
\item the cross-sections as a function of the \pT\ of the $bb$ system are shown in \cref{fig:results:rel:pT_b1b2}
\item the cross-sections as a function of the \pT\ and \mT\ of the $bb4l$ system are shown in \cref{fig:results:rel:pT_ttbar_mT_ttbar};
\item the cross-sections as a function of the \pT\ and mass of the $bbll$ system are shown in \cref{fig:results:rel:pT_bbll_m_bbll}.
\end{itemize}

The compatibility with the data is evaluated by calculating the $\chi^{2}$ and $p$-values, which are summarised in \cref{tab:ParticleLevelChi2_interference_normalised,tab:ParticleLevelChi2_generators_normalised,tab:ParticleLevelChi2_minnlops_normalised}. In general, most of the predictions are not able to describe simultaneously all observables. This was the case in other measurements both in dilepton and semileptonic final states~\cite{TOPQ-2018-15,TOPQ-2018-26}. Among the NLO+PS models, one of the best predictions is provided by \POWHER[7], which is the only one providing a good modelling of both the leading jet and lepton \pt. The PWG+PY8 ($bb4l$) sample also shows a good description of the leading and subleading jets and leptons.

The inclusion of NNLO corrections for the \ttbar\ component of the signal, either by applying the NNLO reweighting to the \ttbar\ simulation or performing a full NNLO+PS simulation via PWG+PY8 \minnlo, described in \cref{sec:datamc:mc_alt_signal}, improves the predictions significantly, which is reflected by the $p$-values increasing systematically for almost all variables. In particular, PWG+PY8 (NNLO rew.) is the only prediction that can describe all the variables simultaneously. This effect is minor for \mbl\ in the $2b$-exclusive phase space, since the NNLO corrections do not significantly change the shape of the \ttbar\ component, which mostly populates the already well-modelled low \mbl\ region.

The models showing the worst compatibility with the data are those with alternative \hdamp\ and \pthard\ variations, which are compatible with only one observable. The other models have intermediate results. No NLO+PS MC can describe the \pt\ of the $bbll$ and $bb4l$ systems, which are more sensitive to higher-order effects and additional radiation.


\begin{table}[t]
\centering
\small
\begin{tabular}[t]{ c| r @{/} l r  | r @{/} l r  | r @{/} l r  } \toprule
\multirow{2}{*}{Observable} & \multicolumn{3}{c|}{\textsc{PWG+PY8}}& \multicolumn{3}{c|}{\textsc{PWG+PY8 (DS)}}& \multicolumn{3}{c}{\textsc{PWG+PY8 ($bb4l$)}}\\
& \multicolumn{2}{c}{$\chi^{2}$/NDF} &  ~$p$-value& \multicolumn{2}{c}{$\chi^{2}$/NDF} &  ~$p$-value& \multicolumn{2}{c}{$\chi^{2}$/NDF} &  ~$p$-value\\
\hline
$\njets$&{\ } 8.7 & 6 & 0.19&{\ } 7.5 & 6 & 0.28&{\ } 20.8 & 6 & <0.01\\
$m^{bbll}$&{\ } 28.1 & 15 & 0.02&{\ } 26.2 & 15 & 0.04&{\ } 23.8 & 15 & 0.07\\
$\mT^{bb4l}$&{\ } 18.9 & 12 & 0.09&{\ } 13.1 & 12 & 0.36&{\ } 12.2 & 12 & 0.43\\
$\PT{bb}$&{\ } 12.1 & 8 & 0.15&{\ } 2.5 & 8 & 0.96&{\ } 6.2 & 8 & 0.62\\
$\PT[1]{\text{jet}}$&{\ } 16.4 & 7 & 0.02&{\ } 16.8 & 7 & 0.02&{\ } 15.1 & 7 & 0.03\\
$\PT[2]{\text{jet}}$&{\ } 4.3 & 7 & 0.75&{\ } 4.5 & 7 & 0.72&{\ } 5.0 & 7 & 0.66\\
$\PT{bb4l}$&{\ } 50.6 & 8 & <0.01&{\ } 50.8 & 8 & <0.01&{\ } 122.9 & 8 & <0.01\\
$\PT{bbll}$&{\ } 36.2 & 10 & <0.01&{\ } 30.5 & 10 & <0.01&{\ } 93.6 & 10 & <0.01\\
$\PT[2]{l}$&{\ } 23.5 & 11 & 0.02&{\ } 19.1 & 11 & 0.06&{\ } 15.5 & 11 & 0.16\\
$\PT[1]{l}$&{\ } 39.4 & 15 & <0.01&{\ } 28.8 & 15 & 0.02&{\ } 21.1 & 15 & 0.13\\
\bottomrule
\end{tabular}
\caption{$\chi^{2}$ and $p$-values quantifying the level of agreement between the measured normalised differential cross-sections and predictions provided by \POWPY[8] with different schemes for computing the $\ttbar/tW$ interference.}
\label{tab:ParticleLevelChi2_interference_normalised}
\end{table}


\begin{table}[t]
\centering
\footnotesize
\begin{tabular}[t]{ c| r @{/} l r  | r @{/} l r  | r @{/} l r  | r @{/} l r  | r @{/} l r  } \toprule
\multirow{2}{*}{Observable} & \multicolumn{3}{c|}{\textsc{PWG+PY8}}& \multicolumn{3}{c|}{\textsc{PWG+PY8 (\hdamp)}}& \multicolumn{3}{c|}{\textsc{PWG+PY8 (\pthard)}}& \multicolumn{3}{c|}{\textsc{PWG+PY8 (recoil)}}& \multicolumn{3}{c}{\textsc{PWG+H7}}\\
& \multicolumn{2}{c}{$\chi^{2}$/NDF} &  ~$p$-value& \multicolumn{2}{c}{$\chi^{2}$/NDF} &  ~$p$-value& \multicolumn{2}{c}{$\chi^{2}$/NDF} &  ~$p$-value& \multicolumn{2}{c}{$\chi^{2}$/NDF} &  ~$p$-value& \multicolumn{2}{c}{$\chi^{2}$/NDF} &  ~$p$-value\\
\hline
$\njets$&{\ } 8.7 & 6 & 0.19&{\ } 7.8 & 6 & 0.25&{\ } 23.0 & 6 & <0.01&{\ } 9.1 & 6 & 0.17&{\ } 22.3 & 6 & <0.01\\
$m^{bbll}$&{\ } 28.1 & 15 & 0.02&{\ } 30.8 & 15 & <0.01&{\ } 29.6 & 15 & 0.01&{\ } 25.5 & 15 & 0.04&{\ } 26.6 & 15 & 0.03\\
$\mT^{bb4l}$&{\ } 18.9 & 12 & 0.09&{\ } 23.3 & 12 & 0.03&{\ } 18.1 & 12 & 0.11&{\ } 16.2 & 12 & 0.18&{\ } 15.3 & 12 & 0.22\\
$\PT{bb}$&{\ } 12.1 & 8 & 0.15&{\ } 31.2 & 8 & <0.01&{\ } 10.3 & 8 & 0.24&{\ } 10.4 & 8 & 0.24&{\ } 8.0 & 8 & 0.44\\
$\PT[1]{\text{jet}}$&{\ } 16.4 & 7 & 0.02&{\ } 34.5 & 7 & <0.01&{\ } 17.5 & 7 & 0.01&{\ } 11.7 & 7 & 0.11&{\ } 10.1 & 7 & 0.18\\
$\PT[2]{\text{jet}}$&{\ } 4.3 & 7 & 0.75&{\ } 13.6 & 7 & 0.06&{\ } 7.1 & 7 & 0.42&{\ } 2.5 & 7 & 0.93&{\ } 1.7 & 7 & 0.98\\
$\PT{bb4l}$&{\ } 50.6 & 8 & <0.01&{\ } 69.5 & 8 & <0.01&{\ } 40.7 & 8 & <0.01&{\ } 46.2 & 8 & <0.01&{\ } 42.7 & 8 & <0.01\\
$\PT{bbll}$&{\ } 36.2 & 10 & <0.01&{\ } 76.4 & 10 & <0.01&{\ } 26.5 & 10 & <0.01&{\ } 31.8 & 10 & <0.01&{\ } 24.0 & 10 & <0.01\\
$\PT[2]{l}$&{\ } 23.5 & 11 & 0.02&{\ } 27.2 & 11 & <0.01&{\ } 25.2 & 11 & <0.01&{\ } 20.6 & 11 & 0.04&{\ } 14.0 & 11 & 0.23\\
$\PT[1]{l}$&{\ } 39.4 & 15 & <0.01&{\ } 53.1 & 15 & <0.01&{\ } 39.6 & 15 & <0.01&{\ } 34.8 & 15 & <0.01&{\ } 19.0 & 15 & 0.21\\
\bottomrule
\end{tabular}
\caption{$\chi^{2}$ and $p$-values quantifying the level of agreement between the measured normalised differential cross-sections and predictions provided by different MC generators.}
\label{tab:ParticleLevelChi2_generators_normalised}
\end{table}


\begin{table}[t]
\centering
\small
\begin{tabular}[t]{ c| r @{/} l r  | r @{/} l r  | r @{/} l r  } \toprule
\multirow{2}{*}{Observable} & \multicolumn{3}{c|}{\textsc{PWG+PY8}} & \multicolumn{3}{c|}{\textsc{PWG+PY8 (NNLO rew.)}}& \multicolumn{3}{c}{\textsc{PWG+PY8 \minnlo}}\\
& \multicolumn{2}{c}{$\chi^{2}$/NDF} &  ~$p$-value& \multicolumn{2}{c}{$\chi^{2}$/NDF} &  ~$p$-value& \multicolumn{2}{c}{$\chi^{2}$/NDF} &  ~$p$-value\\
\hline
$\njets$&{\ } 8.7 & 6 & 0.19&{\ } 5.5 & 6 & 0.48&{\ } 15.2 & 6 & 0.02\\
$m^{bbll}$&{\ } 28.1 & 15 & 0.02&{\ } 25.8 & 15 & 0.04&{\ } 23.8 & 15 & 0.07\\
$\mT^{bb4l}$&{\ } 18.9 & 12 & 0.09&{\ } 17.3 & 12 & 0.14&{\ } 13.0 & 12 & 0.37\\
$\PT{bb}$&{\ } 12.1 & 8 & 0.15&{\ } 4.2 & 8 & 0.84&{\ } 2.9 & 8 & 0.94\\
$\PT[1]{\text{jet}}$&{\ } 16.4 & 7 & 0.02&{\ } 7.1 & 7 & 0.42&{\ } 13.2 & 7 & 0.07\\
$\PT[2]{\text{jet}}$&{\ } 4.3 & 7 & 0.75&{\ } 2.6 & 7 & 0.92&{\ } 6.4 & 7 & 0.49\\
$\PT{bb4l}$&{\ } 50.6 & 8 & <0.01&{\ } 11.9 & 8 & 0.16&{\ } 32.4 & 8 & <0.01\\
$\PT{bbll}$&{\ } 36.2 & 10 & <0.01&{\ } 12.9 & 10 & 0.23&{\ } 16.0 & 10 & 0.10\\
$\PT[2]{l}$&{\ } 23.5 & 11 & 0.02&{\ } 9.2 & 11 & 0.60&{\ } 9.4 & 11 & 0.58\\
$\PT[1]{l}$&{\ } 39.4 & 15 & <0.01&{\ } 11.8 & 15 & 0.70&{\ } 13.4 & 15 & 0.57\\
\bottomrule
\end{tabular}
\caption{$\chi^{2}$ and $p$-values quantifying the level of agreement between the measured normalised differential cross-sections and predictions provided by the nominal  \POWPY[8] set-up,  with the NNLO reweight and \minnlo.}
\label{tab:ParticleLevelChi2_minnlops_normalised}
\end{table}


\FloatBarrier

\subsection{Integrated fiducial cross-section}
\label{sec:results:total}
The integrated fiducial cross-section is measured in both phase spaces, by applying the unfolding procedure described in \cref{sec:analysis_strategy} to a single-bin distribution. In this case, the unfolding matrix is a $1\times1$ matrix, and the unfolding procedure from \cref{eq:unfolding} is equivalent to a simple division of the acceptance-corrected, background-subtracted reconstructed event count in data by the efficiency.

The integrated fiducial cross-sections in the two phase spaces are measured to be
\begin{description}
\item[2$b$-exclusive region] $\sigma_{\text{fid}} = 5.77 \pm 0.01\,\text{(stat.)}\,^{\,+0.27}_{-0.29}\,\text{(syst.)}\pm0.05\text{ (lumi.)}$ pb;
\item[2$b$-inclusive region] $\sigma_{\text{fid}} = 5.97 \pm 0.01\,\text{(stat.)}\,^{\,+0.27}_{-0.30}\text{ (syst.)}\pm0.05\text{ (lumi.)}$ pb.
\end{description}

As shown in \cref{tab:grouped_uncertainties_cross_section}, the measurements in both the $2b$-inclusive and exclusive phase spaces are limited by the uncertainties in the signal modelling and flavour-tagging efficiency. Other significant sources of uncertainties are related to the lepton isolation and jet energy scale.

\begin{table}[tb]
\centering
\renewcommand{\arraystretch}{1.3} %
\resizebox*{!}{!}{
\begin{tabular}{l c c}
\toprule
Phase space                 & \fidExclusive      & \fidInclusive\     \\
\midrule

Fiducial cross-section [pb] & $5.77 $            & $5.97 $            \\
\midrule
Total Uncertainty [\%]      & $^{\,+4.6}_{-5.2}$ & $^{\,+4.6}_{-5.1}$ \\
Statistical [\%]             & $\pm0.2$           & $\pm0.2$           \\
Systematic [\%]            & $^{\,+4.5}_{-5.2}$ & $^{\,+4.6}_{-5.1}$ \\
\midrule
Jets [\%]                   & $^{\,+1.6}_{-1.8}$ & $^{\,+1.8}_{-1.9}$ \\
Pile-up [\%]                & $\pm0.6$ & $^{\,+0.5}_{-0.6}$ \\
Flavour tagging [\%]        & $\pm2.9$ & $\pm3.0$ \\
Background [\%]             & $\pm$0.3           & $\pm$0.3           \\
Leptons and $\etmiss$ [\%]  & $\pm$1.8           & $\pm$1.8           \\
Luminosity [\%]             & $\pm$0.8           & $\pm$0.8           \\
$\ttbar$ modelling [\%]& $\pm$0.7 & $\pm$0.7\\
$tW$ modelling [\%]     & $\pm$0.1              & $\pm$0.1           \\
Generator parameters [\%]   & $^{\,+1.9}_{-2.9}$ & $^{\,+1.7}_{-2.7}$ \\
Parton shower [\%]          & $\pm$1.0           & $\pm$0.8           \\
Hard scattering [\%]        & $\pm$0.7           & $\pm$0.7           \\
\bottomrule
\end{tabular}
}
\caption{Summary of all the uncertainty sources for the total cross-section in the fiducial phase spaces. Only uncertainty classes contributing more than 0.1\% are shown.}
\label{tab:grouped_uncertainties_cross_section}
\end{table}

\begin{figure}[tb]
\centering
\begin{subfigure}[b]{0.48\textwidth}
\centering
\includegraphics[width=\linewidth]{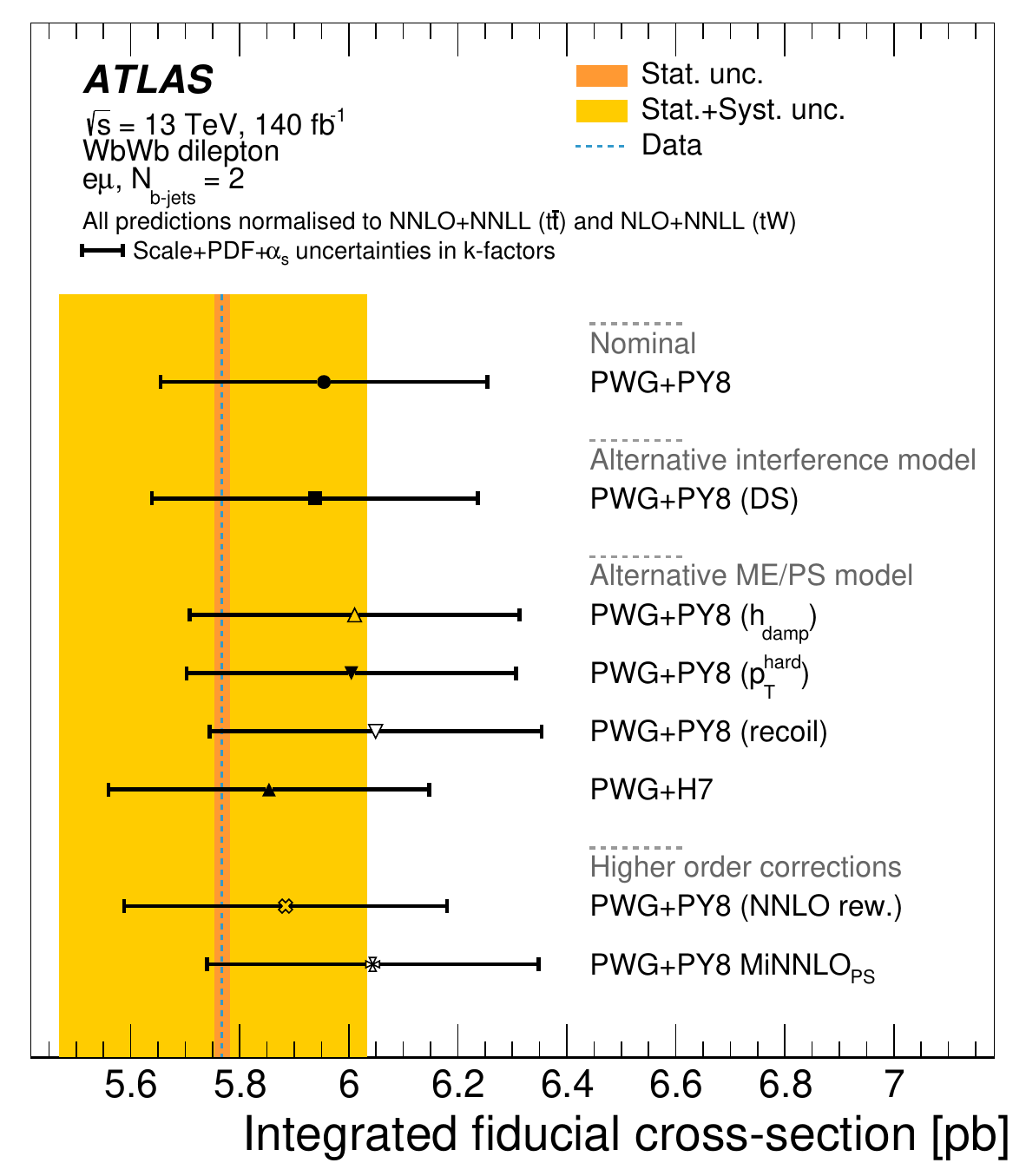}
\caption{}\label{fig:results:total_interference}
\end{subfigure}
\begin{subfigure}[b]{0.48\textwidth}
\centering
\includegraphics[width=\linewidth]{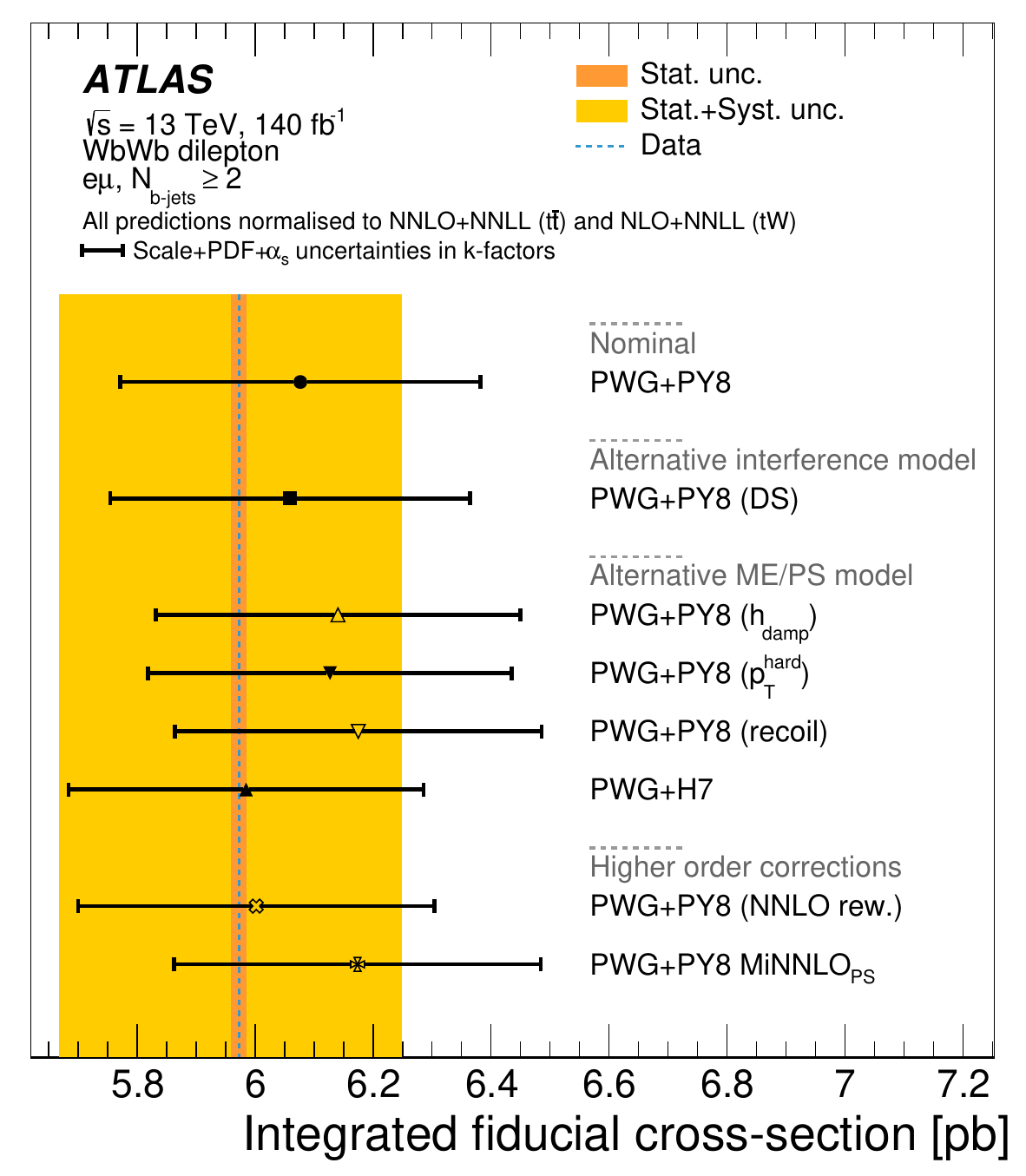}
\caption{}\label{fig:results:total_WbWb}
\end{subfigure}
\caption{Integrated fiducial cross-section measurement in \protect\subref{fig:results:total_interference} the 2$b$-exclusive and \protect\subref{fig:results:total_WbWb} the 2$b$-inclusive regions. The measured cross-sections are compared with theoretical predictions from different MC generators. All the $\ttbar$ and $tW$ theoretical predictions are normalised to the NNLO+NNLL and NLO+NNLL cross-sections respectively.}
\label{fig:results:total}
\end{figure}

The measured cross-sections are compared with theoretical predictions from different MC generators in \cref{fig:results:total}. All the $\ttbar$ and $tW$ theoretical predictions are normalised to the NNLO+NNLL and NLO+NNLL cross-sections respectively, as described in \cref{sec:datamc:mc_nominal_signal}. All differences among the predictions are thus due to how differently the MC generators model the acceptance of the fiducial phase spaces relative to the full phase space, which encompasses the entire kinematic region predicted by the theoretical models. The effect of the theoretical uncertainties, namely scale, PDF and \alphas, on the modelling of the fiducial acceptance is negligible and the dominant source of uncertainty in the predicted fiducial cross-section is due to the correction terms used to correct the NLO cross-sections from the generators to the higher-order predictions ($k$-factors). Since the PWG+PY8 $bb4l$ sample is normalised to the cross-section obtained from the Monte Carlo simulation, the predicted fiducial cross-sections, amounting to $5.40$~pb and $5.49$~pb in the 2$b$-exclusive and 2$b$-inclusive regions, are excluded from the comparison. All the predictions are compatible with the measurement within the quoted uncertainty.

\FloatBarrier


\FloatBarrier

%

%

%
\section{Conclusions}
\label{sec:conclusions}

This study presents the results of a differential cross-section measurement in the \wbwb\ final state using data collected by the ATLAS detector at the LHC at a centre-of-mass energy of $\sqrt{s} = \SI{13}{\TeV}$, corresponding to an integrated luminosity of \SI{140}{\ifb}. The analysis focuses on the fiducial phase spaces defined by requiring at least or exactly two $b$-jets and two leptons with opposite charges and different flavours.

The differential cross-section as a function of the variable most sensitive to the \ttbar\ and $tW$ interference, \mbl, is measured in the $2b$-exclusive region. The results are compared with different schemes to model the interference, with the DR approach showing the best agreement with the data, although it still fails to accurately describe the data in the extreme tail of the distribution. The sample simulating the full $bb4l$ final state shows larger differences in this region. However, a full assessment of this discrepancy's significance requires a complete prescription of theoretical uncertainties for the $bb4l$ model, including scale, PDF, parton-shower, and matching variation. This measurement provides critical input for refining future versions of the $bb4l$ generator as well as enabling improvements in the modelling of the \ttbar\ and $tW$ interference effects.

In the $2b$-inclusive region, the differential cross-sections are measured as a function of kinematic variables related to the leptons, jets, and combinations of the two $b$-jets ($bb$ system), the two $b$-jets and the two charged leptons ($bbll$ system), and the two $b$-jets and all charged and neutral leptons ($bb4l$ system). The results indicate that none of the considered NLO+PS predictions, including the $bb4l$ sample, is able to simultaneously describe all observables in the fiducial phase space. This finding is consistent with previous measurements in both the dilepton and semileptonic final states. The best-performing predictions are from the nominal \POWPY[8], in particular after the inclusion of higher-order corrections to the \ttbar\ sample via the NNLO reweighting and, to a minor extent, \minnlo, and \POWHER[7] models. However, it is important to emphasise that the \POWPY[8] predictions benefit from tuning to data, particularly for the $N_{\text{jets}}$ and $\pt(\ttbar)$ distributions, while the \POWPY[8] \minnlo\ and $bb4l$ predictions represent theoretical predictions without such tuning. Finally, models with alternative \hdamp\ and \pthard\ variations are only compatible with a single observable.

The integrated fiducial cross-sections are also measured in the $2b$-exclusive and $2b$-inclusive regions. The measured cross-sections are found to be $5.77 \pm 0.01\,\text{(stat.)}\,^{\,+0.27}_{-0.30}\,\text{(syst.)}$~pb and $5.97 \pm 0.01\,\text{(stat.)}\,^{\,+0.28}_{-0.31}\,\text{(syst.)}$~pb, respectively. These measurements are primarily limited by uncertainties in the signal modelling and flavour-tagging efficiency, followed by other sources such as lepton isolation and jet energy scale.

This measurement represents a significant improvement with respect to the previous publication~\cite{TOPQ-2017-05}, as the uncertainty is reduced by a factor of two, allowing for better discrimination between models. The sensitivity for most observables is currently limited primarily by uncertainties in signal modelling, including those related to the treatment of $\ttbar/tW$ interference, and by statistical uncertainties in the tails of the distributions. Further improvements are expected with the data collected during the LHC Run~3 and through the use of new MC simulations optimised based on these measurements.

In conclusion, the results of this study provide valuable insights into the modelling of the \wbwb\ final state in the fiducial phase space. The observed discrepancies between the measured cross-sections and theoretical predictions highlight the need for further improvements in the modelling of this process. While this analysis employs separate samples to model the \ttbar\ and $tW$ components of the $WbWb$ final state, it also provides a valuable opportunity to improve and tune the MC simulations that model the full $WbWb$ final state. Such improvements are crucial, as a full simulation of the $WbWb$ final state, with a complete set of systematic uncertainties, would allow for a more accurate disentanglement of the \ttbar\ and $tW$ contributions, thereby reducing the corresponding uncertainties.


%
\section*{Acknowledgements}
%

%
%

%
%

We thank CERN for the very successful operation of the LHC and its injectors, as well as the support staff at
CERN and at our institutions worldwide without whom ATLAS could not be operated efficiently.

The crucial computing support from all WLCG partners is acknowledged gratefully, in particular from CERN, the ATLAS Tier-1 facilities at TRIUMF/SFU (Canada), NDGF (Denmark, Norway, Sweden), CC-IN2P3 (France), KIT/GridKA (Germany), INFN-CNAF (Italy), NL-T1 (Netherlands), PIC (Spain), RAL (UK) and BNL (USA), the Tier-2 facilities worldwide and large non-WLCG resource providers. Major contributors of computing resources are listed in Ref.~\cite{ATL-SOFT-PUB-2025-001}.

We gratefully acknowledge the support of ANPCyT, Argentina; YerPhI, Armenia; ARC, Australia; BMWFW and FWF, Austria; ANAS, Azerbaijan; CNPq and FAPESP, Brazil; NSERC, NRC and CFI, Canada; CERN; ANID, Chile; CAS, MOST and NSFC, China; Minciencias, Colombia; MEYS CR, Czech Republic; DNRF and DNSRC, Denmark; IN2P3-CNRS and CEA-DRF/IRFU, France; SRNSFG, Georgia; BMFTR, HGF and MPG, Germany; GSRI, Greece; RGC and Hong Kong SAR, China; ICHEP and Academy of Sciences and Humanities, Israel; INFN, Italy; MEXT and JSPS, Japan; CNRST, Morocco; NWO, Netherlands; RCN, Norway; MNiSW, Poland; FCT, Portugal; MNE/IFA, Romania; MSTDI, Serbia; MSSR, Slovakia; ARIS and MVZI, Slovenia; DSI/NRF, South Africa; MICIU/AEI, Spain; SRC and Wallenberg Foundation, Sweden; SERI, SNSF and Cantons of Bern and Geneva, Switzerland; NSTC, Taipei; TENMAK, T\"urkiye; STFC/UKRI, United Kingdom; DOE and NSF, United States of America.

Individual groups and members have received support from BCKDF, CANARIE, CRC and DRAC, Canada; CERN-CZ, FORTE and PRIMUS, Czech Republic; COST, ERC, ERDF, Horizon 2020, ICSC-NextGenerationEU and Marie Sk{\l}odowska-Curie Actions, European Union; Investissements d'Avenir Labex, Investissements d'Avenir Idex and ANR, France; DFG and AvH Foundation, Germany; Herakleitos, Thales and Aristeia programmes co-financed by EU-ESF and the Greek NSRF, Greece; BSF-NSF and MINERVA, Israel; NCN and NAWA, Poland; La Caixa Banking Foundation, CERCA Programme Generalitat de Catalunya and PROMETEO and GenT Programmes Generalitat Valenciana, Spain; G\"{o}ran Gustafssons Stiftelse, Sweden; The Royal Society and Leverhulme Trust, United Kingdom.

In addition, individual members wish to acknowledge support from Armenia: Yerevan Physics Institute (FAPERJ); CERN: European Organization for Nuclear Research (CERN DOCT); Chile: Agencia Nacional de Investigaci\'on y Desarrollo (FONDECYT 1230812, FONDECYT 1240864, Fondecyt 3240661, Fondecyt Regular 1240721); China: Chinese Ministry of Science and Technology (MOST-2023YFA1605700, MOST-2023YFA1609300), National Natural Science Foundation of China (NSFC - 12175119, NSFC 12275265); Czech Republic: Czech Science Foundation (GACR - 24-11373S), Ministry of Education Youth and Sports (ERC-CZ-LL2327, FORTE CZ.02.01.01/00/22\_008/0004632), PRIMUS Research Programme (PRIMUS/21/SCI/017); EU: H2020 European Research Council (ERC - 101002463); European Union: European Research Council (BARD No. 101116429, ERC - 948254, ERC 101089007), European Regional Development Fund (SMASH COFUND 101081355, SLO ERDF), Horizon 2020 Framework Programme (MUCCA - CHIST-ERA-19-XAI-00), European Union, Future Artificial Intelligence Research (FAIR-NextGenerationEU PE00000013), Horizon 2020 (EuroHPC - EHPC-DEV-2024D11-051), Italian Center for High Performance Computing, Big Data and Quantum Computing (ICSC, NextGenerationEU); France: Agence Nationale de la Recherche (ANR-21-CE31-0022, ANR-22-EDIR-0002, ANR-24-CE31-0504-01); Germany: Baden-Württemberg Stiftung (BW Stiftung-Postdoc Eliteprogramme), Deutsche Forschungsgemeinschaft (DFG - 469666862, DFG - CR 312/5-2); China: Research Grants Council (GRF); Italy: Istituto Nazionale di Fisica Nucleare (ICSC, NextGenerationEU), Ministero dell'Università e della Ricerca (NextGenEU 153D23001490006 M4C2.1.1, NextGenEU I53D23000820006 M4C2.1.1, NextGenEU I53D23001490006 M4C2.1.1); Japan: Japan Society for the Promotion of Science (JSPS KAKENHI JP22H01227, JSPS KAKENHI JP22H04944, JSPS KAKENHI JP22KK0227, JSPS KAKENHI JP23KK0245, JSPS KAKENHI JP24K23939); Norway: Research Council of Norway (RCN-314472); Poland: Ministry of Science and Higher Education (IDUB AGH, POB8, D4 no 9722), Polish National Science Centre (NCN 2021/42/E/ST2/00350, NCN OPUS 2023/51/B/ST2/02507, NCN OPUS nr 2022/47/B/ST2/03059, NCN UMO-2019/34/E/ST2/00393, UMO-2022/47/O/ST2/00148, UMO-2023/49/B/ST2/04085, UMO-2023/51/B/ST2/00920, UMO-2024/53/N/ST2/00869); Portugal: Foundation for Science and Technology (FCT); Spain: Generalitat Valenciana (ASFAE/2022/008), Ministry of Science and Innovation (MCIN \& NextGenEU PCI2022-135018-2, MICIN \& FEDER PID2021-125273NB, RYC2019-028510-I, RYC2020-030254-I, RYC2021-031273-I, RYC2022-038164-I), Ministerio de Ciencia, Innovación y Universidades/Agencia Estatal de Investigaci\'on (PID2022-142604OB-C22); Sweden: Carl Trygger Foundation (Carl Trygger Foundation CTS 22:2312), Swedish Research Council (Swedish Research Council 2023-04654, VR 2021-03651, VR 2022-03845, VR 2022-04683, VR 2023-03403, VR 2024-05451), Knut and Alice Wallenberg Foundation (KAW 2018.0458, KAW 2022.0358, KAW 2023.0366); Switzerland: Swiss National Science Foundation (SNSF - PCEFP2\_194658); United Kingdom: Leverhulme Trust (Leverhulme Trust RPG-2020-004), Royal Society (NIF-R1-231091); United States of America: U.S. Department of Energy (ECA DE-AC02-76SF00515), Neubauer Family Foundation.

%
%


%
%
%
\printbibliography
\clearpage
 
\begin{flushleft}
\hypersetup{urlcolor=black}
{\Large The ATLAS Collaboration}

\bigskip

\AtlasOrcid[0000-0002-6665-4934]{G.~Aad}$^\textrm{\scriptsize 104}$,
\AtlasOrcid[0000-0001-7616-1554]{E.~Aakvaag}$^\textrm{\scriptsize 17}$,
\AtlasOrcid[0000-0002-5888-2734]{B.~Abbott}$^\textrm{\scriptsize 123}$,
\AtlasOrcid[0000-0002-0287-5869]{S.~Abdelhameed}$^\textrm{\scriptsize 119a}$,
\AtlasOrcid[0000-0002-1002-1652]{K.~Abeling}$^\textrm{\scriptsize 55}$,
\AtlasOrcid[0000-0001-5763-2760]{N.J.~Abicht}$^\textrm{\scriptsize 49}$,
\AtlasOrcid[0000-0002-8496-9294]{S.H.~Abidi}$^\textrm{\scriptsize 30}$,
\AtlasOrcid[0009-0003-6578-220X]{M.~Aboelela}$^\textrm{\scriptsize 45}$,
\AtlasOrcid[0000-0002-9987-2292]{A.~Aboulhorma}$^\textrm{\scriptsize 36e}$,
\AtlasOrcid[0000-0001-5329-6640]{H.~Abramowicz}$^\textrm{\scriptsize 157}$
% ...
\bigskip
\\

% ...
$^{17}$Department for Physics and Technology, University of Bergen, Bergen; Norway.\\
% ...
$^{30}$Physics Department, Brookhaven National Laboratory, Upton NY; United States of America.\\
% ...
$^{36}$$^{(a)}$Facult\'e des Sciences Ain Chock, Universit\'e Hassan II de Casablanca;$^{(b)}$Facult\'{e} des Sciences, Universit\'{e} Ibn-Tofail, K\'{e}nitra;$^{(c)}$Facult\'e des Sciences Semlalia, Universit\'e Cadi Ayyad, LPHEA-Marrakech;$^{(d)}$LPMR, Facult\'e des Sciences, Universit\'e Mohamed Premier, Oujda;$^{(e)}$Facult\'e des sciences, Universit\'e Mohammed V, Rabat;$^{(f)}$Institute of Applied Physics, Mohammed VI Polytechnic University, Ben Guerir; Morocco.\\
% ...
$^{45}$Physics Department, Southern Methodist University, Dallas TX; United States of America.\\
% ...
$^{49}$Fakult\"{a}t Physik, Technische Universit{\"a}t Dortmund, Dortmund; Germany.\\
% ...
$^{55}$II. Physikalisches Institut, Georg-August-Universit\"{a}t G\"ottingen, G\"ottingen; Germany.\\
% ...
$^{104}$CPPM, Aix-Marseille Universit\'e, CNRS/IN2P3, Marseille; France.\\
% ...
$^{119}$$^{(a)}$New York University Abu Dhabi, Abu Dhabi;$^{(b)}$United Arab Emirates University, Al Ain; United Arab Emirates.\\
% ...
$^{123}$Homer L. Dodge Department of Physics and Astronomy, University of Oklahoma, Norman OK; United States of America.\\
% ...
$^{157}$Raymond and Beverly Sackler School of Physics and Astronomy, Tel Aviv University, Tel Aviv; Israel.\\
% ...
$^{a}$ Also at Affiliated with an institute formerly covered by a cooperation agreement with CERN.\\
$^{b}$ Also at An-Najah National University, Nablus; Palestine.\\
$^{c}$ Also at Borough of Manhattan Community College, City University of New York, New York NY; United States of America.\\
$^{d}$ Also at Center for Interdisciplinary Research and Innovation (CIRI-AUTH), Thessaloniki; Greece.\\
$^{e}$ Also at Centre of Physics of the Universities of Minho and Porto (CF-UM-UP); Portugal.\\
$^{f}$ Also at CERN, Geneva; Switzerland.\\
% ...

\end{flushleft}


%

%
%

%

%

\end{document}